\documentclass[preprint2]{aastex}

\usepackage{etex,morefloats}
\usepackage{graphicx,amssymb,amsmath,chemarr}
\usepackage{changepage}
\usepackage{natbib,hyperref}
\usepackage{color}

\shorttitle{A Chemical Network for Lightning \& Life}
\shortauthors{Rimmer \& Helling}

\begin{document}

\title{A Chemical Kinetics Network for Lightning and Life in Planetary Atmospheres}

\author{P.~B. Rimmer\altaffilmark{1} and Ch Helling}
\affil{School of Physics and Astronomy, University of St Andrews,
    St Andrews, KY16 9SS, United Kingdom}

\altaffiltext{1}{pr33@st-andrews.ac.uk}

\begin{abstract}
There are many open questions about prebiotic chemistry in both planetary and exoplanetary
environments.
The increasing 
number of known exoplanets and other ultra-cool, substellar objects has propelled the desire to 
detect life and~prebiotic chemistry outside the solar system. We present an ion-neutral chemical 
network constructed from scratch, \textsc{Stand2015}, that treats hydrogen, nitrogen, carbon 
and oxygen chemistry accurately within a temperature range between 100 K and 30000 K. Formation
pathways for glycine and other organic molecules are included. The network is complete up to 
H6C2N2O3. \textsc{Stand2015} is successfully tested against atmospheric chemistry models for 
\object{HD209458b}, \object{Jupiter} and the present-day \object{Earth} using a simple 1D 
photochemistry/diffusion code. Our results for the early \object{Earth} agree with those of
\citet{Kasting1993} for CO$_2$, H$_2$, CO and O$_2$, but do not agree for water and atomic 
oxygen. We use the network to simulate an experiment where varied chemical 
initial conditions are irradiated by UV light. The result from our simulation is that
more glycine is produced when more ammonia and methane is present. Very
little glycine is produced in the absence of any molecular nitrogen and oxygen. This suggests
that production of glycine is inhibited if a gas is too strongly reducing. Possible applications
and limitations of the chemical kinetics network are also discussed.
\end{abstract}


\section{Introduction}

The potential connection between a focused source of energy and life was first made apparent in 
the Miller-Urey 
experiment \citep{Miller1953}, set to test a hypothesis proposed by \citet{Haldane1928}. 
In this experiment, a gas composed of water vapor, ammonia, methane and molecular hydrogen
was circulated past an electric discharge. After a week's time, various biologically relevant
chemicals had developed, including glycine and alanine, identified with a paper chromatrogram.
A follow-up study of Miller's samples, carried out approximately fifty years later, discovered
a much richer variety of prebiotic compounds than originally thought \citep{Johnson2008}.
Since then, numerous related experiments have been carried out under a variety of conditions
\citep[see][and references therein]{Miller1959,Cleaves2008}.

The input energy source and the initial chemistry have been varied across these different
experiments. An energy source may have been important for the production of prebiotic
species on Earth, because the pathways to formation have considerable activation barriers, often on 
the order of 0.1-1 eV. \citet{Patel2015} generated prebiotic species by exposing HCN and H$_2$S 
to ultraviolet light. The experimental results from \citet{Powner2009} suggest that the 
aqueous synthesis of amino acids, nucleobases and ribose is predisposed, starting from glyceraldehyde
and glycoaldehyde, which they suggest would most likely form through heating and UV irradiation.
Shock synthesis of amino acids due to the atmospheric entry of cometary meteors and micrometeorites
or thunder is also sufficient to overcome these barriers and produce amino acids \citep{Bar-Nun1970}.

The initial chemical conditions are naturally significant to the formation of prebiotic chemistry. 
Of course,
in an environment where hydrogen or carbon were lacking, there would be no complex hydrocarbons.
Nitrogen and phosphorus are also essential to the origins of terrestrial life, although some
scientists, such as \citet{Benner2004}, have speculated that life could occur under very different
chemistries; presently, we lack the ability to explore this possibility. The initial chemical
composition also has an effect on the production of prebiotic chemical species. For example,
hydrogen can be bound in a reducing species, CH$_4$, in an oxidizing species, H$_2$SO$_4$, or
into the neutral species of water (H$_2$O). Both \citet{Schlesinger1983} and \citet{Miyakawa2002}
have found that performing a Miller-Urey like experiment in an oxidizing environment produces
only trace amounts of prebiotic materials, whereas performing the experiment in a reducing
environment produces a great number of prebiotic materials.

The atmosphere of \object{Earth} in its present state is oxidizing ($\approx 21$\%
O$_2$, 78 \% N$_2$). The atmosphere of the \object{Earth} during its first billion years (first
1 Gyr) would have had a very different composition, probably oxidizing or at least
only weakly reducing \citep{Kasting1993}, although \citet{Tian2005} suggest that the Earth's 
atmosphere was once highly reducing. Even if the Earth never possessed a strongly reducing atmosphere,
other planets and moons are known to have both reducing atmospheres and active lighting and UV
photochemistry, 
\object{Jupiter} for example. Extrasolar planets may not simply have diverse compositions,
but also widely varied gas-phase C/O ratios, either intrinsically at formation, as may be the case
with \object{Wasp-12b}, \object{XO-1b}, \object{CoRoT-2b} \citep{Madhu2011,Moses2013}, and possibly
the interior of \object{55 Cancri e} (\citealt{Madhu2012}, but see also \citealt{Nissen2013}); or 
alternatively due to oxygen depletion
into the cloud particles \citep{Bilger2013,Helling2014}. The question of the C/O ratio is not
a settled matter \citep{Benneke2015}.

These diverse planetary and exoplanetary environments provide unique ``laboratories'' within
which to explore prebiotic chemistry. There are many potential drivers for prebiotic
chemistry in planets and exoplanets, from the steep thermal gradients in hot Jupiters and close-in
Super-Earths to the thermal production of organics and complex hydrocarbons in Saturn's storms 
\citep{Moses2015} and photochemical production of complex organics in Titan \citep{Yung1984,Loison2015}.
There is some evidence that cosmic rays drive the formation of hydrogen cyanide in Neptune \citep{Lellouch1994}.
\citet{Molina1999} have proposed pathways to formation of a rich variety of nitriles via cosmic rays in
Titan's atmosphere.

As mentioned above, electric discharges may also be an important source of energy driving
the production of prebiotic species, and are ubiquitous throughout the gas giants.
Discharges in the form of lightning are known to 
occur within our solar system, on \object{Earth}, \object{Jupiter} \citep{Little1999}, 
\object{Saturn} \citep{Dyudina2007}, \object{Uranus} \citep{Zarka1986}, and 
\object{Neptune} \citep{Gurnett1990}. There are some indications of lightning discharges on
\object{Venus} \citep{Taylor1979}, and possibly also in Titan's nitrogen chemistry 
\citep{Borucki1984}, although these traces are still tentative. Lightning is hypothesized to occur 
on exoplanets \citep{Aplin2013,Helling2013b} and brown dwarfs \citep{Helling2013b,Bailey2014}.
Simulated plasma discharges
initiated within Jupiter-like gas compositions suggest that lightning on Jupiter may produce a 
significant amount of trace gases \citep{Borucki1985}.
The comparison between experimental 
rates of production of organic compounds in high-temperature plasmas to chemical equilibrium 
models is unsurprisingly poor \citep{Scattergood1989}, and indicates that a chemical kinetics 
approach will be important in explaining the results of these experiments, Chemical
kinetics seems to be necessary for exploring any of these pathways to the formation of
prebiotic species.
	 
Chemical kinetics models have been applied to planetary and exoplanetary atmospheric conditions
in such a diverse range that it is impractical to provide complete references, so a brief summary
of the work will instead be provided. Photochemical models of the modern Earth have been applied
in the context of 1D models \citep{Owens1985}, up to fully coupled 3D general 
circulation models \citep{Roble1994}, and even within a flexible modular framework that can be
included as a module within other codes
\citep{Sander2005}. The Earth's atmosphere during its first billion years has been extensively 
modeled \citep{Kasting1993,Zahnle1986}. Chemical kinetics models have been applied also to 
\object{Jupiter}'s atmosphere, from the deep atmosphere \citep{Fegley1994,Visscher2010} through the 
stratosphere \citep{Zahnle1995,Moses2005}. The atmosphere of the moon Titan has also been analyzed
using ion-neutral chemical kinetics to better explain the abundance of rich hydrocarbons
in its atmosphere and its stratospheric haze \citep{Yung1984,Keller1998,Lavvas2008a,Lavvas2008b}.

Chemical kinetics models for exoplanetary atmospheres have been typically developed for hot Jupiters,
especially \object{HD189733b} and \object{HD209458b} \citep{Moses2011,Venot2012, Zahnle2009}. 
Almost all of the models for hot Jupiters have been applied only in two dimensions, and so have not 
taken a more complete account of the atmospheric dynamics, instead relying on a
parameterization of vertical mixing using the eddy diffusion coefficient, 
$K_{zz}$ [cm$^2$ s$^{-1}$] \citep[see][their Sec. 4.2]{Lee2015}. \citet{Agundez2014} have taken on 
the ambitious task of coupling a 
chemical kinetics model to 2D dynamics for both \object{HD189733b} and \object{HD209458b}. 
Ion-neutral models have been applied to exoplanets, taking account of photochemistry 
\citep{Lavvas2014}, and 
additionally of cosmic ray ionization \citep{Walsh2011,Rimmer2014}. Chemical
kinetics models have also been 
applied to the extrasolar super-earths \citep{Hu2012,Hu2013,Hu2014}, and have been used to explore 
possible biosignatures on rocky planets \citep{Seager2013a,Seager2013b}. There has also been some
recent investigation into chemistry on helium dominated exoplanets \citep{Hu2015}.

Lightning chemistry has been explored with some basic chemical kinetics models,
e.g. within Earth's mesosphere (\citealt{Luque2009} and \citealt{Parra2013}) and Saturn's
lower ionosphere \citep{Dubrovin2014}. \citet{Dubrovin2014} present interesting
results for Saturn's lower ionosphere, predicting that TLE's within this
region would produce mostly H$_3^+$, what they identify as the primary positive charge carrier
during the duration of the TLE and for sometime after. This would mimic the effect of cosmic ray
ionization. \citet{Parra2013} presented similar results involving terrestrial nitrogen chemistry.
The products of discharge chemistry in the upper part of both hydrogen-rich and nitrogen-rich 
atmospheres 
seem to be similar to the products of cosmic ray chemistry in these same atmospheres.

There are many open questions about prebiotic chemistry in diverse planetary and exoplanetary 
environments, 
as well as in the lab. In this paper, we present a candidate network for exploring UV photochemistry,
cosmic ray chemistry and lightning-driven chemistry, constructed from scratch. We will explore mostly the 
photochemistry
and thermochemistry within this paper, leaving the exploration of lightning-driven chemistry and cosmic ray
chemistry to future work.

The largest task in developing this network has been the collation of a full set of chemical 
reactions that treat both reducing and oxidizing chemistries at temperatures ranging from 100 K
through 30000 K (the approximate peak temperature of lightning, see \citealt{Orville1968,Price1997})
and the selection of rate constants when more than one is published.
Since one interest is the investigation of the formation rate of prebiotic species in diverse
environments, the network is made extensive enough to include the simplest amino acid, glycine.
In this paper, we present this chemical network (\textsc{Stand2015}), and test in a diversity of 
environments. For these tests, we developed a simple 1D photochemistry/diffusion code (\textsc{Argo}).
\textsc{Argo} was developed based on \textsc{Nahoon} \citep{Wakelam2012} by including 
wavelength-dependent photochemistry, cosmic ray transport, water condensation and chemical mixing.

The \textsc{Stand2015} network is presented in Section \ref{sec:Network}. We compare the predictions 
of our network using a simplified 1D photochemistry/diffusion code called \textsc{Argo} 
(Section \ref{sec:model}). The model and network are then combined and tested against other model 
results for \object{HD209458b} and the early \object{Earth}, and compared to
observation for \object{Jupiter} and the present-day Earth in Section \ref{sec:test-cases}.
Finally, in Section \ref{sec:Laboratory} we simulate a Miller-Urey type experiment and 
explore the formation of glycine under various chemical conditions. Section \ref{sec:conclusion} 
contains a short discussion of the results and possible future applications of this model.

\section{The Chemical Network}
\label{sec:Network}

The \textsc{Stand2015} Atmospheric Chemical Network is an H/C/N/O network with 
reactions involving He, Na, Mg, Si, Cl, Ar, K, Ti and Fe, developed from scratch. 
It contains all known reactions for species of up to 6 hydrogen, 2 carbon, 2 nitrogen and 3 oxygen 
atoms, for which a rate constant has been published, as well as a less complete network involving
species with 3+ carbon atoms, 3 nitrogen atoms and/or 4 oxygen atoms. 
A chemical network is effectively a list
of chemical reactions and reaction rate constants. Rate constants are used to calculate the rates 
of production and loss of a particular molecular or ionic species, $P_i$ [cm$^{-3}$ s$^{-1}$] and
$L_i$ [cm$^{-3}$ s$^{-1}$] respectively, and $i$ is enumerated over the list of species. Rate 
constants are of zeroth order (e.g., source terms, $S_i$ [cm$^{-3}$ s$^{-1}$]), first order 
(involving interactions with particles not accounted in the network, such as photons or cosmic rays, 
$k_1$ [s$^{-1}$]), second order (collisions between particle $i$ and other particles within the 
network, $k_2$ [cm$^3$ s$^{-1}$]), or third order (collisions between particle $i$ and other particles, as well as a third body, denoted here as $k_3$ [cm$^6$ s$^{-1}$]). The rates of production and 
loss for a given species, $i$, in terms of rate constants, are generally:
\begin{align}
P_i &= S_i + \sum k_1 \, n_j + \sum k_2 \, n_jn_k + \sum k_3 \, n_{\rm gas} n_jn_k, \label{eqn:production}\\
L_i &= \sum k_1 \, n_i + \sum k_2 \, n_jn_i + \sum k_3 \, n_{\rm gas}n_jn_i. \label{eqn:destruction}
\end{align}
Summation is over all the relevant reactions, some involving species $j$ and/or $k$, that result in the 
production (Eq. (\ref{eqn:production})) or loss (Eq. (\ref{eqn:destruction})) of species $i$. The symbol 
$n_i$ [cm$^{-3}$] denotes the number density of species $i$ and $n_{\rm gas}$ [cm$^{-3}$] denotes the total 
gas number density.

The reaction rate constants have been assembled from various databases. With only a couple 
hundred exceptions, the rate constants for 2-body and 3-body neutral reactions have been assembled
from the NIST Chemical Kinetics Database \citep{NIST2013}. Virtually all of the ion-neutral 
reactions were taken from \citet{Ikezoe1987}. Several rate constants that we have used, relevant 
for terrestrial atmospheric chemistry, are taken from \citet{Sander2011}. The KIDA database provided the rate constants for several dissociative recombination reactions \citep{Wakelam2012}. Coefficients
for the cosmic ray ionization rate constant were taken from the OSU chemical 
network \citep{Harada2010}.

Rate constants were compared to the publicly available networks of 
\citet{Moses2011,Venot2012}, and ion-neutral rate coefficients were checked against the KIDA database
\citep{Wakelam2012}\footnote{\url{http://kida.obs.u-bordeaux1.fr/}}, as well as the OSU 09 2010 
high temperature network 
\citep{Harada2010}\footnote{\url{http://faculty.virginia.edu/ericherb/research.html}}. Some 
further ion-neutral reactions involving the alkali ion 
chemistry were appropriated from \citet{Lavvas2014}. Finally, $\sim 20$ more reactions 
for suspected formation pathways for glycine have been added to the network, from 
\citet{Blago2003,Patel2015}. The full network and references are provided in 
Appendix \ref{app:network}. The following subsections contain brief discussions about the different 
classes of reactions, their rate coefficients and whether reverse reactions have been included.

\subsection{2-Body Neutral-Neutral and Ion-Neutral Reactions}
\label{sec:2-Body}

Two-body neutral-neutral and ion-neutral reactions follow the basic scheme:
\begin{align}
{\rm A + B} &\rightarrow {\rm Y + Z}, \; {\rm and} \\
{\rm A^+ + B} &\rightarrow {\rm Y^+ + Z}.
\end{align}
The rate constants for these reactions are approximated by the Kooij equation \citep{Kooij1893}:
\begin{equation}
k_2 = \alpha \Big(\dfrac{T}{300 \, K}\Big)^{\beta} \, {\rm e}^{-\gamma/T},
\label{eqn:kooij}
\end{equation}
where $T$ [K] is the gas temperature\footnote{Surface chemistry is not considered in this paper,
and the temperature of all chemical species including electrons is set equal to the gas-phase
temperature.}, 
$k_2$ [cm$^3$ s$^{-1}$] is the rate constant, and 
$\alpha$ [cm$^3$ s$^{-1}$], $\beta$ and $\gamma$ are constants characterizing the reaction. All
of these reactions are reversed in our network and we use the rate coefficients for the best 
characterized direction for each reaction, which is typically the exothermic direction. For
neutral-neutral reactions, even when exothermic, there is often a sizable barrier to reaction,
allowing certain elements to be locked into non-equilibrium configurations at low temperatures
effectively for eternity, because the barrier to the lower energy state is too large to be
overcome in the current environment.

Ion-neutral reactions do not typically have barriers in the exothermic direction, and in
many cases the rate constants are altogether temperature independent, closely approximating the 
Langevin approximation. A notable exception are charge exchange reactions,
\begin{equation}	 
{\rm A^+ + B} \rightarrow {\rm B^+ + A},
\label{eqn:charge-exchange}
\end{equation}
which, due to the differences in energy between ionic and neutral ground states, often contains
barriers on the order of a few x 100 K.

The rate constants for the forward reactions are given in Appendix \ref{app:network} with the label 
`2n', reactions 577-1352. These reactions are reversed following the scheme described in 
Appendix \ref{app:reverse}. The ion-neutral reactions are also reversed, are listed in 
Appendix \ref{app:network} with `2i', reactions 1353-2569.

\subsection{3-Body Neutral Reactions, Dissociation Reactions, and Radiative Association Reactions}
\label{sec:3-Body}

Reactions that involve a third body occur primarily in the two forms:
\begin{align}
{\rm A + M} &\rightarrow {\rm Y + Z + M}, \label{eqn:decomp}\\
{\rm A + B + M} &\rightarrow {\rm Z + M}, \label{eqn:combin}
\end{align}
where M represents any third body. Decomposition reactions are well studied at high temperatures,
being important for various combustion processes. Just as in Section \ref{sec:2-Body}, we choose
the reactions best characterized, which in this case often involve endothermic reactions. The
rate coefficients for the majority of these reactions follow the Lindemann 
form \citep{Lindemann1922}. In this form, we first determine the rate constants in the 
low-pressure ($k_0$ [cm$^6$ s$^{-1}$]) and high-pressure ($k_{\infty}$ [cm$^3$ s$^{-1}$]) limits:
\begin{align}
k_0 &= \alpha_0 \Big(\dfrac{T}{300 \, K}\Big)^{\beta_0} \, {\rm e}^{-\gamma_0/T}, \\
k_{\infty} &= \alpha_{\infty} \Big(\dfrac{T}{300 \, K}\Big)^{\beta_{\infty}} \, {\rm e}^{-\gamma_{\infty}/T}.
\end{align}
These are combined with the number density of the neutral third species, [M] [cm$^{-3}$] to determine
the reduced pressure, $p_r = k_0[{\rm M}]/k_{\infty}$, and this can then be utilized to set the
pressure-dependent effective ``two-body'' rate:
\begin{equation}
k_2 = \dfrac{k_{\infty} p_r}{1 + p_r}.
\label{eqn:3-Body}
\end{equation}
Sometimes this expression is multiplied by a dimensionless function $F(p,T)$ to more accurately 
approximate the transition between the low-pressure and high-pressure limits, and this provides
the Troe form \citep{Troe1983}. The coefficients for the Troe form are not explicitly
given.

We favor using the rate constants for three-body combination reactions, and
reversing these reaction to determine the rate of thermal decomposition. In many cases, however, the 
rate constants are unavailable. When we have only the rate coefficients for the
decomposition reactions, we add an additional 500 K barrier to both the decomposition and three
body combination rate constants. This barrier is added in order to limit run-away three body
reactions that can result from reversing decomposition reactions at low temperature.

Additionally, we incorporate a small number of radiative association reactions, of the form:
\begin{equation}
{\rm A + B} \rightarrow {\rm Z + \gamma},
\end{equation}
where $\gamma$ is the radiated photon that carries the excess energy from the association.
We appropriate the Kooij form for this reaction, as with 2-Body Neutral-Neutral reactions, in order
to determine the rate constant $k_{\rm ra}$ [cm$^3$ s$^{-1}$]. We then apply this rate constant,
along with the rate constant for the corresponding three-body reaction, to the adduct form of the 
overall rate constant \citep[][their eq. (B.2)]{Hebrard2013}:
\begin{equation}
k = \dfrac{\big(k_0{\rm [M]}F + k_r\big)k_{\infty}}{k_0{\rm [M]} + k_{\infty}},
\end{equation}
where the function $F$ is from the Troe form of the transition from high to low 
pressure.

The rate constants for the forward reactions are given in Appendix \ref{app:network} with the
labels `2d' for the neutral species and `3i' for ion-neutral species. These reactions are
reversed in the manner described by Appendix \ref{app:reverse}. Reactions 1-420 are reactions of
this type, for which each odd-numbered reaction gives the low-pressure rate constant 
$k_0$ [cm$^6$ s$^{-1}$] and each even-numbered reaction gives the high-pressure rate constant
$k_{\infty}$ [cm$^3$ s$^{-1}$]. Reactions labeled 'ra' are radiative association reactions, numbered 2974-2980.

\subsection{Thermal Ionization \& Recombination Reactions}
\label{sec:thermal-ionization}

A special set of three-body reactions are thermal ionization and three-body recombination
reactions, which proceed by the pair of equations (analogous to Eq's (\ref{eqn:decomp}) and 
(\ref{eqn:combin})):
\begin{align}
{\rm A + M} &\rightarrow {\rm Z^+} + e^- + {\rm M}, \label{eqn:ionize}\\
{\rm A^+} + e^- + {\rm M} &\rightarrow {\rm Z + M}. \label{eqn:recomb}
\end{align}
For which we again use published rates wherever possible for the ionization reactions, (Eq. (\ref{eqn:ionize})), 
but in many cases here use the simple approximation:
\begin{equation}
k_0 = \Bigg(\dfrac{8\pi e^8}{m_ek_BT}\Bigg)^{\!\!1/2}{\rm e}^{-I/k_BT},
\label{eqn:ion-constant}
\end{equation}
where $e = 4.9032 \times 10^{-10}$ esu is the elementary charge, $k_B = 1.38065 \times 10^{-16}$ 
erg/K is the Boltzmann constant, $m_e = 9.1084 \times 10^{-28}$ g is the mass of the electron, 
$I$ is the ionization energy (here in units of erg) which we determine from the change in the Gibbs 
free energy for the reaction. $k_{\infty}$ is then estimated from $k_0$.

Three-body recombination and ionization reactions have been well studied, and in many cases
have well-characterized rate constants. Here we treat the three body recombinations as the reverse 
reactions for the collisional ionization reactions, but the studied
rate coefficients for these reactions generally have a temperature dependence of $T^{-4.5}$, 
at least for $T > 1$ K \citep{Hahn1997}. This creates a problem for reversibility. Using these rates 
will not allow us to
reproduce chemical equilibrium for plasmas and this is largely because we are not properly 
treating the time-dependent plasma conditions in which these rates are often measured.
Many of these rate constants may accurately describe the time to achieve an equilibrium electron 
density in a regime where a strong ionizing source has recently been removed from the environment.

With this in mind, we instead set the recombination rate constants such that, when dissociative
recombination reactions are disabled, the Saha equation is upheld.

These reactions and rate coefficients are also given in 
Appendix \ref{app:network}. The ionization reactions are labeled `ti' and numbered 421-576. 
As with Section \ref{sec:3-Body}, the odd reactions are $k_0$ [cm$^6$ s$^{-1}$] and the even 
numbers are $k_{\infty}$ [cm$^3$ s$^{-1}$].

Finally, we include a series of dissociative recombination reactions, which take the form:
\begin{equation}
{\rm A^+} + e^- \rightarrow {\rm Y + Z}.
\label{eqn:dissoc-recomb-constant}
\end{equation}
These have rate constants parameterized in the form of Eq. (\ref{eqn:kooij}). The reverse reactions 
can in principle be calculated, and their rate constants could be calculated straight-forwardly
using the same principles used for the three-body reactions. This would effectively be analogous
to the rates of three body recombination for any third body, and we do not find that reversing
these reactions changes the results much. When we compare with chemical equilibrium, however,
we disable these reactions. The dissociative recombination reactions are taken only from the
OSU 09 2010 high temperature network \citep{Harada2010}, and shown in Appendix \ref{app:network}, 
numbered 2777-2973, and labeled `dr'.

\subsection{Photochemistry and Cosmic Ray Chemistry}
\label{sec:photochem}

Photochemistry is considered for the species H, H$^-$, He, C, C($^1$D), C($^1$S), N, 
O, O($^1$D), O($^1$S), H$^-$, C$_2$, CH, CN, CO, H$_2$, N$_2$, NO, O$_2$, OH, CO$_2$, H$_2$O, 
HO$_2$, HCN, NH$_2$, NO$_2$, O$_3$, C$_2$H$_2$, H$_2$CO, H$_2$O$_2$, NH$_3$, NO$_3$, CH$_4$, HCOOH, 
HNO$_3$, N$_2$O$_3$, C$_2$H$_4$, C$_2$H$_6$, CH$_3$CHO, C$_4$H$_2$, C$_4$H$_4$, Na, K, and HCl. The photoionization and 
photodissociation cross-sections are taken almost entirely from 
\textsc{PhIDRates}\footnote{\url{phidrates.space.swri.edu}} \citep{Huebner1979,Huebner1992,Huebner2015},
with the exception of C$_4$H$_2$, C$_4$H$_4$ and N$_2$O$_3$, the cross-sections of which are taken from the 
MPI-Mainz UV/VIS Spectral Atlas\footnote{\url{http://satellite.mpic.de/spectral_atlas/index.html}} 
\citep{Keller2013}.

\begin{figure}
\epsscale{1.0}
\plotone{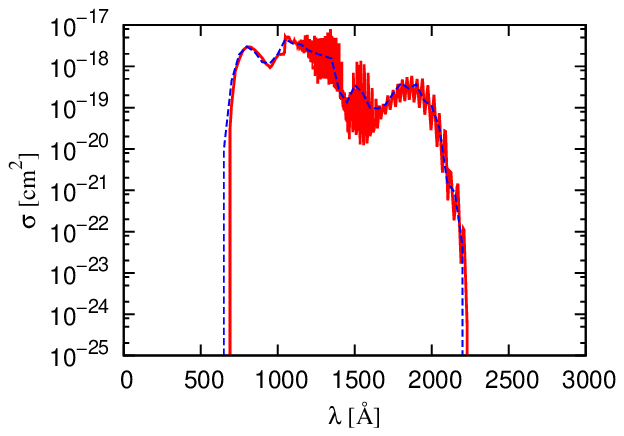}
\caption{The photodissociation cross sections of NH$_3$ $\rightarrow$ $^1$NH + H$_2$, 
$\sigma$ [cm$^2$], as a function of wavelength, $\lambda$ (\AA), from \textsc{PhIDRates} 
\citep[original data from][red line]{McNesby1962,Schurath1969}. The data is compared to our binned
fit (blue line).\label{fig:sigma}}
\end{figure}

We divide the cross-sections between 200 bins each $\approx 50$ \AA\ wide. A comparison 
between our binned cross-sections and the raw cross-sections from \textsc{PhIDRates} is plotted 
for an example reaction (Figure \ref{fig:sigma}). The cross-sections, both in the database and here,
of the form $\sigma(\lambda)$ with $\sigma$ in units cm$^2$ and wavelength in units of \AA. 
The resolution for the UV cross-sections is fairly low, and cannot encapsulate
the fine structure of the UV emission lines or the
UV cross-sections. This is especially important when treating ionospheres of gas giants, since, e.g.
the fine structure in the H$_2$ bands leave small spectral windows through which photons can penetrate
and effectively ionize deeper in the atmosphere. Such a low resolution spectrum will effectively
close these windows and underestimate the ion production in the ionosphere \citep{Kim1994,Kim2014}.
High resolution is also a important for capturing where the UV flux and cross sections both peak; a low 
resolution
cross section can in this case underestimate the destruction rate of the species with this resonant
photochemical cross-section. As can be seen below, these issues do not significantly affect the comparisons
of this model for HD209458b, Jupiter or Earth. For photoionization deep in the atmosphere, 
where high resolution is essential, the network itself need not be modified. The transport of UV photons 
line by line would need to be calculated.

The tabulated chemical cross-sections are combined with $F(\lambda,z)$ [photons cm$^{-2}$ s$^{-1}$ 
\AA$^{-1}$], the radiant flux density onto a unit sphere (hereafter called the actinic flux)
located at atmospheric height, $z$ [cm], to determine the photochemical rate constants,
\begin{equation}
k_{{\rm ph},i}(z) = \tau_f\int_{1 \AA}^{10^4 \AA} \sigma_i \, F(\lambda, z) \; d\lambda,
\label{eqn:photo-rates}
\end{equation}
where $i$ is indexed over the molecules listed above, for which photochemistry is considered.
$\tau_f$ is a dimensionless parameter representing the fraction of time (over a period much
longer than the longest characteristic time scale for the atmosphere) the particular atmospheric
region is irradiated; for tidally locked planets, $\tau_f = 1$ (dayside) or $0$ (nightside), 
the diurnal average for a rotating planet is $\tau_f = 1/2$. The photoionization and 
photodissociation reactions are listed in Appendix \ref{app:network}, reactions 
numbered 2570-2693, and labeled `pi' for photoionization reactions \& `pd' for 
photodissociation reactions.

Cosmic ray ionization and dissociation is parameterized by $\zeta$ \citep{Rimmer2013}, to treat 
both direct ionization by galactic cosmic
rays and ionization by secondary particles produced in air showers. The cosmic ray ionization
rate depends on the chemical species in question, since different species will have different 
chemical cross-sections for the photons produced by cosmic rays, and this is accounted for by 
multiplying $\zeta(z)$ by a constant $\kappa_{{\rm CR},i}$ such that:
\begin{equation}
k_{{\rm CR},i}(z) = \kappa_{{\rm CR},i} \zeta(z).
\label{eqn:cosmic-ray}
\end{equation}
We treat low energy cosmic rays ($E < 1$ GeV) for these objects as though they have
been significantly shielded by the
astrospheres of the host stars, and therefore set the fitting parameters for the incident cosmic
ray flux to $\alpha = 0.1$ and $\gamma = -1.3$ in the equation for the flux of cosmic ray
particles:
\begin{equation}
 j(E) =
\begin{cases}
  j(E_1) \bigg( \dfrac{p(E)}{p(E_1)} \bigg)^{\gamma}, & \text{if } E > E_2 \\
  j(E_1) \bigg( \dfrac{p(E_2)}{p(E_1)} \bigg)^{\gamma} \; \bigg( \dfrac{p(E)}{p(E_2)} \bigg)^{\alpha}, & \text{if } E_{\rm cut} < E < E_2 \\
  0, & \text{if } E < E_{\rm cut}
\end{cases}
\label{eqn:init-flux}
\end{equation}
where $p(E) = \frac{1}{c}\sqrt{E^2 + 2EE_0}$, $E_0 = 9.38 \times 10^8$ eV, $E_1 = 10^9$ eV,
and $E_2 = 2 \times 10^8$ eV, and the flux at $E_1$ is set to $j(E_1) = 0.22$ cm$^{-2}$ s$^{-1}$
sr$^{-1}$ (GeV/nucleon)$^{-1}$. All of these parameters except $\alpha$ are observationally
well-constrained \citep{Indriolo2009}. For a demonstration of how $\alpha$ affects the cosmic
ray spectrum, and a discussion of the Monte Carlo transport we use for cosmic rays of energy $< 1$ GeV,
see \citet{Rimmer2012,Rimmer2013}. For ionization rate by cosmic rays of energy $> 1$ GeV,
$Q_{\rm HECR}$ [cm$^{-3}$ s$^{-1}$], we use the
analytical method of \citet{Velinov2008}.

Cosmic ray reactions are listed in Appendix \ref{app:network}, numbered 2694-2776, and labeled
`cr'.

\subsection{Test for Chemical Equilibrium}
\label{sec:tce}

\begin{figure*}
\epsscale{2.0}
\plotone{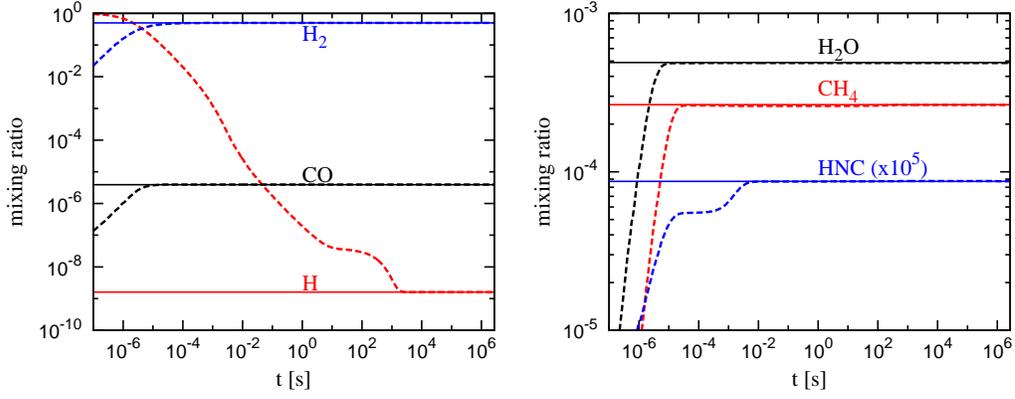}
\caption{Mixing ratios as a function of time [s] at 1 bar and 1000 K (dashed lines), compared to
chemical equilibrium (solid lines) for H$_2$, H, CO, CH$_4$ and H$_2$O.
\label{fig:Chemical-Equilibrium}}
\end{figure*}

At sufficiently high temperatures and pressures, a gas should rapidly settle into chemical
equilibrium. An important test for a chemical network is that its steady state solution
converges to the chemical equilibrium solution.
To perform this test of our network, we solve the chemical kinetics at a single 
($T, p$) point, using
the rate constants from the \textsc{Stand2015}
network, disabling the cosmic ray reactions, photochemistry and dissociative recombination. 
We compute a time-dependent solution of the equation
\begin{equation}
\dfrac{dn_i}{dt} = P_i - L_i.
\label{eqn:simple-kinetics}
\end{equation}
We solve this equation for $T = 1000$ K and $p = 1$ bar, with solar abundances
from \citet{Asplund2009}. We compare our results to chemical equilibrium calculations using the
Burcat polynomials \citep{Burcat2005}, and plot our comparisons in Figure 
\ref{fig:Chemical-Equilibrium} and find excellent agreement. This agreement is not surprising;
we have used the same thermochemical data to reverse our reactions, and only include reversed
reactions in this test, so once the system achieves steady state, computationally achievable
at this pressure and temperature, the chemistry has effectively settled into equilibrium.

\begin{figure}
\epsscale{1.0}
\plotone{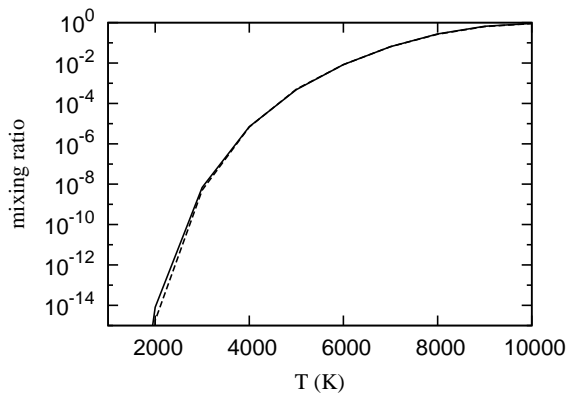}
\caption{Mixing ratio as a function of temperature. The solid line is from the Saha equation and 
the dashed line is the result from our model calculation.\label{fig:Saha}}
\end{figure}

We also compare our electron number density to the electron number density achieved using the Saha 
equation, this time at a pressure of $10^{-4}$ bar and over a range of temperatures from 1000 K to 
10000 K. This comparison is plotted in Figure \ref{fig:Saha}. The comparison is virtually perfect
when $T \gtrsim 2000$ K, unsurprising given the way the three-body recombination reactions 
are calculated (see Section \ref{sec:thermal-ionization}). At $\sim 1000$ K, our results diverge 
from the Saha equation. This is because the integrator does not reliably calculate mixing ratios 
below $\sim 10^{-30}$. Indeed, at this stage, the electron number density achieves $\sim 10^{-300}$
cm$^{-3}$ while the H$^+$ number density rests at $\sim 10^{-60}$ cm$^{-3}$, producing significant
charge balance errors. These large errors in the charge balance fluctuate, and only appear when the
ionization fraction is $\lesssim 10^{-30}$, at which point ion-neutral chemistry is inconsequential.

\section{1-D Photochemistry/Diffusion Code}
\label{sec:model}

We have developed a simple 1D photochemistry/diffusion code (\textsc{Argo}), for the purposes of
testing the \textsc{Stand2015} network. The required inputs for \textsc{Argo} are: 
\begin{itemize}
\item $(p,T)$ profile of the atmosphere.
\item Vertical eddy diffusion ($K_{zz}$ [cm$^2$ s$^{-1}$]) profile of the atmosphere \citep[see
discussion in][]{Lee2015}.
\item Atmospheric elemental abundances.
\item Boundary conditions at top and bottom of the $p,T$ profile.
\item Actinic flux\footnote{The actinic flux is the radiance integrated over all angles, expressing 
flow of energy through a unit sphere. There are subtle differences between the actinic flux and the
spectral irradiance, see \citet{Madronich1987}.} at the top of the atmosphere.
\item Chemical Network (in our case, \textsc{Stand2015}).
\item Initial chemical composition
\end{itemize}
All of these inputs except the chemical composition are fixed.

\begin{figure}
\epsscale{1.0}
\plottwo{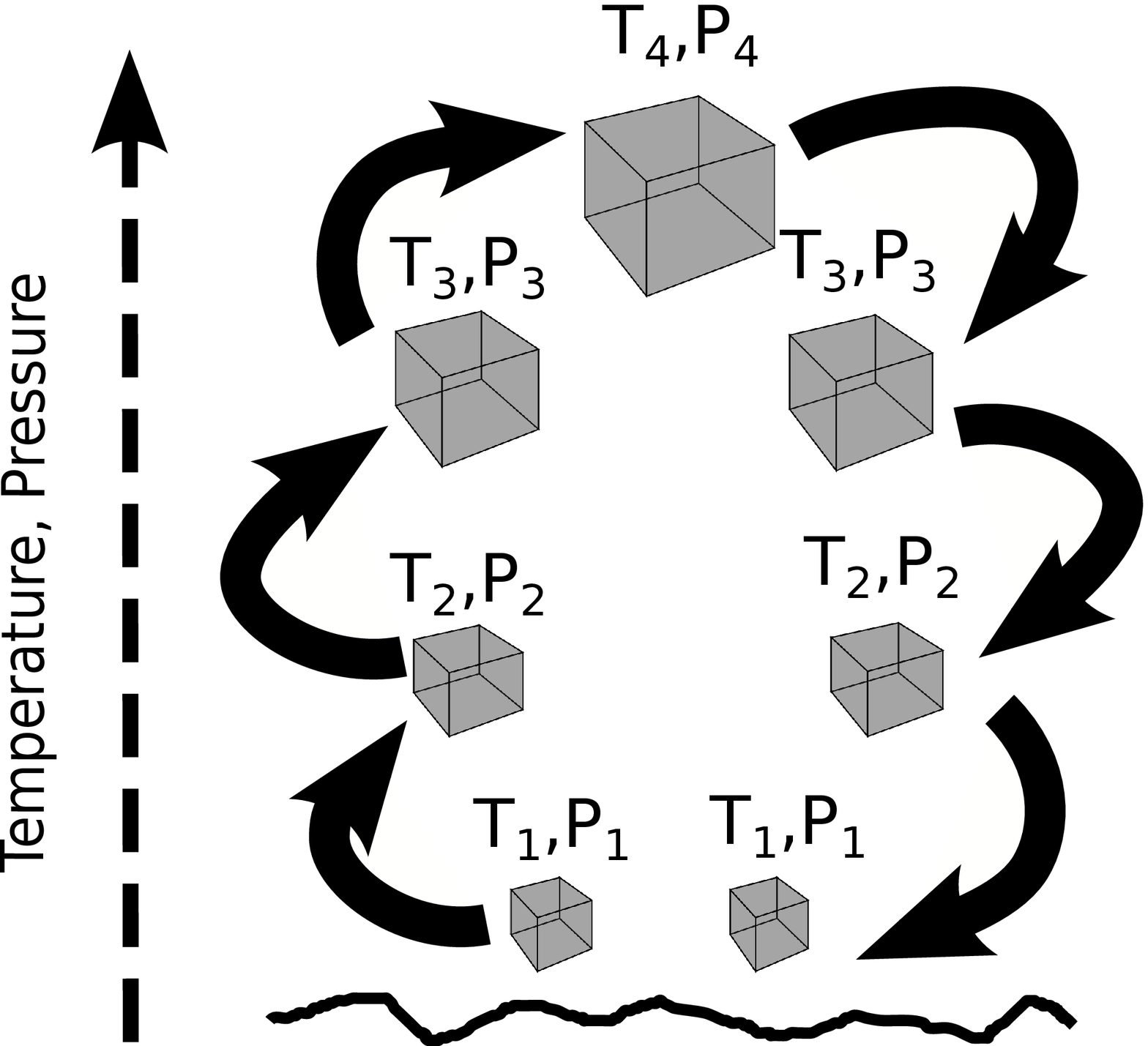}{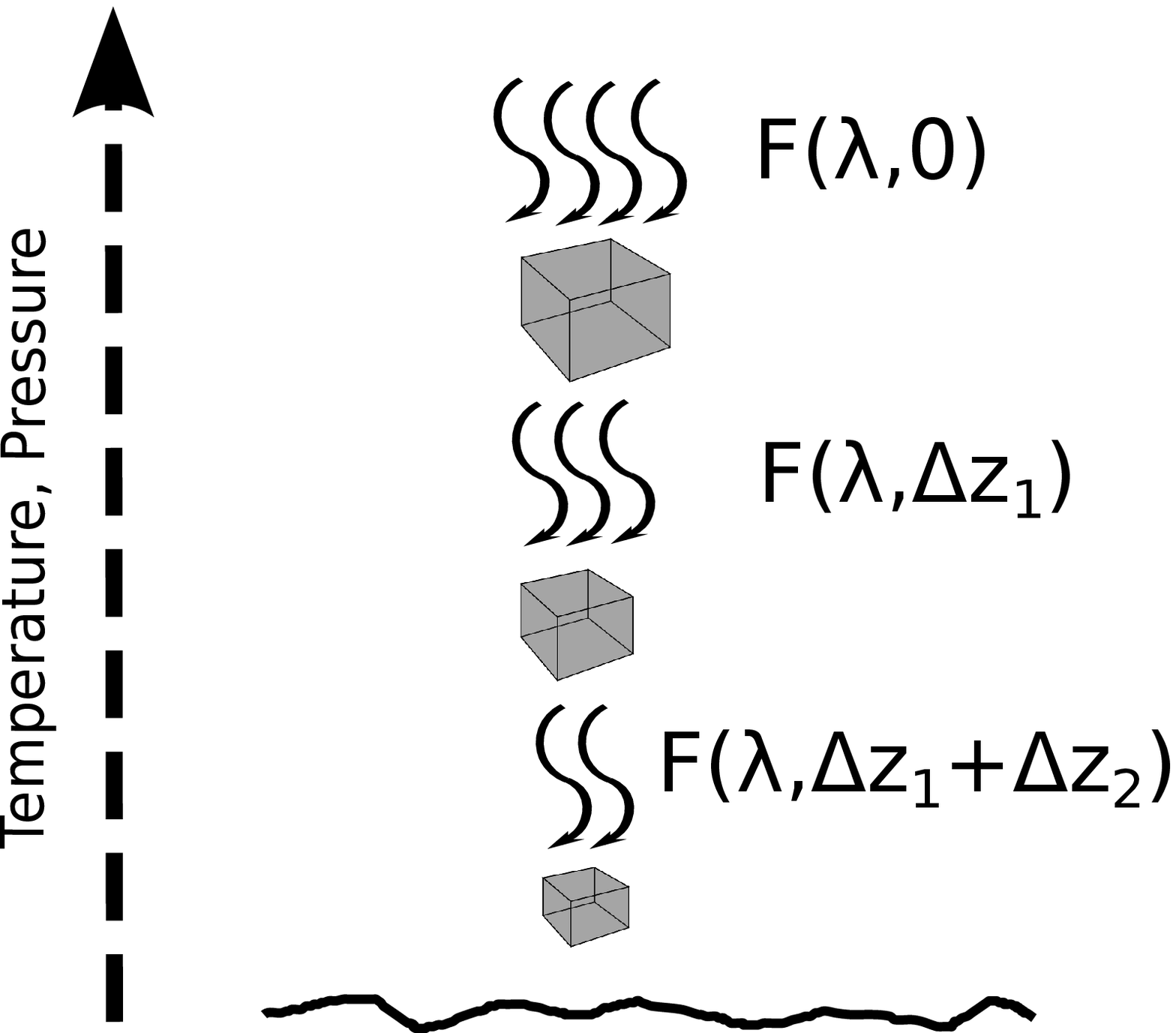}
\caption{Picture representation of the model. The picture on the left represents the motion of the
single parcel from the bottom of the atmosphere, $T_1,P_1$, up to the top of the atmosphere, 
$T_4,P_4$, and then back down, see Section \ref{sec:Equation}. Once this journey is completed, we 
irradiate the atmosphere,
by stacking up the parcel at different times, when it was located at different parts of the
atmosphere. The picture on the right represents the calculation of the depth-dependent actinic flux
discussed in Section 
\ref{sec:radtran}. Only photons of wavelength between 1-10000 \AA\ are considered Figure 
\ref{fig:FlowChart} gives a flowchart for the calculation. \label{fig:cartoon}}
\end{figure}

With these inputs, \textsc{Argo} solves molecular transport in a fully Lagrangian manner, 
similar to \citet{Alam2008} and \citet{Zahnle1995}. The model consists of two
parts: (1) A chemical transport model (Section \ref{sec:Equation}) and (2) calculation
of the photochemical and cosmic ray chemical rate constants from cross-sections and a depth-dependent
actinic flux (Section \ref{sec:radtran}). A conceptual illustration is shown in 
Figure \ref{fig:cartoon}.

\subsection{The Continuity Equations for Chemical Species}
\label{sec:Equation}
The coupled 1D continuity equations describing the time-dependent vertical atmospheric chemistry 
are:
\begin{equation}
\dfrac{\partial n_i}{\partial t} = P_i - L_i - \dfrac{\partial \Phi_i}{\partial z},
\label{eqn:continuity}
\end{equation}
where $n_i$ [cm$^{-3}$] is the number density of species $i$, and $i = 1,...,N_s$, and
$N_s$ is the total number of species. $P_i$ [cm$^{-3}$ s$^{-1}$] is the rate of production and 
$L_i$ [cm$^{-3}$ s$^{-1}$] is the rate of loss of species $i$. The right-most term is the vertical 
change 
in flux $\Phi_i$ [cm$^{-2}$ s$^{-1}$], and represents the flux due to both 
eddy ($K$ [cm$^2$ s$^{-1}$]) and molecular diffusion ($D$ [cm$^2$ s$^{-1}$]) respectively, 
related as \citep[][their Eq. (15.14)]{Banks1973},
\begin{align}
\Phi_i =& -K \Big[\dfrac{\partial n_i}{\partial z} + n_i \Big(\dfrac{1}{H_0} + 
\dfrac{1}{T}\dfrac{dT}{dz}\Big)\Big] \notag\\ 
&- D \Big[\dfrac{\partial n_i}{\partial z} + 
n_i \Big(\dfrac{1}{H_i} + \dfrac{1 + \alpha_T}{T}\dfrac{dT}{dz}\Big) \Big],
\label{eqn:diffusion}
\end{align}
where $H_0$[cm] is the pressure scale height of the atmosphere at $z$ [cm], 
$H_i$[cm] is the molecular scale 
height of the atmosphere for species $i$, and $\alpha_T$ is the thermal diffusion factor 
\citep{Banks1973,Yung1999,Zahnle2006,Hu2012}. For molecular diffusion coefficients, we adopt 
Chapman-Enskog theory \citep{Enskog1917,Chapman1991}. Eddy diffusion coefficients are either 
determined empirically, as with \object{Earth} and \object{Jupiter}, or are derived from global 
circulation models, as is the case for \object{HD209458b}.

In Eq. (\ref{eqn:diffusion}), the terms dealing with eddy diffusion and molecular diffusion
are separated out, clarifying four regions that Eq's (\ref{eqn:continuity}),
(\ref{eqn:diffusion}) describe. {\bf(1)} Deep within the atmosphere, where pressures and 
temperatures are sufficiently large, the thermochemistry dominates,
and the equation simplifies to Equation (\ref{eqn:simple-kinetics}). The 
atmospheric chemical composition converges to
chemical equilibrium or at least to some stable quasi-equilibrium. {\bf (2)} Higher in the 
atmosphere, the eddy diffusion may dominate, and the species are quenched, their abundance mixed 
evenly over a wide range of the atmosphere at time-scales shorter than the chemical time-scales. 
{\bf (3)} Above this region,
molecular diffusion may dominate, and at that point, species lighter than the mean molecular mass
of the atmospheric gas will rise up, and species heaver than the mean molecular mass will settle
down, and the chemistry will largely be determined by the individual scale heights of the
atmospheric constituents. {\bf (4)} Non-equilibrium processes, such as photochemistry or cosmic 
ray chemistry, may create a fourth region, the composition of which is determined by irreversible 
chemical reactions.

Since the purpose of this paper is to introduce a new chemical kinetics network for lightning and
prebiotic processes, our focus is not on the atmospheric dynamics (for this, see \citealt{Lee2015}). 
We therefore apply a simple 
approximation to Eq. (\ref{eqn:continuity}), inspired by \citet{Alam2008}. We first cast 
Eq. (\ref{eqn:continuity}) in a Lagrangian formulation, and consider Eddy diffusion to be moving 
small parcels of the gas vertically. We follow a single parcel as 
it moves up from the lower boundary of the temperature profile, and then 
returns down again. In reality, the parcel 
would be jostled in all three dimensions as it makes a complex journey up to the top of the 
atmosphere, but 1D transport models are unable to capture this effect in full.

The differential diffusion of molecules into and out of the parcel requires a different approach. 
The discrete formulas used by \citet[][their Eq. 9]{Hu2012} in the Lagrangian frame are:
\begin{align}
\dfrac{\partial n_{i,j}}{\partial t} &= P_{i,j} - L_{i,j} n_{i,j} - d_{j + 1/2}\dfrac{n_{{\rm gas},j+1/2}}{n_{{\rm gas},j+1}}n_{i,j+1}\notag\\
&-\Big(d_{j+1/2}\dfrac{n_{{\rm gas},j+1/2}}{n_{{\rm gas},j}} - d_{j-1/2}\dfrac{n_{{\rm gas},j-1/2}}{n_{{\rm gas},j}} \Big)n_{i,j}\notag\\
&+d_{j-1/2}\dfrac{n_{{\rm gas},j-1/2}}{n_{{\rm gas},j-1}}n_{i,j-1}. \label{eqn:discrete}
\end{align}
Here, $j$ represents the parcel being followed, $j-1$ the parcel directly beneath $j$, 
$j + 1$ the parcel above $j$, and $j \pm 1/2$ an arithmetic average between $j$ and
$j \pm 1$. $n$ without any $i$ subscript represents $n_{\rm gas}$ at the relevant 
parcel, and
\begin{equation}
d_{j \pm 1/2} = \dfrac{D_{j \pm 1/2}}{2(\Delta z)^2}\Big[\dfrac{(\bar{m} - m_i)g\Delta z}{k_B T_{j \pm 1/2}} - \dfrac{\alpha_T}{T_{j \pm 1/2}}(T_{j \pm 1} - T_j)\Big].
\label{eqn:disc-diffusion}
\end{equation}
$\bar{m}$[g] denotes the mean molecular mass of the atmosphere at $z$ and $m_i$[g] the mass of 
species $i$.

Both the third and last terms on the R.H.S. of Eq. (\ref{eqn:discrete}) do not depend on $n_i$ and 
can therefore
be treated as source terms, $P_i$. The fourth term can be treated as a term in $L_i$, such that 
molecules ``destroyed'' by this reaction are ``banked'', ${\rm A} \rightarrow {\rm BA}$. The 
``banked'' molecules re-enter the parcel at a rate determined by the third and last terms on the 
R.H.S. of the 
equation, thus conserving mass throughout the parcel's travels. Violations of this 
conservation do not appear here, but can be accounted for via further reactions, settling, 
condensation and evaporation, outgassing and escape, discussed in Appendix \ref{app:outgassing}. 
Although it is straight-forward to handle atmospheric escape with this method, we do not do so for 
any of the test cases in this paper.

Equation (\ref{eqn:discrete}) is solved within \textsc{Argo} in the same numerical manner as 
\textsc{Nahoon}
\citep{Wakelam2012}, by the implicit time-dependent Gear method as incorporated by the Livermore
Solver for Ordinary Differential Equations (DLSODE) \citep{Gear1971,Brown1989}. 

\subsection{Calculating the XUV and Cosmic Ray Flux}
\label{sec:radtran}

Once the fluid parcel has completed the atmospheric profile, the solar XUV 
actinic flux from
1 \AA\ to 10000 \AA\ as a function of depth, $z$ [cm]\footnote{The depth for this model extends from 
$z = 0$, the bottom of the temperature profile for the planet in question, to $z = z_{\rm top}$, the 
top of the profile.} 
and wavelength $\lambda$ [\AA] is calculated.
We consider both the direct and approximate diffusive actinic flux. 
The local height-dependent actinic flux is calculated without any iteration on the local temperature.
The cross-sections for various photochemical reactions (Sect. \ref{sec:photochem}), are 
multiplied by each vertical step $(\Delta z)_j$ [cm], where $(\Delta z)_j$ is the size of the step at 
height $z_j$. The total optical depth as a function of the wavelength takes the form
\begin{equation}
\tau(\lambda,z) = \Sigma_j \big[(\Delta z)_j \Sigma_i \sigma_i n_{ij} \big] + \tau_s,
\label{eqn:optical-depth}
\end{equation}
where $i$ is summed over all species for which 
photoabsorption is considered (see Section \ref{sec:photochem} for a list of these species). 
$\tau_s$ is the 
optical depth due to Rayleigh scattering, and the actinic flux as a function 
of depth is defined as \citep{Hu2012}:
\begin{equation}
F(\lambda,z) = F(\lambda,z_{\rm top})e^{-\tau(\lambda,z)/\mu_0} + F_{\rm diff},
\label{eqn:actinic-flux}
\end{equation}
where $\mu_0 = \cos \theta$, where $\theta$ is the stellar zenith angle; we set $\mu_0 = 1/2$ 
for all calculations
within this paper \citep[see][their Fig. 7]{Hu2012}. $F_{\rm diff}$ denotes the actinic flux
of the diffusive radiation, determined using the $\delta$-Eddington 2-stream method \citep{Toon1989}.
Once the actinic flux is calculated, the photochemical rates are determined as in Section 
\ref{sec:photochem}. Once the depth dependent flux, $F(z,\lambda)$ [cm$^{-2}$ s$^{-1}$ \AA$^{-1}$], 
is determined for all layers, the parcel's path through the atmospheric profile is repeated, 
now accounting for the photochemistry. The cosmic ray ionization rate, $\zeta(z)$ 
[s$^{-1}$] is likewise calculated in a depth-dependent manner following \citet{Rimmer2013} and 
incorporated into the chemistry (Sec. \ref{sec:photochem}).

\begin{figure}
\epsscale{1.0}
\plotone{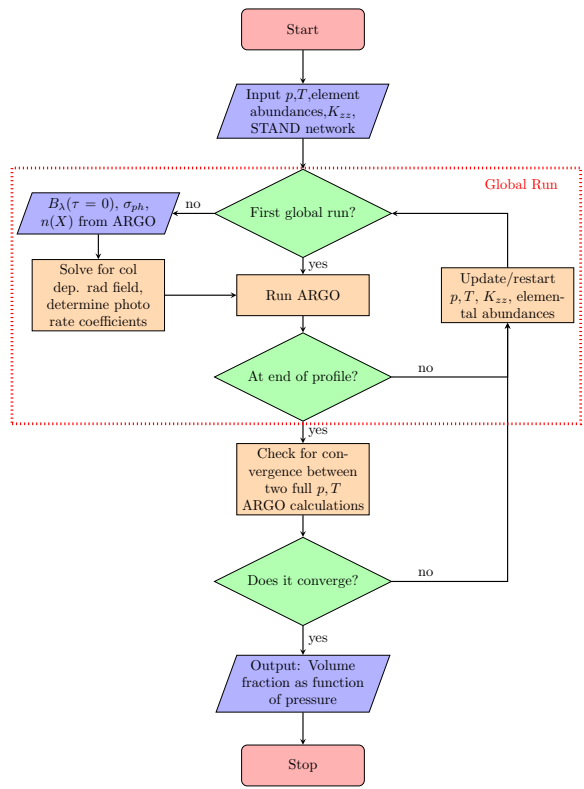}
\caption{A flow-chart representation for the program.\label{fig:FlowChart}}
\end{figure}

A new depth-dependent composition is constructed, then applied to 
Eq. (\ref{eqn:optical-depth}) to solve again for $F(z,\lambda)$. 
The value of $\zeta(z)$ does not change significantly between iterations.
This process is 
repeated until the results converge; i.e. until the profile from the previous global calculation 
(transport + depth-dependent flux) agrees to within 1\% the profile from the current global 
calculation. The number of repetitions depends on the parameters, but is typically between 5 and 12
global iterations. This iterative process is represented as a flow-chart in 
Figure \ref{fig:FlowChart}.

This method is both simple and functional, requiring relatively little computational resources. 
It is also
straight-forward to adapt to diverse chemical environments, since it does not require the selection
of ``fast'' and ``slow'' chemistry to ease computational speed. These strengths do not come without a
cost: The simplistic dynamics does not transition as smoothly from the eddy diffusion regime to the
molecular diffusion regime as the Eulerian formulation, and
can result in steep changes over a handful of height-steps. 

\subsection{Testing the Atmospheric Transport Model for Molecular Diffusion}
\label{sec:diffusion}

In order to benchmark the \textsc{Stand2015} chemical network in different planetary atmospheres, 
we test the molecular diffusion within \textsc{Argo}. We consider a 1D isothermal gas 
under a constant surface gravity, $g = 10^3$ cm/s$^2$, with temperature $T = 300$ K, at hydrostatic
equilibrium. The gas is initially composed of carbon and hydrogen atoms, each with a mixing ratio of 
$X_0({\rm C}) = n({\rm C})/n_{\rm gas} = 0.5$ and $X_0({\rm O}) = 0.5$ throughout. 
All chemistry is disabled. It is expected that the heavier species, carbon, will
settle into the atmosphere, and the lighter species, hydrogen, will rise up, until they stratify.
The analytic solution to this system is well-known. The
mixing ratio should be determined by the scale-heights of the individual species such that, for the 
carbon abundance:
\begin{equation}
X({\rm C}) = \dfrac{X_0({\rm C})e^{-z/H_{\rm C}}}{X_0({\rm H})e^{-z/H_{\rm H}} + X_0({\rm C})e^{-z/H_{\rm C}}},
\label{eqn:analytic}
\end{equation}
where $X({\rm C})$ is the final steady state carbon mixing ratio, and $H_{\rm H}$ [cm] and 
$H_{\rm C}$ [cm] are the atmospheric scale heights for the hydrogen and carbon.

\begin{figure}
\epsscale{1.0}
\plotone{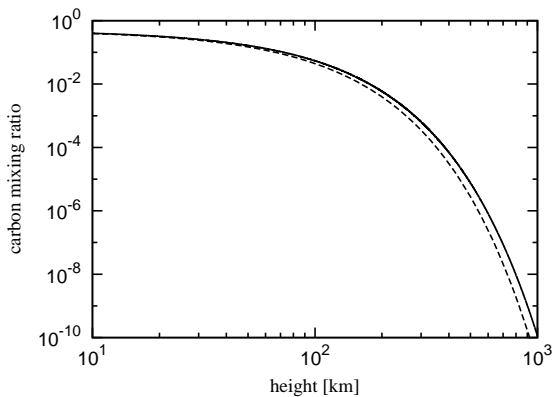}
\caption{The carbon mixing ratio as a function of atmospheric height [km]. We test for diffusion, 
with chemistry turned off, for carbon atoms and hydrogen atoms in a gas at hydrostatic equilibrium 
for an
isothermal gas ($g = 10^3$ cm s$^{-2}$, $T = 300$ K). The solid line is the result from 
\textsc{Argo} and the dashed line is the analytic result (Eq. (\ref{eqn:analytic})). 
\label{fig:Diffusion-Test}}
\end{figure}

The code is run until steady state is achieved, when the carbon in
the very upper atmosphere diffuses into the lower atmosphere. The steady-state mixing ratio, as
a function of height is compared the analytic mixing ratio, Eq. (\ref{eqn:analytic}), in Figure 
\ref{fig:Diffusion-Test}. The 
comparison is reasonable through the extent of the atmosphere.

\section{Testing the Network for Planetary Environments}
\label{sec:test-cases}

The \textsc{Stand2015} network contains chemical reactions for an H/C/N/O gas, and
including both highly reducing to highly oxidizing atmospheres, and for a temperature range of
100 K to 30000 K. The network should then be tested for a variety of planetary 
atmospheres with
different chemical compositions, from the (probably) oxidizing atmosphere of the early Earth to
the highly reducing atmosphere of Jupiter. The large range of temperatures is tested for
the irradiated exoplanet \object{HD209458b}. 
We also test our model against the height dependent measurements of select trace species within
the atmosphere of the present-day \object{Earth}. It would be interesting to apply our model to Titan,
due to its rich nitrile and organic chemistry. Titan's atmosphere is a very rich and complex environment,
and it is important to account for these complexities when modeling Titan. Titan has upper atmospheric
hazes, temperatures low enough to condense several molecular species, and ionization and dissociation
by energetic particles including cosmic rays, Saturn magnetospheric particles, solar wind protons and
interplanetary electrons. As useful as a study of the atmosphere of Titan would be for exploring
Miller-Urey-like chemistry \citep{Waite2007}, such a model is beyond the scope of this paper.
The boundary conditions for these various objects
are given in Section \ref{sec:boundary-conditions}. We then compare our results to the results
from other chemical kinetics models and, where possible, with observations, for 
\object{HD209458b} (Section \ref{sec:HD209-test}), \object{Jupiter} (Section \ref{sec:Jupiter-test}), and the \object{Earth} (Section \ref{sec:Earth-test}).

\subsection{Boundary Conditions for Three Test Cases: HD209458b, Jupiter and the Earth}
\label{sec:boundary-conditions}

Below, we compare the results of our chemical kinetics to other results for \object{HD209458b}
and also for \object{Jupiter} and the \object{Earth}. Each of these objects has different
boundary conditions and parameters. These conditions and parameters include the temperature
profile of the object's atmosphere, the eddy diffusion profile, the elemental abundances,
the initial composition at the lower boundary of the atmospheric profile, and the unattenuated 
UV flux. For HD209458b, the conditions at the lower boundary of the atmospheric profile rapidly 
develop from the prescribed initial conditions toward chemical equilibrium. For Jupiter and the early
Earth, the composition at the lower boundary is stable over the dynamical 
time-scale ($dn_i(z = 0)/dt \approx 0$), and so the initial composition effectively acts as a 
lower boundary condition. The assumed elemental abundances and initial conditions at the lower 
boundary of the atmospheric profile are given in Table \ref{tab:initial-conditions}.

We take \object{HD209458b} to have solar elemental abundances throughout its atmosphere, and set
the initial conditions at the lower boundary of the atmosphere to be entirely atomic. The initial
composition hardly matters here, since the composition quickly settles to chemical equilibrium at
such a high temperature and pressure. The temperature profile and eddy diffusion profile for 
\object{HD209458b} are both taken from \citet{Moses2011}, so that we can directly compare
results.

\begin{figure}
\epsscale{1.0}
\plotone{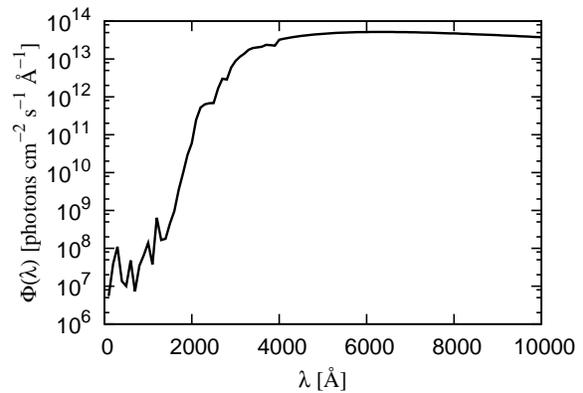}
\caption{The solar flux used in our model [photons cm$^{-2}$ s$^{-1}$ \AA$^{-1}$], as a function
of wavelength, $\lambda$ [\AA], taken from \citet{Huebner1979,Huebner1992,Huebner2015}.
Weighted versions of this flux are used for HD209458b and Jupiter. This flux is
used, unadjusted, for the early Earth.
\label{fig:solar-flux}}
\end{figure}

Since \object{HD209458} is a G0 star, we use the solar UV flux.
The unattenuated solar UV flux at 1 AU is obtained from the \textsc{SORCE} data
\citep{Rottman2006} for 1 \AA\ - 350 \AA\, and 1150 \AA\ - 10000 \AA\, with data from 
\textsc{PhIDRates} for the 350 - 1150 \AA\ range. The binned flux we use is plotted in 
Figure \ref{fig:solar-flux}. This flux is adapted to \object{HD209458b} by multiplying the 
solar UV flux
by a factor of $(d_{\oplus}/d_p)^2$, where $d_{\oplus}$ [AU] is the distance from the Earth 
to the Sun and $d_p \approx 0.047$ AU is the approximate distance between \object{HD209458b} and
its host star. 
This may not be the most accurate approximation 
to the UV behavior of \object{HD209458}, since it might have quite different activity from our sun
\citep{Tu2015}. 

For \object{Jupiter}, we use the temperature and eddy profiles from \citet{Moses2005}. 
For consistency, we set the initial conditions at the lower boundary of Jupiter's
atmosphere to be the same as \citet{Moses2005}; see Table \ref{tab:initial-conditions}. The 
solar UV spectrum
at 1 AU is used for \object{Jupiter}, although multiplied by a factor of $(d_{\oplus}/d_J)^{-2}$, 
where $d_J \approx 4.5$ AU is the square of the distance between the sun and \object{Jupiter}.

For the present-day \object{Earth}, we use the measured surface mixing ratios from the US Standard
Atmosphere 1976 (see Table\ref{tab:initial-conditions}) and the temperature profile from
\citet{Hedin1987,Hedin1991}, Fig. \ref{fig:Earth-Profile}. We use the present day solar flux at
1 AU as our incident UV flux.

We use the same chemical lower boundary conditions as from \cite{Kasting1993} for the
atmosphere of the early \object{Earth} (Table \ref{tab:initial-conditions}). The temperature 
profile for the early Earth is assumed to be the same as that of the present Earth 
\citep{Hedin1987,Hedin1991}, Fig. \ref{fig:Earth-Profile}. The UV field used for this model is 
that of the young Sun calculated using the scaling relationships of \citet{Ribas2005} for 
wavelengths between 1 \AA\ -- 1200 \AA\ and the UV field of the solar analogue $\kappa^1$ Cet 
above 1200 \AA\ \citep{Ribas2010}.

\subsection{HD209458b}
\label{sec:HD209-test}

\object{HD209458b} was first observed by \citet{Henry2000}, and is one of a growing number of Hot 
Jupiters to have a measured spectrum, via transit \citep[e.g.][]{Queloz2000}, and also in 
emission \citep[e.g.][]{Knutson2008}. Various molecular species have been tentatively identified 
in the spectrum, such as TiO \citep{Desert2008}, water \citep{Madhu2009,Swain2009,Beaulieu2010}, 
CO, CO$_2$ and methane features \citep{Madhu2009,Swain2009}. \object{HD209458b} has been extensively 
modeled, with retrieval modeling \citep{Madhu2009}, and with hydrodynamic global circulation 
models \citep{Showman2008}. This planet has also been a popular target for non-equilibrium chemistry
models such as those of \citet{Liang2003,Zahnle2009,Moses2011,Venot2012,Agundez2014} and 
\citet{Lavvas2014}.

\begin{figure}
\epsscale{1.2}
\plotone{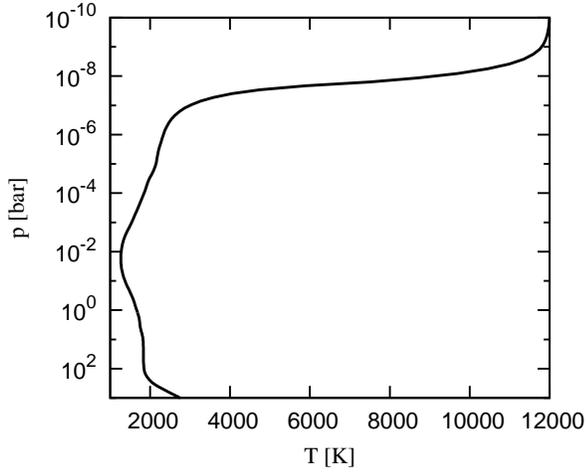}
\caption{Temperature profile for HD209458b, $T$ [K] as a function of $p$ [bar], as used
by \cite{Moses2011}.\label{fig:HD209-profile}}
\end{figure}

We have chosen the atmosphere of \object{HD209458b} as one candidate for benchmarking
our results because it is well characterized and has been the subject of several 
non-equilibrium chemistry models, and it has a very high temperature even among Hot
Jupiters. An additional benefit to \object{HD209458b} is its suspected temperature inversion 
\citep[][although this is debated, see also \citealt{Schwarz2015}]{Knutson2008}, which
allows us to test our chemistry at very high temperatures both at both high and low pressures.
The thermal profile of \object{HD209458b} from \citet{Moses2011}
is shown in Figure \ref{fig:HD209-profile}. The local gas-phase temperature $T > 2000$ K both when 
$p > 100$ bar and when the gas-phase pressure, $p < 10^{-4}$ bar. This is a wide parameter space 
relevant for ion-neutral chemistry initiated via thermal ionization. 

We compare our results to the predictions of two different chemical kinetics models.
(1) We compare our results to the results of \citet{Moses2011} with the ion-neutral
chemistry disabled. (2) We compare the ionic abundances for our most abundant ions to the results 
of \citet{Lavvas2014}. Also in this case, we disable cosmic ray chemistry in order to draw a 
better comparison to the ion-neutral chemistry. 

\begin{figure*}
\epsscale{2.0}
\plotone{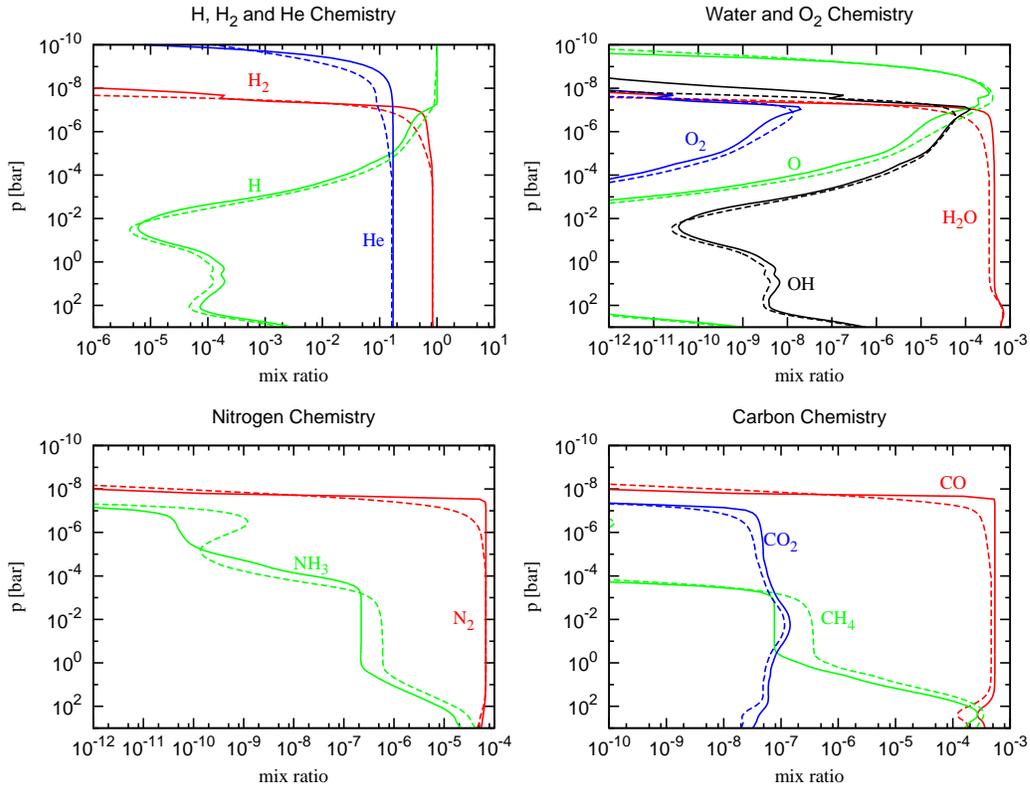}
\caption{Mixing ratios for various chemical species as a function of pressure, $p$ [bar]. 
A comparison between our model (solid lines) and that of \citet[][dashed lines]{Moses2011},
for H/H$_2$ chemistry, water and O$_2$ chemistry, nitrogen chemistry and carbon chemistry
in the atmosphere of HD209458b. \label{fig:HD209-compare}}
\end{figure*}

We compare our network and transport model to \citet{Moses2011} by examining the volume mixing
ratios of major neutral species: H, H$_2$, He (hydrogen/helium chemistry), OH, H$_2$O, 
O and O$_2$ (oxygen/water chemistry), N$_2$ and NH$_3$ (nitrogen chemistry), and CO, CH$_4$ and 
CO$_2$ (carbon chemistry). See Figure \ref{fig:HD209-compare}. These species were chosen because
they are abundant and, in the case of H$_2$ and N$_2$, play an important role in the 
non-equilibrium chemistry. N$_2$ provides the reservoir 
for the transition between N$_2 \rightleftharpoons $ NH$_3$. Other species were chosen because they
contribute to features observed in transit spectroscopy, (e.g. CO$_2$). The molecules CO 
and H$_2$O do both. Helium was chosen because its mixing ratio is not significantly affected by
the chemistry. It changes with pressure due to molecular diffusion, and so it provides a useful 
comparison between our dynamical calculations and those of \citet{Moses2011}.

The transition of carbon between CO and CH$_4$, and nitrogen between N$_2$ and NH$_3$ is very 
sensitive to non-equilibrium chemistry, as ${\rm CH_4 \approx CO}$ when $p \sim 100$ bar and 
$T \sim 2000$ K. As the pressure decreases
rapidly while the temperature remains relatively high ($T > 1000$ K), the thermochemical
equilibrium ratio for CH$_4$/CO plummets, approaching $10^{-7}$ at 0.1 bar in the \object{HD209458b}
atmosphere. The
time it takes the carbon to meander from CH$_4$ to CO, however, becomes significantly
longer than the relevant dynamical timescales (for \object{HD209458b}, this time-scale is prescribed 
by the eddy diffusion coefficient, see \citealt{Bilger2013}), and the CH$_4$ and CO abundances are 
quenched. The same sort of process governs the transition of nitrogen from N$_2$ to NH$_3$.

The pathways for both CH$_4 \rightleftharpoons $ CO and N$_2 \rightleftharpoons $ NH$_3$
interconversions are not well understood. In both cases, the paths competing with one another are
often circuitous, and tend to be regulated by one of several reactions encountered along the
journey, a slow rate-limiting step \citep{Moses2014}. The time-scale of the transition between 
species is almost
entirely set by the rate by which that single reaction proceeds. As discussed in 
Section \ref{sec:Network}, rate coefficients can be frustratingly uncertain, with different
estimations sometimes varying by more than an order of magnitude. For example, compare the
rate experimental and theoretical rate constants for C$_2$H$_6$ $\rightarrow$ CH$_3$ + CH$_3$
(\citealt{2009YAN/GOL} and \citealt{2005KIE/SAN}, respectively).
The path that one 
believes regulates
these central transitions can be very different depending on what rate coefficients are used.

An illustrative example is the reaction\\ 
${\rm CH_3 + H_2O} \rightarrow {\rm CH_3OH + H}$. 
\citet{Hidaka1989} has determined the rate for ${\rm CH_3 + H_2O \rightarrow \; Products}$,
Reaction (\ref{eq:CH3OH}), proceeds 
with a barrier of $\approx 2670$ K \citep[see][for a discussion on this reaction]{Visscher2010}. 
With reasonable assumptions of the branching ratios
for this reaction, namely that the branching ratios do not change much with temperature, one would 
set the same barrier to ${\rm CH_3 + H_2O \rightarrow CH_3OH + H}$, as done by \citet{Venot2012}.
However, \citet{Moses2011} carried out quantum chemical calculations for this reaction using 
MOLPRO and estimate
a barrier for this particular branch of $\approx 10380$ K, much larger than the activation energies
of the other branches. 
With the smaller barrier, the path carbon takes from CH$_4$ to CO proceeds as:
\begin{align}
{\rm H_2 + M} &\rightarrow {\rm H + H + M} \notag\\
{\rm CH_4 + H} &\rightarrow {\rm CH_3 + H_2} \notag\\
{\rm \mathbf{CH_3 + H_2O}} &\rightarrow {\rm \mathbf{CH_3OH + H}} \label{eq:CH3OH}\\
{\rm CH_3OH + H} &\rightarrow {\rm CH_2OH + H_2} \notag\\
{\rm CH_2OH + M} &\rightarrow {\rm H_2CO + H_2 + M} \notag\\
{\rm H_2CO + H} &\rightarrow {\rm HCO + H_2} \notag\\
{\rm HCO + H} &\rightarrow {\rm CO + H_2} \notag\\
{\rm HCO + M} &\rightarrow {\rm CO + H + M}\notag\\
\cline{1-2}
{\rm CH_4 + H_2O} &\rightarrow {\rm CO + 3H_2}.
\end{align}
We adopt the rates of \citet{Moses2011} for this pathway, as well as the smaller rate coefficient
for the three-body reaction H$_2$O + CH$_2$ + M $\rightarrow$ CH$_3$OH. An examination of our 
results would reveal that, as with \citet{Venot2012}, the transition of carbon from CH$_4$ to CO is 
much more efficient than with \citet{Moses2011}. We have examined the rates at which reactions
proceed in our network and find {\it another} formation pathway:
\begin{align}
{\rm H_2 + OH} &\leftrightarrow {\rm H_2O + H} \notag\\
{\rm OH + O} &\rightarrow {\rm O_2 + H} \notag\\
{\rm CH_4 + H} &\rightarrow {\rm CH_3 + H_2} \notag\\
{\rm CH_3 + H} &\rightarrow {\rm CH_2 + H_2} \notag\\
{\rm CH_2 + O_2} &\rightarrow {\rm COOH + H} \notag\\
{\rm \mathbf{COOH + H_2O}} &\rightarrow {\rm \mathbf{CH_2O_2 + OH}} \label{eqn:COOH}\\
{\rm CH_2O_2 + M} &\rightarrow {\rm CO_2 + H_2 + M} \notag\\
{\rm CO_2 + H} &\rightarrow {\rm CO + OH}\notag\\
\cline{1-2}
{\rm CH_4 + O} &\rightarrow {\rm CO + 2H_2},
\end{align}
The atomic oxygen arises from thermal dissociation of OH or photodissociation of H$_2$O
followed by diffusion downward.
This pathway is critically dependent on Reaction (\ref{eqn:COOH}). To our knowledge, the three-body
rate coefficient for this reaction has not been determined. This reaction has
instead appeared in our network as the reverse reaction of 
${\rm CH_2O_2 + OH \rightarrow COOH + H_2O}$, for which we use an estimate based on reaction
energetics \citep{Mansergas2006}. This pathway is highly uncertain, and removing it makes up the
majority of the difference between our results and those of \citet{Moses2011} for methane between
1 -- $10^{-4}$ bar. We suspect further differences owe to our different thermochemical constants
and the use of slightly different solar abundances.

The path of nitrogen from NH$_3$ to N$_2$ is considerably more uncertain. The path is believed to 
roughly 
follow from NH$_3$ to NH via hydrogen abstraction, which will in turn react with another NH$_X$
species to form N$_2$H$_Y$. This species will be destroyed either by reacting with hydrogen or
via thermal decomposition, to form N$_2$. The reactions N$_2$H$_{X+2} \rightarrow$ NH$_2$ + NH$_X$ 
involve large uncertainties, which result in variations of the NH$_3$ quenched abundance by
an order of magnitude. We find, similar to \citet{Moses2011}, that:
\begin{align}
{\rm H_2 + M} &\rightarrow {\rm H + H + M} \notag\\
{\rm NH_3 + H} &\rightarrow {\rm NH_2 + H_2} \notag\\
{\rm NH_2 + H} &\rightarrow {\rm NH + H_2} \notag\\
{\rm \mathbf{NH_2 + NH}} &\rightarrow {\rm \mathbf{N_2H_2 + H}} \label{eqn:N2H2}\\
{\rm N_2H_2 + H} &\rightarrow {\rm NNH + H_2} \notag\\
{\rm NNH + M} &\rightarrow {\rm N_2 + H + M} \notag\\
\cline{1-2}
{\rm 2NH_3} &\rightarrow {\rm N_2 + 3H_2}
\end{align}
with Reaction (\ref{eqn:N2H2}) as the rate limiting step. The profile we have for NH$_3$
deviates considerably from the results of \citet{Moses2011}, but this is for large part due to a
difference in the nitrogen thermochemistry and initial abundances at high pressures propagating
up through the atmosphere. Figure \ref{fig:HD209-compare} shows that
our quenching height is in both cases higher than for \citet{Moses2011}, suggesting that
the nitrogen in NH$_3$ migrates to N$_2$ more slowly in our network, even overtaking 
\citet{Moses2011} at $\sim 10^{-4}$ bar, but that we start with less NH$_3$ than \citet{Moses2011}.
The increase in NH$_3$ abundance at $\sim 5 \times 10^{-6}$ bar is due to a formation path
for NH$_3$ in \citet{Moses2011} that is less efficient in our network.

\begin{figure}
\epsscale{1.1}
\plotone{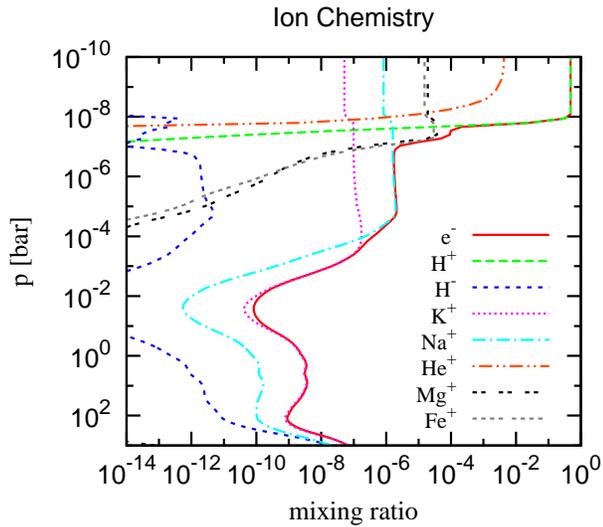}
\caption{Mixing ratios for the dominant ionic species as a function of pressure, $p$ [bar]
for the atmosphere of HD209458b. 
\label{fig:HD209-Ion}}
\end{figure}

We conclude this section with a brief discussion of the most neutral ions, in comparison with
\citet{Lavvas2014}. We have plotted the most abundant ions in Figure \ref{fig:HD209-Ion}. Note 
that, for this paper, $n_{\rm gas}$ is a sum of all neutral gas particles, cations, ions and 
electrons, so the mixing ratio of ions cannot increase above unity. This plot allows a direct 
comparison to \citet[][their Fig's 5 \& 6]{Lavvas2014}. In our model, K$^+$ is the most 
abundant ion deep within the atmosphere, followed by Mg$^+$ and Fe$^+$. \citet{Lavvas2014} does not 
consider these species, but they don't seem to affect the abundances of other ions very much deep 
within the 
atmosphere. When the pressure delves to $10^{-2}$ bar, 
K$^+$ deviates considerably between our results and those of \citet{Lavvas2014}. This is likely due 
to the inclusion of several other ions in our model that
become dominant charge carriers at this height, including several complex hydrocarbon ions, of the
form ${\rm C_nH_m^+}$. This indicates that ion-neutral chemistry can be significantly influenced
by the variety of ions and neutral species under consideration. This will be especially true for
the potassium chemistry. Our network contains a small number of potassium-bearing species. Including
new species and reactions could significantly affect the degree of ionization. It will be interesting
to discover how an expanded potassium and sodium chemistry affects the overall ion-neutral chemistry
and the resulting abundances of trace species.

Between $10^{-3}$ and $10^{-4}$ bar, Na$^+$ overtakes K$^+$ as the dominant positive charge carrier,
and remains so until $\sim 10^{-7}$ bar. This transition, the ratios between the ions, and the
abundances of the ions, are nearly identical between our model and that of \citet{Lavvas2014}. Within
the thermosphere of \object{HD209458b}, there are some small discrepancies between our model and
\citet{Lavvas2014} for He$^+$, and quite large discrepancies for C$^+$ which we suggest are owing
to the non-Alkali photochemistry that \citet{Lavvas2014} include, but that we have not included
here. 

\subsection{Jupiter}
\label{sec:Jupiter-test}

The atmosphere of \object{Jupiter} is divided into three regions: (1) the troposphere, where the
gas-phase temperature $T$ decreases with atmospheric height, (2) the stratosphere, where $T$ is 
roughly constant with increasing height, and (3) the thermosphere, where $T$ increases with height. 
In this
section, we consider the chemical composition of Jupiter's stratosphere.
The stratosphere of Jupiter is rich in hydrocarbons, owing to its large gas-phase C/O ratio, 
because the majority of the oxygen is locked in water ice and then gravitationally
settling to below the tropopause. This is predicted to lead to a ${\rm C/O} \sim 2 \times 10^6$ 
\citep{Moses2005} in the absence of external sources of H$_2$O and CO$_2$ 
\citep{Feucht1997,Moses2000a,Moses2000b} such as Shoemaker-Levy 9 \citep{Cavalie2012}. Jupiter's
stratosphere provides an
extreme example of how surface deposition can radically affect the C/O ratio, an effect more recently 
predicted for exoplanets and brown dwarfs \citep{Bilger2013,Helling2014}.
The high C/O ratio, in combination with the large 
abundance of hydrogen (H$_2$ and CH$_4$ are the two most abundant volatiles in the stratosphere 
and lower thermosphere), means that the stratosphere of Jupiter is strongly 
reducing \citep{Strobel1983}.

\citet{Fouchet2000} have observed ethane and acetylene in Jupiter's stratosphere.
Ethylene has also been observed by \citet{Bezard2001}. The stratospheric chemistry of Jupiter has 
been modeled by several groups, including \citet{Gladstone1996} and \citet{Moses2005}. We adopt
the lower boundary conditions and temperature profile that \citet{Moses2005} used and model the 
carbon-oxygen chemistry in the stratosphere of Jupiter, ignoring
the nitrogen chemistry (most of the nitrogen will be locked in NH$_3$ ice). Boundary conditions
are discussed in Section \ref{sec:boundary-conditions}.

Our lower boundary is set to be identical to \citet{Moses2005}. These boundary conditions are 
somewhat artificial; the carbon
budget is controlled by the photochemistry and the dynamics. There is no effective destruction
pathway for the stable hydrocarbons, but the time-scale for their formation is often competing with
the dynamical time-scales. In the thermosphere, $\sim 10^{-7} - 10^{-8}$ bar, these hydrocarbons are
lost through photodissociation and photoionization as well as molecular diffusion. At the base,
the chemistry is halted once the dynamical timescale is reached, effectively treating the bottom
boundary as an open boundary through which the hydrocarbons would continue to diffuse.
In reality, the complex hydrocarbons are carried into Jupiter's deep atmosphere, where the
high temperatures and pressures dissociate these hydrocarbons, and force the carbon budget to return
to chemical equilibrium values: CH$_4$ with trace amounts of CO and other species. 
\citet[][their Fig. 6]{Visscher2010} demonstrate how the carbon budget is set deep within Jupiter's 
atmosphere; we do not model this region.

\begin{figure}
\epsscale{1.2}
\plotone{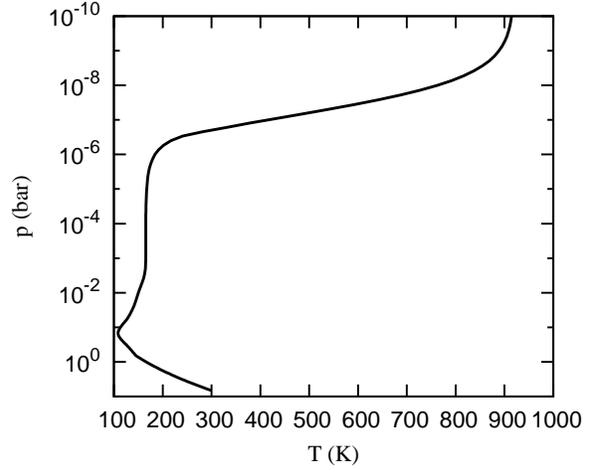}
\caption{Temperature profile for Jupiter, $T$ [K] as a function of $p$ [bar]
\citep{Moses2005}. \label{fig:Jupiter-Profile}}
\end{figure}

\begin{figure*}
\epsscale{2.0}
\plotone{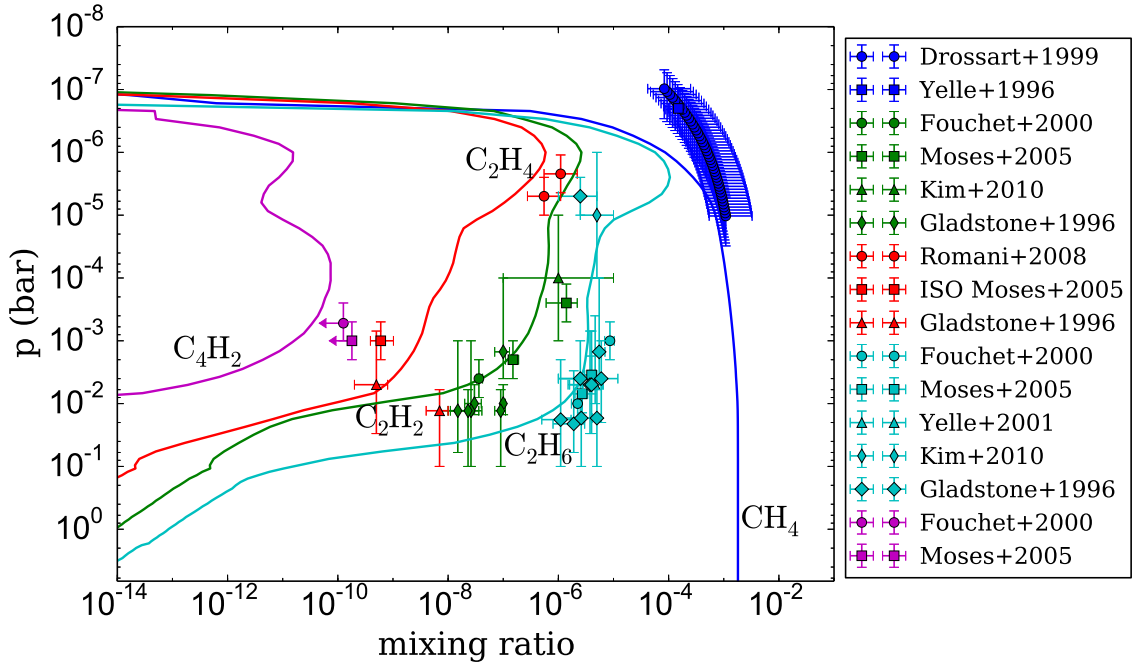}
\caption{Mixing ratios for various chemical species as a function of pressure, $p$ [bar]. A
comparison between our model (solid lines) and that of various observations, for
complex hydrocarbons in the stratosphere of Jupiter. \label{fig:Jupiter-compare}}
\end{figure*}

\begin{figure}
\epsscale{1.2}
\plotone{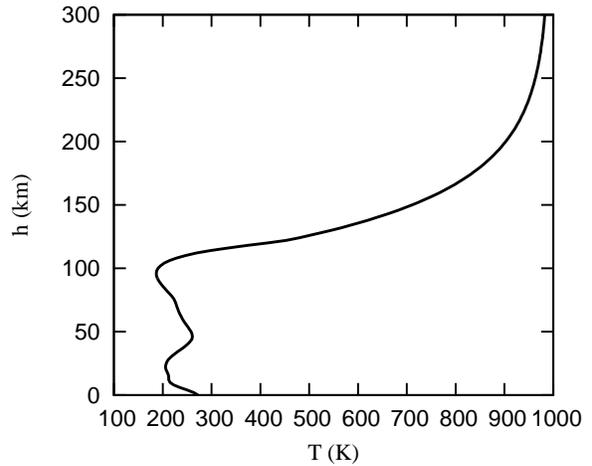}
\caption{Temperature profile used for the Early Earth chemistry, temperature [K] vs. height [km]. 
This profile is a synthetic profile for the Earth's atmosphere generated with the MSIS-E-90 
model for the date 2000/1/1 \citep{Hedin1987,Hedin1991}.\label{fig:Earth-Profile}}
\end{figure}

With these reactions removed from the network, we ran the network using the temperature and $K_zz$
profiles from \citet{Moses2005}, shown in Figure \ref{fig:Jupiter-Profile}. Comparisons between 
our results and a representative set of observations for the depth dependent mixing 
ratios, for the species CH$_4$, C$_2$H$_2$, C$_2$H$_4$, C$_2$H$_6$ and C$_4$H$_2$, are shown in 
Figure \ref{fig:Jupiter-compare}. The observations for CH$_4$ are taken from \citet{Drossart1999} 
and \citet{Yelle1996}, C$_2$H$_2$ observations are from \citet{Fouchet2000}, \citet{Moses2005} and
\citet{Kim2010}, C$_2$H$_4$ observations are from \citet{Romani2008} and \citet{Moses2005}, 
C$_2$H$_6$ observations are from \citet{Fouchet2000}, \citet{Moses2005}, \citet{Yelle2001} and 
\citet{Kim2010}, and the C$_4$H$_2$ observations are from \citet{Fouchet2000} and \citet{Moses2005}.
We also incorporate observations for C$_2$H$_2$, C$_2$H$_4$ and C$_2$H$_6$ from 
\citet{Gladstone1996} and references therein.

Many of the published observations do not include error bars in atmospheric
pressure. Additionally, there may seasonal
in the pressure-temperature structure and the location of the homopause, which adds uncertainty to 
our predictions
as a function of pressure. To account for these sources of uncertainty, we place error bars
for the pressure at a factor of two above and below the published observations when errors in pressure
were not given. These errors in pressure are of the same order as observations where errors in
pressure are given.
We do not compare our results for oxygen-bearing species, because the abundances of these species
are expected to be greatly enhanced in the stratosphere by the addition of an external source of
oxygen, such as Shoemaker-Levy 9.

The differences between our results and those of other models arise 
primarily because of different photochemistries and different rate constants, especially for the
re-formation of methane after its photodissociation,
\begin{align}
{\rm CH_3 + H_2} &\rightarrow {\rm CH_4 + H}, \; \text{and} \\
{\rm CH_3 + H + M} &\rightarrow {\rm CH_4 + M}.
\end{align}
Differences between Jovian photochemical models can result in very large discrepancies between
stratospheric abundances of complex hydrocarbons. The differences between \citet{Gladstone1996} and
\citet{Moses2005} span several orders of magnitude in some cases 
\citep[see][their Fig. 14]{Moses2005}. 

Both ethane and acetylene agree reasonably well between our model and the observations, and 
the results for C$_4$H$_2$ lie more than a factor of five below the observational 
upper limits. Our predictions for the location of the methane homopause do not agree very well
with observations.
We use the eddy diffusion coefficient from Model C in \citet{Moses2005}, and either this or
the use of the Chapman-Enskog diffusion coefficient for Methane may
be the source of the discrepancy. Our results are similar to the Model C results of 
\citet[][their Fig. 14]{Moses2005}. The molecule with the largest discrepancy between the two models 
is ethylene (C$_2$H$_4$), with the largest discrepancy between our predictions and the 1 millibar
observations (ignoring the observation from \citealt{Gladstone1996} that predicts a mixing ratio
of $\sim 10^{-8}$). In our model, the primary path of formation for ethylene follows from the 
photodissociation of ethane (Reaction 2679 in the network),
\begin{equation}
{\rm C_2H_6 + \gamma} \rightarrow {\rm C_2H_4 + H_2},
\end{equation}
and ethane is formed from CH$_4$ following paths to formation like this one:
\begin{align}
2\big({\rm CH_4 + \gamma} &\rightarrow {\rm ^1CH_2 + H_2}\big), \notag\\
2\big({\rm ^1CH_2 + H_2} &\rightarrow {\rm CH_3 + H}\big), \notag\\
{\rm CH_3 + CH_3 + M} &\rightarrow {\rm C_2H_6 + M}; \notag\\
\cline{1-2}
{\rm 2CH_4 + 2\gamma} &\rightarrow {\rm C_2H_6 + 2H}.  \label{eqn:ethane-Jupiter}
\end{align}
These differences may be resolved by a more careful accounting of pressure-dependent branching
ratios, such as those of:
\begin{equation}
{\rm H + C_2H_5} \rightarrow {\rm CH_3 + CH_3}
\end{equation}
from \citet{Loison2015}. We use the Kooij form for these reactions (Section \ref{sec:2-Body}),
which does not account for the effect that pressure has on the rate constant.

Ion-neutral chemistry also makes a contribution, via the formation of C$_2$H$_4$ from the reaction
\begin{align}
{\rm CH_5^+ + C_2H_2} &\rightarrow {\rm C_2H_3^+ + CH_4} \notag\\
{\rm C_2H_3^+} + e^- + {\rm M} &\rightarrow {\rm C_2H_3 + M} \notag\\
{\rm C_2H_3 + CH_4} &\rightarrow {\rm C_2H_4 + CH_3},
\end{align}
and CH$_5^+$ forms from a series of reactions starting with the photoionization of CH$_3$ and
then a series of hydrogen abstractions, ${\rm CH_x^+ + H_2} \rightarrow {\rm CH_{x+1}^+ + H}$.
It should be emphasized that this is not the primary formation pathway for ethylene, but it is
an important path of formation in our chemistry and makes some contribution to the mixing ratios
at 1 millibar.

Finally, there is a large discrepancy for CO, but this is not due to differences in the chemistry.
Rather, this results from \citet{Moses2005} injecting CO, CO$_2$ and H$_2$O into Jupiter's 
stratosphere.
Inclusion of this external source of oxygen-bearing species is justified by a number of data-model
comparisons mentioned at the beginning of this section. We neglected to include these external
sources, and therefore oxygen-bearing species, especially H$_2$O and CO$_2$ (not shown) fail
to agree with observations. Our results therefore suggest that some external source of oxygen-bearing
species is necessary to explain the H$_2$O and CO$_2$ observations in Jupiter's stratosphere.

\subsection{The Earth}
\label{sec:Earth-test}

The Earth's atmosphere is well studied, and the profiles of trace species are
well constrained, and the formation and destruction of these species is controlled
by photochemistry and deposition. Comparing our results to the present day Earth
atmosphere therefore provides a comprehensive test of our chemical network 
(Section \ref{sec:Earth-today}). Additionally,
the connection between lightning-driven and NO$_x$ chemistry\footnote{Referring primarily 
to NO and NO$_2$ chemistry.} has been extensively studied
with experiments, observations and models, and provides a useful regime in which to compare
the results of \textsc{Stand2015} applied to a lightning shock model 
(Section \ref{sec:lightning}). It is important
to find out what our model predicts in habitable environments before the onset of life, and
so we apply our model to the Early Earth (Section \ref{sec:early-Earth}).

\subsubsection{Present Day Earth Atmosphere}
\label{sec:Earth-today}

The best understood planetary atmosphere, in terms of both models and
observations, is the atmosphere of the present day Earth. Earth's atmosphere
has been studied \emph{in situ}, with the use of countless balloon experiments
used to measure various trace elements, and remotely, with satellite measurements.
Models of Earth's atmosphere range from simple to complex, both dynamically (1D diffusion
to 3D global circulation models) and chemically (from treating only oxygen and hydrogen
chemistry to modeling the transport and chemistry of chlorofluorocarbons and biological
aerosols). \citet{Seinfeld2006} provide a useful introduction and
review to the subject.

\begin{figure}
\epsscale{1.0}
\plotone{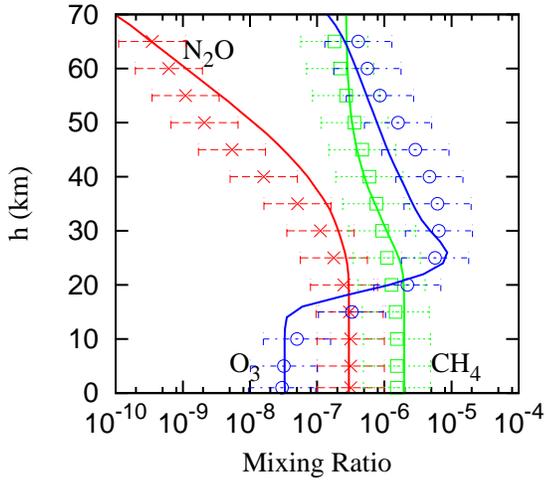}
\caption{Mixing ratios of ozone, methane and nitrous oxide as a function of atmospheric height [km] for the atmosphere of the present-day Earth.
The lines are produced by our model and the points are taken from globally averaged measurements 
\citep{Massie1981}. Errors are set to an order of magnitude to account for diurnal and latitudinal variations.
\label{fig:O3-compare}}
\end{figure}

\begin{figure}
\epsscale{1.0}
\plotone{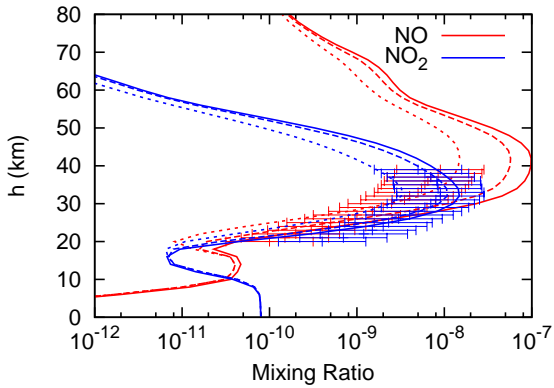}
\caption{Mixing ratios of NO and NO$_2$ as a function of atmospheric height [km]for the atmosphere of the present-day Earth.
The lines are produced by our model and the points are taken from balloon measurements 
\citep{Sen1998}. Errors are set to an order of magnitude to account for diurnal and latitudinal variations. We also show the results from suppressing the rate constant for Reaction 1300 in the network by a factor of
2 (dashed) and a factor of 10 (dotted).
\label{fig:NO-compare}}
\end{figure}

\begin{figure}
\epsscale{1.0}
\plotone{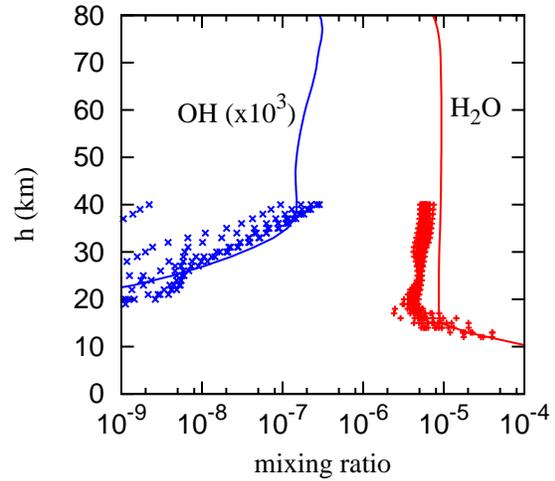}
\caption{Mixing ratios of OH and H$_2$O as a function of atmospheric height [km] for the atmosphere of the present-day Earth.
The lines are produced by our model and the points are taken from balloon measurements at various latitudes,
heights, and times \citep{Kovalenko2007}.
\label{fig:H2O-compare}}
\end{figure}

Our interest is in validating our photochemical network to the present-day Earth,
and not in coupling Earth's geochemistry to its atmospheric chemistry. We
therefore make some simplifying assumptions when we set our boundary conditions.
We compare our model to the contemporary Earth by setting the lower
boundary conditions, temperature profile and external UV field as given in 
Section \ref{sec:boundary-conditions}. We present these comparisons for
O$_3$, CH$_4$ and N$_2$O (Figure \ref{fig:O3-compare}), NO and NO$_2$
(Figure \ref{fig:NO-compare}) and OH and H$_2$O (Figure 
\ref{fig:H2O-compare}).

The data for O$_3$, CH$_4$ and N$_2$O is taken from the 
globally averaged mixing ratios from \citet{Massie1981}. Following 
\citet{Hu2012}, we apply error bars spanning an order of magnitude in
mixing ratio to reflect the temporal and spatial variations. Our model
fits the measured CH$_4$ to within the error bars throughout the 
atmosphere. The O$_3$ predicted by the model deviates from the data with
errors at 15 km, and the N$_2$O deviates from the data with errors
between 40 km -- 55 km. This may be due to an over-estimation of the
optical depth. If more UV photons in the model penetrated through to
$\sim 10$ km, the O$_3$ mixing ratios would be enhanced at 15 km, and the
N$_2$O mixing ratios would be destroyed more efficiently deeper in the
atmosphere.

The data for NO and NO$_2$ is taken from balloon observations at 35
deg N in 1993 \citep{Sen1998}, and here also we apply error bars spanning
an order of magnitude to reflect spatial and temporal variations. As with
\citet{Hu2012}, we seem to overpredict the abundance of NO in the upper
atmosphere (30-40 km). We find that this overprediction is due to
Reaction 1300 in the network:
\begin{align}
{\rm N_2O + O(^1D)} &\rightarrow {\rm NO + NO}; \\
k &= 7.25 \times 10^{-11} \; {\rm cm^3 \, s^{-1}}. \notag
\end{align}
We use the rate suggested by the JPL Chemical Kinetics and Photochemical Data
for Use in Atmospheric Studies \citep{Sander2011}. If the rate constant for this reaction
is decreased by a factor somewhere between 2 and 10, we come into much better agreement at 30-40 km, 
and worse agreement between 20-30 km (see Figure \ref{fig:NO-compare}).

Finally, the data from OH and H$_2$O was taken from
balloon measurements at various latitudes and heights in 2005 
\citep{Kovalenko2007}. We plot each individual datapoint without error bars
in order to represent the observed variations; changes at other points
of the globe or at other times of the year or day may lead to more significant
variations in the abundances. The H$_2$O predictions are within a factor of 
five of the observed water abundance, and our OH predictions lie within the
measurements, indicating that the model correctly reproduces the water and
OH mixing ratios.

\subsubsection{Lightning Shock Model and NO$_x$ chemistry}
\label{sec:lightning}

\begin{figure}
\epsscale{1.0}
\plotone{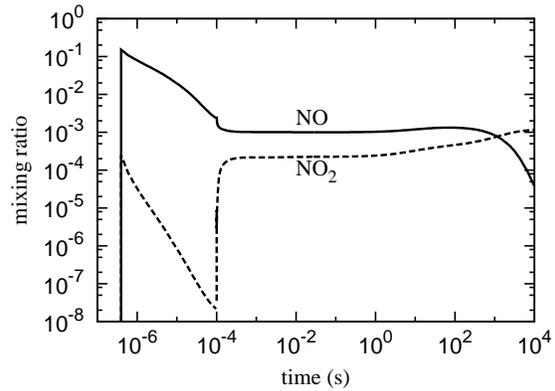}
\caption{Mixing ratios of NO (solid) and NO$_2$ (dashed) vs time [s] in a simulation of a lightning shock on a parcel of gas with an Earth-like atmospheric composition, initially 
at 300 K and 1 bar. The temperature and pressure vary as a function of time as described by
\citet{Orville1968}, until $10^{-4}$ s, at which time conditions are returned to 300 K and 1 bar,
and the system is allowed to further evolve.\label{fig:lightning}}
\end{figure}

It is also useful to to the model's NO$_x$ lightning-driven chemistry in the present day 
atmosphere. For this purpose, we apply a simple shock model in order to explore the formation of 
NO$_x$ species due to lightning at a single small region in the atmosphere. We employ the temperature
and pressure calculations of \citet[][his Fig's 1 and 3]{Orville1968} and the time-scaled results
of \citet[][their Fig's 2 and 3]{Jebens1992}, fitting these to an exponential function.
We use the following functions of temperature and pressure:
\begin{align}
T(t) &= {\rm 300 \, K} + ({\rm 29800.0 \, K}) \, e^{-t/({\rm 55.56 \, \mu s})};\\
P(t) &= {\rm 1.0 \, bar} + ({\rm 7.0 \, bar}) \, e^{-t/({\rm 5.88 \, \mu s})}.
\end{align}a
We 
start with present day atmospheric chemistry at the base of the troposphere, except without the
N$_2$O, NO and NO$_2$ species, and with $T = 300$ K and $p = 1$ bar. The shock occurs at 1 ns, and
is allowed to evolve until 0.1 ms. At this point the calculation is terminated, and another calculation
initiated using for its initial conditions the final conditions of the shock model, except with
temperature and pressure returned to 300 K and 1 bar, respectively. This model is run until $10^4$ s
and results are shown in Figure \ref{fig:lightning}.

We find that the NO$_x$ species are formed in our model thermally by the 
Zel'dovich mechanism \citep{Zeldovich1966}:
\begin{align}
{\rm O_2 + M} &\rightleftharpoons {\rm O + O + M}, \\
{\rm N_2 + M} &\rightleftharpoons {\rm N + N + M}, \\
{\rm O + N_2} &\rightarrow {\rm NO + N}, \\
{\rm N + O_2} &\rightarrow {\rm NO + O}, \\
\cline{1-2}
{\rm O_2 + N_2} &\rightarrow {\rm 2NO}.
\end{align}
We compare our NO yield to the lightning discharge experiments performed by \citet{Navarro2001}. 
We use for our NO mixing ratio the values found before the end 
of the shock ($10^{-4}$ s in Fig. \ref{fig:lightning}), between $10^{-2}$ and $10^{-3}$, to 
\cite [][their Eq. 4]{Navarro2001}. We find that:
\begin{align}
P({\rm NO}) &\approx \big(2.4 \times 10^{22} \, {\rm K/J}\big) \dfrac{X({\rm NO})}{T_f}\notag\\ 
            &\approx 2-20 \times 10^{16} \, {\rm molecules \; J^{-1}},
\end{align}
where $T_f$ [K] is the ``freeze-out'' temperature after which the NO mixing ratio does not change 
appreciably over the time-scale of the experiment, which we set to 1000 K (the approximate temperature
of our model at $t \approx 10^{-4}$ s). This is consistent with the production of NO in the ``hot core''
region of the experiment. This is also roughly consistent with the literature values for NO production
of $10^{17}$ molecules J$^{-1}$ \citep{Borucki1984,Price1997}.

This is an order-of-magnitude comparison between the code and lightning experiments and models,
and for a more complete comparison will need to be applied to a model atmosphere, where diffusion and
photochemistry together will further process the NO$_x$ species. We plan to do this in a future paper.

\subsubsection{The Early Earth}
\label{sec:early-Earth}

The presence of life and the evolution of the sun both have radically altered \object{Earth}'s 
atmospheric chemistry. \citet{Oparin1957} and \citet{Miller1959} thought that the atmosphere of the 
early \object{Earth}\footnote{``early Earth'' in this context means the \object{Earth} in its 
first 1 Gyr} was largely reducing, dominated by methane, ammonia and molecular hydrogen. 
\citet{Kasting1993} made a compelling case that prebiotic formation of hydrogen would be too slow 
to allow for much molecular hydrogen in the atmosphere of the early \object{Earth}. Furthermore,
a major constituent in the early Earth atmosphere needs to be a strong greenhouse gas, in order to
compensate for the cooler young sun. The atmospheric chemistry of the early Earth is difficult
to determine, and a severe lack of data results in many possible early Earth chemistries. As
an illustrative example, \citet{Tian2005} argue that hydrogen escape was less efficient during
the first 1 Gyr as was previously thought\footnote{The debate is ongoing 
\citep{Claire2006,Catling2006}.}. If \citet{Tian2005} are correct, then Earth's early 
atmospheric composition could have been reducing.

We present a model of the atmosphere of the early \object{Earth}, using the same lower boundary
conditions as shown in \citet[][his Fig. 1]{Kasting1993}, and a temperature profile for
the present Earth 
\citep{Hedin1987,Hedin1991}\footnote{\url{http://omniweb.sci.gsfc.nasa.gov/vitmo/}}, 
shown in Figure \ref{fig:Earth-Profile}.
The lower boundary conditions used for the early Earth are
given in Section \ref{sec:boundary-conditions}. We treat outgassing using the deposition
method (Appendix \ref{app:outgassing}). 

\begin{figure}
\epsscale{1.2}
\plotone{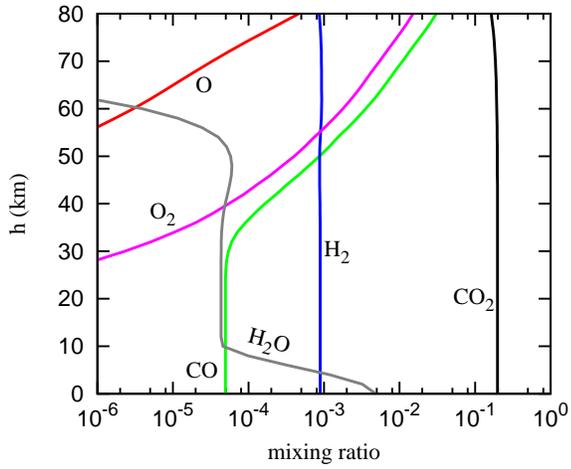}
\caption{Mixing ratios for O, H$_2$, CO, O$_2$, H$_2$O and CO$_2$, as a function of height [km], 
for early Earth photochemistry. These results can be compared to the results of 
\citet[][his Fig. 1]{Kasting1993}. \label{fig:Earth-compare}}
\end{figure}

We compare our results to those of \citet[][esp his Fig. 1]{Kasting1993}. Our results
are presented in Figure \ref{fig:Earth-compare}. The results compare reasonably well for
CO and O$_2$, but not for H$_2$O and O. The CO abundance begins to increase at 30 km, 10 km
higher than for \citet{Kasting1993}, and achieves a mixing ratio of $\approx 5 \times 10^{-3}$ at 60 km,
which is within a factor of 2 of \citet{Kasting1993}. The O$_2$ likewise begins to rise above
a mixing ratio of $10^{-6}$ 10 km higher in the atmosphere, and also achieves a mixing ratio of 
$\approx 2.5 \times 10^{-3}$, again within a factor of 2 of \citet{Kasting1993}. The water vapor profile
is quite different, however. Instead of falling below a mixing ratio of $10^{-6}$ at 10 km, the H$_2$O
mixing ratio in our model levels out at 5 $\times$ 10$^{-4}$, increasing slightly at $\sim 50$ km before
plummeting. Also, the oxygen mixing ratio only reaches $\approx 3 \times 10^{-6}$, approximately two orders
of magnitude below the mixing ratio predicted by \citet{Kasting1993}. These differences may be due to
the different assumed young solar UV fields between ourselves and \citet{Kasting1993}, but we suspect 
that the differences are more likely due either to differences in the water condensation or the temperature 
profiles used. This seems especially likely for atomic oxygen, which is primarily destroyed by the reaction:
\begin{equation}
{\rm O + H_2O} \rightarrow {\rm OH + OH},
\end{equation}
in spite of the sizeable $7640$ K barrier. When the water vapor drops off at $\sim 55$ km,
this destruction route becomes unviable, and the atomic oxygen mixing ratio rapidly increases.
a
\section{Glycine Formation in a Laboratory Environment}
\label{sec:Laboratory}

The formation of glycine, among several other amino acids, amines and nucleotides, has been 
investigated for a variety of chemical compositions, from reducing \citep{Miller1953} to 
oxidizing \citep{Schlesinger1983,Miyakawa2002,Cleaves2008}, and exploring various energy 
sources \citep[see][and references therein]{Miller1959}. In a recent experiment,
HCN and H$_2$S were exposed to UV light (peak frequency 2540 \AA), resulting in the formation
of numerous complex prebiotic compounds \citep{Patel2015}. The techniques used in this experiment
afforded the experimenters to track the pathways of formation for these various species.

Prebiotic species are produced in smaller concentrations within a more oxidizing 
environments \citep{Miller1959}. {Methane has been found to be important for the formation
of prebiotic compounds} 
{\citep{Schlesinger1983,Miyakawa2002}. The correlation between reducing}
chemistry and the efficient production of prebiotic molecules, combined with compelling evidence 
that the atmosphere of the early Earth was oxidizing \citep{Kasting1993}, would suggest that 
other processes were responsible for producing the prebiotic chemical inventory on Earth. 
This process is hypothesized to have taken place within hydrothermal vents 
\citep[e.g.][]{Ferris1992}, on the surfaces of crystals \citep{Vijayan1980}, or possibly within the 
interstellar medium \citep[e.g.][]{Greenberg1995}.

\citet{Cleaves2008} have repeated Miller's experiment in a reducing environment, and discovered that
amino acids can be efficiently produced in such environments, but that nitrites (e.g. HONO) destroy 
these species as quickly as they are produced. Adding ferrous iron, in the form of FeO or FeS$_2$ 
(in the form of pyrite surfaces) effectively removes the nitrites and allows the amino acids to
survive.

We explore the formation of glycine in the context of a weak radiating source. An unattenuated 
monochromatic beam of light at $\lambda_0 = 1000$ \AA\ is applied with an intensity of 
$\approx 2 \times 10^{-3}$
erg cm$^{-2}$ s$^{-1}$, corresponding to a flux of $F_0 = 10^8$ photons cm$^{-2}$ s$^{-1}$. This flux
is applied to Eq. (\ref{eqn:photo-rates}) such that:
\begin{align}
k_{{\rm ph},i}(z) &= \int_{{\rm 1 \AA}}^{{\rm 10^4 \AA}} \sigma_i(\lambda) \, F_0 \delta(\lambda - \lambda_0) \; d\lambda; \notag\\
&= F_0 \sigma_i(\lambda_0),
\end{align}
where $\delta$ is the Dirac delta function.

The
formation pathways for glycine have not been rigorously determined, although there are some
proposed pathways. We include four possible pathways to glycine formation in our network. First,
we include glycine formation via the three body interaction of various species. These reactions
have significant barriers, and so will only occur efficiently at rather high temperatures.
The reactions are:
\begin{align}
{\rm C_2H_4 + HNO_2} &\rightarrow {\rm NH_2CH_2COOH}, \\
{\rm C_2H_5 + NO_2} &\rightarrow {\rm NH_2CH_2COOH}, \\
{\rm CH_3NO + H_2CO} &\rightarrow {\rm NH_2CH_2COOH},
\end{align}
with rate constants set equal to the three-body formation for analogous chemical species 
(e.g. CH$_2$COOH). Also included is the ion-neutral pathway proposed for interstellar formation for 
glycine from \citet{Charnley1997},
\begin{align}
{\rm CH_6NO^+ + HCOOH} &\rightarrow {\rm C_2H_6NO_2^+ + H_2O}, \\
{\rm C_2H_6NO_2^+} + e^- &\rightarrow {\rm NH_2CH_2COOH + H}.
\end{align}
And finally, the formation of glycine by a possible pathway similar to that suggested by 
\citet{Patel2015},
\begin{equation}
{\rm CH_3NO + CH_3O} \rightarrow {\rm NH_2CH_2COOH + H}
\end{equation}
is included.

\begin{figure}
\epsscale{1.2}
\plotone{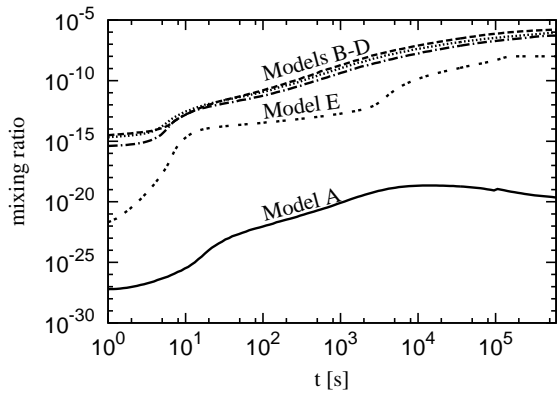}
\caption{Mixing ratio of glycine as a function of time, for five lab simulations, 
labelled Models A-E, with parameters given in Table \ref{tab:glycine} and described in Section
\ref{sec:Laboratory}.\label{fig:glycine}}
\end{figure}

Additionally, we include FeO and reactions between FeO and nitrites. We also inject our gas with
HCOOH in order to facilitate the ion-neutral formation pathway; it is likely that there are other
presently unknown paths of formation for formic acid. We run this network for a set of 
five different initial compositions given
in Table \ref{tab:glycine}, labeled Model A - E. Model A is a strongly reducing
environment, with only the gases NH$_3$, CH$_4$, H$_2$ and H$_2$O, FeO and HCOOH (Model A). We
transition to a more reducing environment in the successive models
(Models B, C, D). Finally, for Model E, we run the experiment starting solely from CO$_2$, N$_2$,
H$_2$O, FeO and HCOOH. We run all models using the unattenuated UV flux, at 1 bar pressure 
and 300 K
temperature. The model is run to $t \approx$ 1 week. Our results are plotted in 
Figure \ref{fig:glycine}.

Moving from Model B to E, less and less glycine is formed, falling from a mixing ratio
of $10^{-6}$ for Model B to $10^{-8}$ for Model E. This is what is expected from the Miller-Urey
experiments performed for various chemical compositions: as the chemistry becomes less reducing,
it becomes more difficult to form prebiotic molecules.

More interesting is Model A. If all N$_2$
and CO$_2$ are removed, certain formation pathways to NO$_2$, HNO$_2$ and especially H$_2$CO are
inhibited. Additionally, HCNO forms more slowly from HCN, and especially the ionic form, CHNO$^+$
(in its various permutations) is difficult to form without some excess unbonded atomic nitrogen
or oxygen present in the gas. Model A produces virtually no glycine. We traced this back to the key 
reactions:
\begin{align}
{\rm N_2} + \gamma  &\rightarrow {\rm N_2^+} + e^-, \notag\\
{\rm N_2^+} + e^- &\rightarrow {\rm N + N}, \notag\\
{\rm CO_2} + \gamma &\rightarrow {\rm CO + O}
\end{align}
the same formation pathway for amines in the early Earth as suggested by \citet{Zahnle1986}. In our 
case, however, the atomic nitrogen and oxygen are both important in completing the formation of 
HCNO and its isomers.

\section{Conclusion}
\label{sec:conclusion}

In this paper, we have presented a gas-phase chemical network, \textsc{Stand2015}. The 
photochemistry/diffusion code, \textsc{Argo}, was used to test the network. We have shown that the 
predictions from 
\textsc{Stand2015} converge to chemical equilibrium under the appropriate conditions and also that
the molecular diffusion modeled by \textsc{Argo} makes a reasonable approximation to analytical
calculations of molecular diffusion for an isothermal gas in hydrostatic equilibrium. We have
compared our model results (\textsc{Stand2015}+\textsc{Argo}) to chemical kinetics models for
\object{HD209458b}, \object{Jupiter} and the \object{Earth}. For \object{Jupiter}, we found
that ion-neutral chemistry may provide significant alternative pathways to formation of various
hydrocarbons, especially ethylene (C$_2$H$_4$).

Finally, we numerically simulate a Urey-Miller-like experiment\footnote{The experiment we simulate 
is more 
like that of \citet{Patel2015}.} under various initial chemistries. We found that, in an artificial 
environment, when derivatives of FeO and pyrite (FeS$_2$) can destroy nitrites in the presence 
of a reservoir of formic acid, the formation of glycine is considerable also in reducing 
environments, 
approaching a mixing ratio of $\sim 10^{-6}$. For an environment more similar to the atmosphere of 
the early \object{Earth}, the mixing ratio drops to $\sim 10^{-8}$. Surprisingly, for a gas without 
any CO$_2$, O$_2$ or N$_2$, virtually no glycine is formed. If this result is robust for various other
energy sources (shocks, thermal energy, etc.) and for other prebiotic species, this would 
suggest that the early \object{Earth} chemistry should not be too strongly reducing, else the 
formation of glycine and other prebiotic species would be severely inhibited.

This network has limitations. It has only been tested for 1D atmosphere 
models, with non-self-consistent temperature profiles. Using this network within a global circulation
model is presently unrealistic, but a reduced version of this network, constructed specifically for
given atmospheres, could in principle be employed in 2D or 3D atmosphere simulations. Sulfur
chemistry has been shown to play an important role in the formation of prebiotics, and is an
essential constituent in volcano plumes. The inclusion of sulfur chemistry will be a natural
next step to take the model. Additionally, the models of prebiotic chemistry should consider the
formation of species other than glycine. The formation of ribose (C$_5$H$_{10}$O$_5$) of 
nucleotides, such as adenine (C$_5$H$_5$N$_5$), and of phosphorus-bearing species should
also be included to more fully encapsulate the formation of the prebiotic chemical 
reservoir. 

One serious problem with this network,
and indeed with any chemical kinetics network, is the uncertainty in rate coefficients. The effects
of this uncertainty can be estimated using sensitivity analysis \citep[e.g., within][]{Venot2012},
but can ultimately only be resolved slowly as better experimental and theoretical determinations
of the reaction rates are made available. More accurate determinations,
especially of the reaction rates for the nitrogen chemistry, would be extremely helpful.
This network
and model provide a window into a detailed analysis of prebiotic chemistry, but much work must still
be done in order to accurately predict the full budget of prebiotic molecules in the variety of
environments in which they may occur.

\acknowledgments

Both authors gratefully acknowledge the support of the ERC Starting Grant \#257431.
P.~B.~R. is grateful to J.~I. Moses for several helpful discussions about the network and model
comparisons, to G. Laibe for his help with the Lagrangian numerical methods, and to C.~R. Stark
for his help understanding prebiotic formation in plasma environments. Both authors are grateful
for the anonymous referee whose report has helped significantly improve this paper. They also 
express thanks to Ian Taylor at St Andrews for his help with computational resources. Finally, they
acknowledge the National Institute of Standards and Technology, the databases of which were
essential to the completion of this project. 

\appendix

\section{List of Species, Reactions and Rates}
\label{app:network}

The purpose of this appendix is to explicitly lay out the content of the chemical network
itself. We list the species considered in the network and the reactions.

The species include the elements H/C/N/O, and the network includes a complete chemistry for
molecules and ions of up to 2 carbon, 6 hydrogen, 2 nitrogen and 3 oxygen atoms. The different
chemical kinetics for various neutral molecular isomers is included as completely as possible, 
although much about branching ratios for reactions is presently not well understood. A list of
all the neutral species is given in Table \ref{tab:list-of-species}. This table lists the species
considered and includes the formula as used in the network, the standard formula, the
name of the molecule and the source we used for the thermochemical data. In some cases, the
chemical formula in the network is different from the standard chemical formula. This is because
we incorporated our own method for distinguishing isomers, in order to make sure that we did not
incorporate the same molecule under two different formulas. 

This list also includes some species with the elements Na, Mg, Si, Cl, K, Ti and Fe. The chemistry 
attempts to include only the dominant species with these elements, in which they would be present 
in the gas phase. These species are generally only present in the gas-phase for very high 
temperatures (generally $> 1000$ K). For cooler objects, these species are typically ignored. The 
noble gases He and Ar are included, both for the sake of completeness, and because they can play 
an important role in organic ion-neutral chemistry through charge-exchange reactions.

Ions are also included, and a list of the ionic species is given in Table \ref{tab:list-of-ions}.
In this case, the uncertainty in reaction rates and branching ratios is much more severe, and
so we made no attempt at present to distinguish isomers of ionic species. 

It is difficult to determine which rate constants to use for a specific reaction, since there are
often many to choose from, and they do not always agree well with each other. We employed the
following method for determining which rate constant to include in our network, after plotting
all the rate constants versus temperature over a range of 100 K to 30000 K:
\begin{enumerate}
\item If there exists only one published rate constant for a given reaction, we use that
value.
\item Reject all rate constants that become unrealistically large at extreme temperature.
\item Choose rate constants that agree with each other over the range of validity.
\item If the most recent published rate constant disagrees with (3), and the authors give
convincing arguments for why the previous rates were mistaken, we use the most recently published
rate.
\end{enumerate}
The full list of forward reactions and rate constants determined by this method comprise the
\textsc{Stand2015} network and are given in Table  \ref{tab:STAND-network}. 
Reverse reactions are not explicitly shown; when reactions are
reversible, bidirectional arrows are shown. When they are irreversible, or simply not reversed
in the network, only unidirectional arrows are shown. Table \ref{tab:STAND-network} additionally
includes a full list of the references for the rate constants used for each given reaction.

\section{Reversing Reactions}
\label{app:reverse}

For reverse reactions, we follow the prescription given by \citet{Burcat2005}. 
For the reaction:
\begin{equation}
{\rm A + B} \rightarrow {\rm C + D + E},
\label{eqn:reversing}
\end{equation}
there is a rate constant, $k_f$. We resolve to determine the reverse rate constant, $k_r$,
for the reaction:
\begin{equation}
{\rm C + D + E} \rightarrow {\rm A + B}.
\end{equation}
Note that the number of species is different between the r.h.s. and l.h.s. of 
Eq. (\ref{eqn:reversing}). We denote this difference in number of reactants and products
($n_{\rm react}$ and $n_{\rm prod}$, respectively) by 
$\Delta \nu$, which in our case $= n_{\rm prod} - n_{\rm react} = 2 - 3 = -1$.
We then solve for the reaction rate constant as \citep[][their Eq. (6)]{Burcat2005},
\begin{align}
K_c =& (RT)^{-\Delta \nu}\, \exp\Big(\Delta a_1 \big(\log T - 1\big) + \dfrac{\Delta a_2T}{2}+\dfrac{\Delta a_3T^2}{6}\notag\\ 
&+ \dfrac{\Delta a_4T^3}{12} + \dfrac{\Delta a_5T^4}{20}-\dfrac{\Delta a_6}{T} + \Delta a_7\Big), \label{eqn:rate-constant}
\end{align}
where $R = 8.314472$ J mol$^{-1}$ K$^{-1}$ is the gas constant, and 
$\Delta a_i = a_i({\rm C + D + E}) - a_i({\rm A + B})$ for $1 \leq i \leq 7$ are the
NASA thermodynamics coefficients, which \citet{Burcat2005} describes and tabulates.
It is important to emphasize here, as done by \citet{Visscher2011}, the multiplicative factor
$(RT)^{-\Delta \nu}$, which in our example would be $1.38065 \times 10^{-22}\,T$.

The Burcat values for the NASA coefficients have been used for all possible species
(see Tab. \ref{tab:list-of-species}. For some species, however, the coefficients had to be
 obtained from other sources. For sources with elements Na, Mg, Si, Cl, K, Ti and Fe, the
Burcat values were sparse, so we made use instead of the NASA-CEA values 
\citep{McBride1993,Gordon1999}, which use 9-coefficient polynomials, so we fit them to a
series of 7-coefficient polynomials for various temperature ranges. We do the same for
the monatomic gases and ions at high temperatures $6000$ K $< T < 20000$ K, using fits to the
polynomials provided by \citet{Gordon1999}. For some species, the thermodynamic properties have
not been determined. In these cases, for neutral species we use Benson's additivity method as 
described by \citet{Cohen1993}.

Benson's additivity method can be naturally combined with the NASA and Burcat polynomial 
coefficients using the experimental values for the small alkanes listed within 
\citet{Cohen1993}. For the arbitrary alcohol from \citet{Cohen1993}, we use methanol, 
and for the arbitrary ether, dimethyl ether.
The Benson coefficients are:
\begin{align}
P_i &= \frac{1}{2}a_i({\rm C_2H_6}), \\
S_i &= a_i({\rm C_3H_8}) - a_i({\rm C_2H_6}), \\
D_i &= a_i({\rm C_2H_6O}) - a_i({\rm C_2H_6}), \\
F_i &= a_i({\rm CH_3OH}) - a_i({\rm CH_4}).
\end{align}
Here, $a_i(X)$ denotes the seven coefficients, $i = 1,...,7$ for species $X$. The coefficients
for fundamental bonds are calculated using these coefficients as follows:
\begin{align}
a_i([{\rm C-H}]) &= \frac{1}{2}P - \frac{1}{4}S, \\
a_i([{\rm C-C}]) &= \frac{3}{2}S - P, \\
a_i([{\rm C-O}]) &= \frac{1}{2}D + \frac{3}{4}S - \frac{1}{2}P, \\
a_i([{\rm O-H}]) &= F - \frac{1}{2}D - \frac{1}{2}S.
\end{align}
The values for these bonds are given in Table \ref{tab:Benson}. The values for [N-H], [N-C] and
[N-O] are similarly determined.

It has been suggested by \citet{Lias1988} and \citet{Cohen1993} that using Benson's additivity
method to determine the thermodynamic properties of ions, or at least strongly of strongly 
polarizing groups, can lead to large errors, because the thermodynamic properties of ions do
not depend linearly on their length, although \citet{Hammerum1999} have had some success applying
Benson's method to ions. 

We found, by investigating the thermodynamic properties of ionic species tabulated 
by \citet{Burcat2005}, that the thermodynamic properties of ions do depend nonlinearly but 
predictably based on size. We therefore placed all the known thermodynamic properties of ions into 
a database, and have extrapolated to calculate the thermodynamic properties for the undetermined 
ions.

\section{Outgassing, Condensation, Evaporation and Escape}
\label{app:outgassing}

Boundary conditions play a key role in determining the atmospheric compositions of planets. For
rocky planets, these  boundary conditions are set by outgassing and escape into the exosphere. At 
temperatures $\lesssim 1500$ K, metals such as silicates, iron, corundum, begin to condense out of 
the exoplanet and brown dwarf atmospheric gas phase. At much lower temperatures, when, various 
other species (e.g. water, ammonia, methane) may also condense out. It is important for comparison 
to previous models to consider both the atmospheric boundary conditions and atmospheric condensation.

As discussed in Section \ref{sec:Equation}, there exist, in addition to the \textsc{Stand2015}
reactions, a series of ``banking'' reactions for all major species, that collect particles and 
reintroduce them to the fluid parcel at a rate determined by the diffusion time-scales. 
The very bottom banking reaction can be set to act effectively as an outgassing rate. 
Imagine a particular reservoir
for a species, A. This reservoir is outgassing into the atmosphere with a flux, $\Phi({\rm A})$ 
[cm$^{-2}$ s$^{-1}$]. This can be accounted by first taking a reservoir concentration of A, which 
for a large reservoir will be $\gg [{\rm A}] (z = 0)$, the bottom of the atmosphere. For a reservoir 
that will not be appreciably depleted over the chemical-dynamical time-scale of the atmosphere, the 
rate is simply:
\begin{equation}
P_{\rm out}({\rm A}) = \dfrac{\Phi({\rm A})}{\Delta z}.
\label{eqn:outgassing-large-reservoir}
\end{equation}
For a finite reservoir, we can place the reservoir concentration into the bottom ``bank'' for the 
species in question, and the $t = 0$ flux, $\Phi({\rm A},0)$, and concentration ($[{\rm BA}]$) can 
be used to determine the rate of outgassing,
\begin{equation}
P_{\rm out}({\rm A}) = L({\rm BA}) \, [{\rm BA}],
\label{eqn:outgassing-small-reservoir}
\end{equation}
where
\begin{equation}
L({\rm BA}) = \dfrac{\Phi({\rm A},0)}{[{\rm BA}(t = 0)]\,\Delta z}.
\label{eqn:flux-estimate}
\end{equation}
These approximations are not used anywhere in this paper. For \object{HD209458b}, we simply start 
with solar elemental abundances, with everything in atomic form at the bottom of the atmosphere. For 
both \object{Jupiter} and \object{Earth}, we start with fixed lower boundary conditions.

Condensation or evaporation of species A can be treated by the reactions (``JA'' represents
``A in condensate form''):
\begin{align}
{\rm A} &\rightarrow {\rm JA}, & \text{for condensation}; \\
{\rm JA} &\rightarrow {\rm A}, & \text{for evaporation}.
\end{align}
This physical process is treated in two ways in this paper for \object{Earth} and \object{Jupiter}. 
The first method is by considering supersaturation concentrations, above which the species in 
question condenses out and below which the species in question will evaporate. This method is given 
by \citep{Hamill1977,Toon1981,Hu2012}, and has the form (for species A):
\begin{align}
P &= \dfrac{[{\rm A}]}{t_c}, \;\;\;\;\;\; L = \dfrac{[{\rm JA}]}{t_c} \\
t_c &= \dfrac{m_{\rm A}v_{th}}{4 \rho_{\rm nuc}}\dfrac{n_{\rm gas} - n_c(T,p)}{a}
\label{eqn:condense-timescale}
\end{align}
where $m_{\rm A}$ [g] is the mass of the condensing species, $v_{th}$ is the thermal velocity of the 
gas, $\rho_{\rm nuc}$ [g cm$^{-3}$] is the material density of the condensation seed, 
$n_{\rm gas}$ [cm$^{-3}$] 
is the density of the gas, and $n_c$ [cm$^{-3}$] is the saturation number density, at the given 
temperature and pressure, and $a$ [cm] is the average radius of the nucleation site. We consider 
condensation only for low temperatures, so $n_c = p_v/k_B T$, where $p_v$ [dyn cm$^{-2}$] is the 
vapor pressure, and is estimated using the relatively simple Antoine equation:
\begin{equation}
\log p_v = A - \dfrac{B}{C + T},
\label{eqn:vapor-pressure}
\end{equation}
where $A$, $B$ and $C$ are all parameters taken from the tabulated NIST chemistry 
webbook\footnote{\url{http://webbook.nist.gov/chemistry/}}.

Alternatively, one can use the method commonly used in the astrochemical context 
\citep{Hasegawa1992,Caselli1998}, where
\begin{equation}
L \, {\rm[s^{-1}]}  = \pi a^2 v_{th} n_{\rm nuc}
\label{eqn:deposition}
\end{equation}
and
\begin{equation}
P \, = \nu_0 [{\rm JA}] \, e^{-E_D/k_B T}.
\label{eqn:desorption}
\end{equation}
Here, $E_D$ is the desorption energy, an empirically determined quantity, taken from 
\citet{Garrod2008}. The frequency, 
\begin{equation}
\nu_0 \, {\rm [Hz]} = \Big(\dfrac{2n_sE_D}{\pi^2 m_{\rm A}}\Big)^{1/2},
\label{eqn:harmonic-oscillator}
\end{equation}
is the characteristic frequency of the surface. The number of sites is estimated, also empirically, 
by the relation $n_s = 1.5 \times 10^{15} \, {\rm cm^2} \; \big(a/a_0 \big)^2$, where 
$a_0 = 0.1$ $\mu$m. The advantage of this approximation is that it is identical to the form 
generally used for complex surface chemistry in protoplanetary disks. This would allow one to take 
the results from disk chemistry and utilize them straightforwardly in atmospheric outgassing 
models.

It is worth pointing out that exponentiating Eq. (\ref{eqn:vapor-pressure}), dividing by $k_BT$, and 
then placing the resulting form of $n_c$ into Eq. (\ref{eqn:condense-timescale}), yields a form:
\begin{equation}
\dfrac{P}{L} \sim \dfrac{\rm Const.}{v_{th}}e^{T_c/T}
\end{equation}
where $T_c/T = B/(C + T)$ from Eq. (\ref{eqn:vapor-pressure}). The two forms are therefore analogous 
parameterizations, with the same temperature dependence, but the saturation approach is dependent on 
the parameterized vapor pressure, and the deposition approach is parameterized by the number of 
nucleation sites and the binding energy of the nucleation particle. 

Neither the supersaturation method nor the deposition method explain where the condensation seeds
first arise. It is assumed that the condensation seeds are already present, and therefore that
condensation occurs whenever the supersaturation ratio, $S \gtrsim 1$. In some environments like 
Earth, the condensation seeds come in the form of sand or ash particles, and the supersaturation
ratio for water to condense is very small, $S \approx 1.01$. If the seed particles are not already
present in the atmosphere, they must form within the gas phase by the growth from small to large, 
complex clusters. This requires a supersaturation ratio
$S \gg 100$, which only occurs when $T \ll T_c$ \citep{Helling2013}. \citet{Zsom2012} explore the
microphysics of water condensation and cloud formation for Earth and Earth-like planets

None of these reactions appear in the generic kinetic network, because their inclusion is atmosphere 
dependent. Condensation is not considered at all for \object{HD209458b} because it is too hot, but 
is considered for \object{Earth} and \object{Jupiter} for water. Ammonia and methane condensation 
can also be considered for \object{Jupiter} and methane and other condensation should be considered 
for even colder planets, such as \object{Uranus} and \object{Neptune}.

\bibliographystyle{apj}

\begin{thebibliography}{}

\bibitem[{Ackerman(1971)}]{Ackerman1971}
Ackerman, M. 1971, in Mesospheric Models and Related Experiments (Springer),
  149

\bibitem[{Adachi {et~al.}(1981)Adachi, Basco, \& James}]{1981ADA/BAS}
Adachi, H., Basco, N., \& James, D. 1981, Int J Chem Kin, 13, 1251

\bibitem[{Adam {et~al.}(2005)Adam, Hack, Zhu, Qu, \& Schinke}]{2005ADA/HAC}
Adam, L., Hack, W., Zhu, H., Qu, Z., \& Schinke, R. 2005, JChPh, 122, 114301

\bibitem[{Adams {et~al.}(1970)Adams, Bohme, \& Ferguson}]{2-0075}
Adams, N., Bohme, D., \& Ferguson, E. 1970, JChPh, 52, 5101

\bibitem[{Adams \& Smith(1976{\natexlab{a}})}]{2-2002}
Adams, N., \& Smith, D. 1976{\natexlab{a}}, JPhB, 9, 1439

\bibitem[{Adams \& Smith(1976{\natexlab{b}})}]{2-2001}
---. 1976{\natexlab{b}}, IJMIP, 21, 349

\bibitem[{Adams \& Smith(1977)}]{2-2004}
---. 1977, CPL, 47, 383

\bibitem[{Adams \& Smith(1978)}]{2-2010}
---. 1978, CPL, 54, 530

\bibitem[{Adams {et~al.}(1978{\natexlab{a}})Adams, Smith, \& Grief}]{2-2005}
Adams, N., Smith, D., \& Grief, D. 1978{\natexlab{a}}, IJMIP, 26, 405

\bibitem[{Adams {et~al.}(1978{\natexlab{b}})Adams, Smith, \& Grief}]{2-2008}
---. 1978{\natexlab{b}}, IJMIP, 26, 405

\bibitem[{Adams {et~al.}(1980)Adams, Smith, \& Paulson}]{4-032}
Adams, N.~G., Smith, D., \& Paulson, J.~F. 1980, JChPh, 72, 288

\bibitem[{Adamson {et~al.}(1997)Adamson, DeSain, Curl, \& Glass}]{1997ADA/DES}
Adamson, J., DeSain, J., Curl, R., \& Glass, G. 1997, JPCA, 101, 864

\bibitem[{Aders \& Wagner(1973)}]{1973ADE/WAG}
Aders, W.-K., \& Wagner, H.~G. 1973, Berich Bunsen Gesell, 77, 712

\bibitem[{{Ag{\'u}ndez} {et~al.}(2014){Ag{\'u}ndez}, {Parmentier}, {Venot},
  {Hersant}, \& {Selsis}}]{Agundez2014}
{Ag{\'u}ndez}, M., {Parmentier}, V., {Venot}, O., {Hersant}, F., \& {Selsis},
  F. 2014, \aap, 564, A73

\bibitem[{{Alam} \& {Lin}(2008)}]{Alam2008}
{Alam}, J.~M., \& {Lin}, J.~C. 2008, MWRv, 136, 4653

\bibitem[{Albritton {et~al.}(1983)Albritton, Dotan, Streit, Fahey, Fehsenfeld,
  \& Ferguson}]{4-258}
Albritton, D., Dotan, I., Streit, G., {et~al.} 1983, JChPh, 78, 6614

\bibitem[{Allison {et~al.}(1996)Allison, Lynch, Truhlar, \&
  Gordon}]{1996ALL/LYN}
Allison, T.~C., Lynch, G.~C., Truhlar, D.~G., \& Gordon, M.~S. 1996, JPhCh, 100, 13575

\bibitem[{Almatarneh {et~al.}(2005)Almatarneh, Flinn, \& Poirier}]{2005ALM/FLI}
Almatarneh, M., Flinn, C., \& Poirier, R. 2005, CaJCh,
  83, 2082

\bibitem[{Alvarez \& Moore(1994)}]{1994ALV/MOO}
Alvarez, R.~A., \& Moore, C.~B. 1994, JPhCh, 98,
  174

\bibitem[{Anastasi \& Hancock(1988)}]{1988ANA/HAN}
Anastasi, C., \& Hancock, D.~U. 1988, FaTr II, 84, 1697

\bibitem[{Andersson {et~al.}(2003)Andersson, Markovic, \& Nyman}]{2003AND/MAR}
Andersson, S., Markovic, N., \& Nyman, G. 2003, JPCA, 107, 5439

\bibitem[{Anglada(2004)}]{2004ANG}
Anglada, J.~M. 2004, JAChS, 126, 9809

\bibitem[{Anicich {et~al.}(1976)Anicich, Huntress, \& Futrell}]{0-038}
Anicich, V.~G., Huntress, W.~T., \& Futrell, J.~H. 1976, CPL, 40, 233

\bibitem[{Anicich {et~al.}(1986)Anicich, Huntress, \& McEwan}]{4-482}
Anicich, V.~G., Huntress, W.~T., \& McEwan, M.~J. 1986, JPhCh, 90, 2446

\bibitem[{{Aplin}(2013)}]{Aplin2013}
{Aplin}, K.~L. 2013, {Electrifying Atmospheres: Charging, Ionisation and
  Lightning in the Solar System and Beyond} (Springer Science \& Business Media)

\bibitem[{Arenas {et~al.}(2000)Arenas, Marcos, L{\'o}pez-Toc{\'o}n, Otero, \&
  Soto}]{2000ARE/MAR}
Arenas, J.~F., Marcos, J.~I., L{\'o}pez-Toc{\'o}n, I., Otero, J.~C., \& Soto,
  J. 2000, JChPh, 113, 2282

\bibitem[{Armentrout {et~al.}(1978)Armentrout, Berman, \& Beauchamp}]{4-115}
Armentrout, P., Berman, D., \& Beauchamp, J. 1978, CPL,
  53, 255

\bibitem[{{Asplund} {et~al.}(2009){Asplund}, {Grevesse}, {Sauval}, \&
  {Scott}}]{Asplund2009}
{Asplund}, M., {Grevesse}, N., {Sauval}, A.~J., \& {Scott}, P. 2009, \araa, 47,
  481

\bibitem[{Atkinson {et~al.}(1989)Atkinson, Baulch, Cox, Hampson, Kerr, \&
  Troe}]{1989ATK/BAU}
Atkinson, R., Baulch, D., Cox, R., {et~al.} 1989, JPCRD, 18, 881

\bibitem[{Atkinson {et~al.}(1992)Atkinson, Baulch, Cox, Hampson, Kerr, \&
  Troe}]{1992ATK/BAU}
---. 1992, AtmEn A, 26, 1187

\bibitem[{Atkinson {et~al.}(1997)Atkinson, Baulch, Cox, Hampson, Kerr,
  Rossi, \& Troe}]{1997ATK/BAU}
---. 1997, JPCRD, 26, 521

\bibitem[{Atkinson {et~al.}(2001)Atkinson, Baulch, Cox, Crowley, Hampson,
  Kerr, Rossi, \& Troe}]{2001ATK/BAU}
---. 2001, Summary of Evaluated Kinetic \& Photochemical Data for Atmospheric Chemistry

\bibitem[{Atkinson {et~al.}(2004)Atkinson, Baulch, Cox, Crowley, Hampson,
  Hynes, Jenkin, Rossi, \& Troe}]{2004ATK/BAU}
Atkinson, R., Baulch, D., Cox, R., {et~al.} 2004, ACP, 4, 1461

\bibitem[{Atkinson {et~al.}(1973)Atkinson, Finlayson, \& Pitts}]{1973ATK/FIN}
Atkinson, R., Finlayson, B., \& Pitts, J. 1973, JAChS, 95, 7592

\bibitem[{{Atreya} {et~al.}(2003){Atreya}, {Mahaffy}, {Niemann}, {Wong}, \&
  {Owen}}]{Atreya2003}
{Atreya}, S.~K., {Mahaffy}, P.~R., {Niemann}, H.~B., {Wong}, M.~H., \& {Owen},
  T.~C. 2003, \planss, 51, 105

\bibitem[{Ausloos(1975)}]{2-6031}
Ausloos, P., 1975, Interactions between Ions and Molecules,
  ed. P.~Ausloos (Plenum Press), 489

\bibitem[{Ausloos \& Lias(1981)}]{4-065}
Ausloos, P., \& Lias, S. 1981, JAChS, 103,
  3641

\bibitem[{Avramenko \& Krasnen'kov(1966)}]{1967AVR/KRA}
Avramenko, L., \& Krasnen'kov, V. 1966, Russ Chem B, 15, 394

\bibitem[{Back \& Griffiths(1967)}]{Back1967}
Back, R., \& Griffiths, D. 1967, JChPh, 46, 4839

\bibitem[{Badnell {et~al.}(2005)Badnell, Bautista, Butler, Delahaye, Mendoza,
  Palmeri, Zeippen, \& Seaton}]{TOPbase}
Badnell, N.~R., Bautista, M., Butler, K., {et~al.} 2005, MNRAS, 360, 458

\bibitem[{{Bailey} {et~al.}(2014){Bailey}, {Helling}, {Hodos{\'a}n}, {Bilger},
  \& {Stark}}]{Bailey2014}
{Bailey}, R.~L., {Helling}, Ch., {Hodos{\'a}n}, G., {Bilger}, C., \& {Stark},
  C.~R. 2014, \apj, 784, 43

\bibitem[{Baker {et~al.}(1971)Baker, Baldwin, \& Walker}]{1971BAK/BAL}
Baker, R., Baldwin, R., \& Walker, R. 1971 in International Symposium on Combustion (Elsevier), 291

\bibitem[{Baker {et~al.}(1969)Baker, Kerr, \& Trotman-Dickenson}]{1969BAK/KER}
Baker, R., Kerr, J., \& Trotman-Dickenson, A. 1969, J Chem Soc A, 390

\bibitem[{Baldwin {et~al.}(1961)Baldwin, Booth, \& Brattan}]{1961BAL/BOO}
Baldwin, R.~R., Booth, D., \& Brattan, D. 1961, CaJCh, 39,
  2130

\bibitem[{Baldwin {et~al.}(1984)Baldwin, Keen, \& Walker}]{1984BAL/KEE}
Baldwin, R.~R., Keen, A., \& Walker, R.~W. 1984, FaTr I, 80, 435

\bibitem[{Baldwin {et~al.}(1971)Baldwin, Langford, Matchan, Walker, \&
  Yorke}]{1971BAL/LAN}
Baldwin, R.~R., Langford, D., Matchan, M., Walker, R., \& Yorke, D. 1971 in International Symposium on Combustion (Elsevier), 251

\bibitem[{{Banks} \& {Kockarts}(1973)}]{Banks1973}
{Banks}, P.~M., \& {Kockarts}, G. 1973, {Aeronomy} (Elsevier)

\bibitem[{Bar-Nun {et~al.}(1970)}]{Bar-Nun1970}
{Bar-Nun}, A., {Bar-Nun}, N., Bauer, S.~H., \& Sagan, C. 1970, Science, 168.3930, 470

\bibitem[{Barassin {et~al.}(1983)Barassin, Barassin, \& Thomas}]{4-390}
Barassin, J., Barassin, A., \& Thomas, R. 1983, IJMIP, 49, 51

\bibitem[{Barfield {et~al.}(1972)Barfield, Koontz, \& Huebner}]{Barfield1972}
Barfield, W., Koontz, G., \& Huebner, W. 1972, JQSRT, 12, 1409

\bibitem[{Barnett {et~al.}(1987)Barnett, Marston, \& Wayne}]{1987BAR/MAR}
Barnett, A.~J., Marston, G., \& Wayne, R.~P. 1987, FaTr II, 83, 1453

\bibitem[{Barsuhn \& Nesbet(1978)}]{Barsuhn1978}
Barsuhn, J., \& Nesbet, R. 1978, JChPh, 68, 2783

\bibitem[{Bass {et~al.}(1976)Bass, Ledford, \& Laufer}]{Bass1976}
Bass, A.~M., Ledford, A.~E., \& Laufer, A.~H. 1976, J Res NBS A Phys Chem,
  80A, 143

\bibitem[{Batt {et~al.}(1975)Batt, McCulloch, \& Milne}]{1975BAT/MCC}
Batt, L., McCulloch, R., \& Milne, R. 1975, Int J Chem Kinet, 7, 1

\bibitem[{Batt {et~al.}(1977)Batt, Milne, \& McCulloch}]{1977BAT/MIL}
Batt, L., Milne, R., \& McCulloch, R. 1977, Int J Chem Kin, 9, 567

\bibitem[{Batt \& Rattray(1979)}]{1979BAT/RAT}
Batt, L., \& Rattray, G. 1979, Int J Chem Kin, 11, 1183

\bibitem[{Bauer {et~al.}(1985)Bauer, Becker, \& Meuser}]{1985BAU/BEC}
Bauer, W., Becker, K., \& Meuser, R. 1985, Berich Bunsen Gesell, 89, 340

\bibitem[{Bauerle {et~al.}(1995)Bauerle, Klatt, \& Wagner}]{1995BAU/KLA}
Bauerle, S., Klatt, M., \& Wagner, H. 1995, Berich Bunsen Gesell, 99, 870

\bibitem[{Baulch {et~al.}(2005)}]{Baulch2005}
Baulch, D., Bowman, C.~T., Cobos, C.~J., {et~al.} 2005, JPCRD, 34, 757

\bibitem[{Baulch {et~al.}(1992)Baulch, Cobos, Cox, Esser, Frank, Just, Kerr,
  Pilling, Troe, Walker, {et~al.}}]{1992BAU/COB}
Baulch, D., Cobos, C., Cox, R., {et~al.} 1992, JPCRD, 21, 411

\bibitem[{Baulch {et~al.}(1994)Baulch, Cobos, Cox, Frank, Hayman, Just, Kerr,
  Murrells, Pilling, Troe, {et~al.}}]{1994BAU/COB}
---. 1994, JPCRD, 23, 847

\bibitem[{Baulch {et~al.}(1982)Baulch, Cox, Crutzen, Hampson, Kerr, Troe,
  Watson, {et~al.}}]{Baulch1982}
Baulch, D., Cox, R., Crutzen, P., {et~al.} 1982, JPCRD, 11, 327

\bibitem[{Baulch {et~al.}(1981)Baulch, Duxbury, Grant, \&
  Montague}]{1981BAU/DUX}
Baulch, D., Duxbury, J., Grant, S., \& Montague, D. 1981, Evaluated kinetic
  data for high temperature reactions. Volume 4. Homogeneous gas phase
  reactions of halogen-and cyanide-containing species, Tech. rep., DTIC
  Document

\bibitem[{Beaty \& Patterson(1965)}]{0-070}
Beaty, E., \& Patterson, P. 1965, PhRv, 137, A346

\bibitem[{{Beaulieu} {et~al.}(2010){Beaulieu}, {Kipping}, {Batista}, {Tinetti},
  {Ribas}, {Carey}, {Noriega-Crespo}, {Griffith}, {Campanella}, {Dong},
  {Tennyson}, {Barber}, {Deroo}, {Fossey}, {Liang}, {Swain}, {Yung}, \&
  {Allard}}]{Beaulieu2010}
{Beaulieu}, J.~P., {Kipping}, D.~M., {Batista}, V., {et~al.} 2010, \mnras, 409,
  963

\bibitem[{Becker {et~al.}(1992{\natexlab{a}})Becker, Rahman, \&
  Schindler}]{1992BEC/RAH}
Becker, E., Rahman, M., \& Schindler, R. 1992{\natexlab{a}}, Berich Bunsen Gesell, 96, 776

\bibitem[{Becker {et~al.}(1992{\natexlab{b}})Becker, K{\"o}nig, Meuser, Wiesen,
  \& Bayes}]{1992BEC/KON}
Becker, K., K{\"o}nig, R., Meuser, R., Wiesen, P., \& Bayes, K.~D.
  1992{\natexlab{b}}, J Photoch Photobio A, 64, 1

\bibitem[{{Benneke}(2015)}]{Benneke2015}
{Benneke}, B. 2015, arXiv:1504.07655

\bibitem[{Benner {et~al.}(2004)Benner, Ricardo, \& Carrigan}]{Benner2004}
Benner, S.~A., Ricardo, A., \& Carrigan, M.~A. 2004, Curr Opin Chem Biol, 8, 672

\bibitem[{Benson(1994)}]{1994BEN}
Benson, S.~W. 1994, Int J Chem Kin, 26, 997

\bibitem[{Bergeat {et~al.}(1998)Bergeat, Calvo, Daugey, Loison, \&
  Dorthe}]{1998BER/CAL}
Bergeat, A., Calvo, T., Daugey, N., Loison, J.-C., \& Dorthe, G. 1998, JCPA, 102, 8124

\bibitem[{Bergeat {et~al.}(2009)Bergeat, Moisan, M{\'e}reau, \&
  Loison}]{2009BER/MOI}
Bergeat, A., Moisan, S., M{\'e}reau, R., \& Loison, J.-C. 2009, CPL, 480, 21

\bibitem[{Bethe \& Salpeter(1957)}]{Bethe1957}
Bethe, H.~A., \& Salpeter, E.~E. 1957, Quantum mechanics of one-and
  two-electron atoms, Vol.~63 (Springer Berlin)

\bibitem[{Betowski {et~al.}(1975)Betowski, Payzant, Mackay, \& Bohme}]{2-6019}
Betowski, D., Payzant, J., Mackay, G.~I., \& Bohme, D. 1975, CPL, 31, 321

\bibitem[{{B{\'e}zard} {et~al.}(2001){B{\'e}zard}, {Drossart}, {Encrenaz}, \&
  {Feuchtgruber}}]{Bezard2001}
{B{\'e}zard}, B., {Drossart}, P., {Encrenaz}, T., \& {Feuchtgruber}, H. 2001,
  \icarus, 154, 492

\bibitem[{Bierbaum {et~al.}(1976)Bierbaum, DePuy, Shapiro, \& Stewart}]{2-9000}
Bierbaum, V.~M., DePuy, C., Shapiro, R., \& Stewart, J.~H. 1976, JAChS, 98, 4229

\bibitem[{Bierbaum {et~al.}(1984)Bierbaum, Grabowski, \& DePuy}]{4-088}
Bierbaum, V.~M., Grabowski, J.~J., \& DePuy, C.~H. 1984, JPhCh, 88, 1389

\bibitem[{Biggs {et~al.}(1995)Biggs, Canosa-Mass, Fracheboud, Shallcross, \&
  Wayne}]{1995BIG/CAN}
Biggs, P., Canosa-Mass, C.~E., Fracheboud, J.-M., Shallcross, D.~E., \& Wayne,
  R.~P. 1995, FaTr, 91, 817

\bibitem[{{Bilger} {et~al.}(2013){Bilger}, {Rimmer}, \& {Helling}}]{Bilger2013}
{Bilger}, C., {Rimmer}, P., \& {Helling}, Ch. 2013, \mnras, 435, 1888

\bibitem[{{Blagojevic} {et~al.}(2003){Blagojevic}, {Petrie}, \&
  {Bohme}}]{Blago2003}
{Blagojevic}, V., {Petrie}, S., \& {Bohme}, D.~K. 2003, \mnras, 339, L7

\bibitem[{Blake \& Jackson(1969)}]{1969BLA/JAC}
Blake, P., \& Jackson, G. 1969, J Chem Soc B, 94

\bibitem[{Blitz {et~al.}(1997)}]{Blitz1997}
Blitz, M., Johnson, D., Pilling, M., {et~al.} 1997, JCS(FaTr), 93, 1473

\bibitem[{Bogan \& Hand(1978)}]{1978BOG/HAN}
Bogan, D.~J., \& Hand, C.~W. 1978, JPhCh, 82, 2067

\bibitem[{Bogdanchikov {et~al.}(2004)Bogdanchikov, Baklanov, \&
  Parker}]{2004BOG/BAK}
Bogdanchikov, G., Baklanov, A., \& Parker, D. 2004, CPL,
  385, 486

\bibitem[{B{\"o}hland {et~al.}(1985)B{\"o}hland, D{\~o}bẽ, Temps, \&
  Wagner}]{1985BOH/DOB}
B{\"o}hland, T., D{\~o}bẽ, S., Temps, F., \& Wagner, H.~G. 1985, Berichte Bunsen Gesell, 89, 1110

\bibitem[{Bohme {et~al.}(1970)Bohme, Adams, Mosesman, Dunkin, \&
  Ferguson}]{2-0073}
Bohme, D., Adams, N., Mosesman, M., Dunkin, D., \& Ferguson, E. 1970, JChPh, 52, 5094

\bibitem[{Bohme \& Fehsenfeld(1969)}]{2-0055}
Bohme, D.~K., \& Fehsenfeld, F. 1969, CaJCh, 47, 2717

\bibitem[{Bohme {et~al.}(1971)Bohme, Lee-Ruff, \& Young}]{2-6004}
Bohme, D.~K., Lee-Ruff, E., \& Young, L.~B. 1971, JAChS, 93, 4608

\bibitem[{Bohme \& Mackay(1981)}]{4-223}
Bohme, D.~K., \& Mackay, G.~I. 1981, JAChS,
  103, 2173

\bibitem[{Bohme {et~al.}(1980)Bohme, Mackay, \& Schiff}]{4-122}
Bohme, D.~K., Mackay, G.-I., \& Schiff, H. 1980, JChPh,
  73, 4976

\bibitem[{Bohme {et~al.}(1974)Bohme, Mackay, Schiff, \& Hemsworth}]{2-6017}
Bohme, D.~K., Mackay, G., Schiff, H., \& Hemsworth, R. 1974, JChPh, 61, 2175

\bibitem[{Bohme {et~al.}(1979)Bohme, Mackay, \& Tanner}]{4-468}
Bohme, D.~K., Mackay, G., \& Tanner, S. 1979, JAChS, 101, 3724

\bibitem[{Bohme \& Raksit(1985)}]{4-236}
Bohme, D.~K., \& Raksit, A.~B. 1985, CaJCh, 63, 3007

\bibitem[{Bohme {et~al.}(1982)Bohme, Raksit, \& Schiff}]{4-142}
Bohme, D.~K., Raksit, A., \& Schiff, H. 1982, CPL, 93, 592

\bibitem[{Bolden {et~al.}(1970)Bolden, Hemsworth, Shaw, \& Twiddy}]{2-8000}
Bolden, R., Hemsworth, R., Shaw, M., \& Twiddy, N. 1970, JPhB, 3, 45

\bibitem[{Bolden \& Twiddy(1972)}]{2-8002}
Bolden, R., \& Twiddy, N. 1972, FaDi, 53, 192

\bibitem[{Borisov {et~al.}(1977)Borisov, Zamanskii, Potmishil, Skachkov, \&
  Foteenkov}]{1977BOR/ZAM}
Borisov, A., Zamanskii, V., Potmishil, K., Skachkov, G., \& Foteenkov, V. 1977,
  Kinet. Catal. (USSR), 18

\bibitem[{Borucki {et~al.}(1985)Borucki, Kenzie, McKay, Duong, \&
  Boac}]{Borucki1985}
Borucki, W., Kenzie, R.~M., McKay, C., Duong, N., \& Boac, D. 1985, Icarus, 64,
  221

\bibitem[{{Borucki} {et~al.}(1984){Borucki}, {McKay}, \&
  {Whitten}}]{Borucki1984}
{Borucki}, W.~J., {McKay}, C.~P., \& {Whitten}, R.~C. 1984, \icarus, 60, 260

\bibitem[{Bose \& Candler(1996)}]{1997BOS/CAN}
Bose, D., \& Candler, G.~V. 1996, JChPh, 104, 2825

\bibitem[{Boughton {et~al.}(1997)Boughton, Kristyan, \& Lin}]{1997BOU/KRI}
Boughton, J., Kristyan, S., \& Lin, M. 1997, CP, 214, 219

\bibitem[{Bowers {et~al.}(1969)Bowers, Elleman, \& King}]{0-012}
Bowers, M., Elleman, D., \& King, J. 1969, JChPh,
  50, 4787

\bibitem[{Bozzelli {et~al.}(1994)Bozzelli, Chang, \& Dean}]{1994BOZ/CHA}
Bozzelli, J.~W., Chang, A.~Y., \& Dean, A.~M. 1994 in International Symposium on Combustion (Elsevier), 965

\bibitem[{Bozzelli \& Dean(1989)}]{1989BOZ/DEA}
Bozzelli, J.~W., \& Dean, A.~M. 1989, JPhCh, 93,
  1058

\bibitem[{Bozzelli \& Dean(1990)}]{1990BOZ/DEA}
---. 1990, JPhCh, 94, 3313

\bibitem[{Bozzelli \& Dean(1995)}]{1995BOZ/DEA}
---. 1995, Int J Chem Kin, 27, 1097

\bibitem[{Bravo-P{\'e}rez {et~al.}(2002)Bravo-P{\'e}rez, Alvarez-Idaboy,
  Cruz-Torres, \& Ru{\'\i}z}]{2002BRA/ALV}
Bravo-P{\'e}rez, G., Alvarez-Idaboy, J.~R., Cruz-Torres, A., \& Ru{\'\i}z,
  M.~E. 2002, JCPA, 106, 4645

\bibitem[{Breen \& Glass(1971)}]{1970BRE/GLA}
Breen, J., \& Glass, G. 1971, Int J Chem Kin, 3, 145

\bibitem[{Brion {et~al.}(1979)Brion, Tan, Van~der Wiel, \& Van~der
  Leeuw}]{Brion1979}
Brion, C., Tan, K., Van~der Wiel, M., \& Van~der Leeuw, P.~E. 1979, JESRP, 17, 101

\bibitem[{Broad \& Reinhardt(1976)}]{Broad1976}
Broad, J.~T., \& Reinhardt, W.~P. 1976, PhRvA, 14, 2159

\bibitem[{Brown \& Hindmarsh(1989)}]{Brown1989}
Brown, P.~N., \& Hindmarsh, A.~C. 1989, ApMaC, 31, 40

\bibitem[{Browning \& Fryar(1973)}]{Browning1973}
Browning, R., \& Fryar, J. 1973, JPhB, 6, 364

\bibitem[{Brownsword {et~al.}(1996)Brownsword, Gatenby, Herbert, Smith,
  Stewart, \& Symonds}]{1996BRO/GAT}
Brownsword, R., Gatenby, S., Herbert, L., {et~al.} 1996, FaTr, 92, 723

\bibitem[{Brunetti \& Liuti(1975)}]{1975BRU/LIU}
Brunetti, B., \& Liuti, G. 1975, ZPC, 94,
  19

\bibitem[{{Bryukov} {et~al.}(2006){Bryukov}, {Dellinger}, \&
  {Knyazev}}]{2006BRY/DEL}
{Bryukov}, M.~G., {Dellinger}, B., \& {Knyazev}, V.~D. 2006, JPCA, 110, 936

\bibitem[{Bryukov {et~al.}(2004)Bryukov, Knyazev, Lomnicki, McFerrin, \&
  Dellinger}]{2004BRY/KNY}
Bryukov, M.~G., Knyazev, V.~D., Lomnicki, S.~M., McFerrin, C.~A., \& Dellinger,
  B. 2004, JCPA, 108, 10464

\bibitem[{Bulatov {et~al.}(1980)Bulatov, Buloyan, Cheskis, Kozliner, Sarkisov,
  \& Trostin}]{1982BUL/BUL}
Bulatov, V., Buloyan, A., Cheskis, S., {et~al.} 1980, CPL,
  74, 288

\bibitem[{Burcat \& Ruscic(2005)}]{Burcat2005}
Burcat, A., \& Ruscic, B. 2005, Third millenium ideal gas and condensed phase
  thermochemical database for combustion with updates from active
  thermochemical tables

\bibitem[{Burt {et~al.}(1970)Burt, Dunn, McEwan, Sutton, Roche, \&
  Schiff}]{2-6000}
Burt, J., Dunn, J., McEwan, M., {et~al.} 1970, JChPh,
  52, 6062

\bibitem[{B{\"u}ttrill {et~al.}(1974)B{\"u}ttrill, Kim, \&
  Huntress}]{1-124}
B{\"u}ttrill, S., Kim, J., \& Huntress, W. 1974, J. Chem. Phys, 61, 2122

\bibitem[{{Cairns} \& {Samson}(1965)}]{Cairns1965}
{Cairns}, R.~B., \& {Samson}, J.~A.~R. 1965, \jgr, 70, 99

\bibitem[{Calvert \& Pitts(1966)}]{Calvert1966}
Calvert, J., \& Pitts, J. 1966, Photochemistry (Wiley: New York)

\bibitem[{Campomanes {et~al.}(2001)Campomanes, Men{\'e}ndez, \&
  Sordo}]{2001CAM/MEN}
Campomanes, P., Men{\'e}ndez, I., \& Sordo, T.~L. 2001, JPCA, 105, 229

\bibitem[{Canosa {et~al.}(1979)Canosa, Penzhorn, \& Von~Sonntag}]{1979CAN/PEN}
Canosa, C., Penzhorn, R.-D., \& Von~Sonntag, C. 1979, Berich Bunsen Gesell, 83, 217

\bibitem[{Canosa-Mas {et~al.}(1988)Canosa-Mas, Smith, Toby, \&
  Wayne}]{1988CAN/SMI}
Canosa, C., Smith, S.~J., Toby, S., \& Wayne, R.~P. 1988, FaTr II, 84, 263

\bibitem[{Caridade {et~al.}(2005)Caridade, Rodrigues, Sousa, \&
  Varandas}]{2005CAR/ROD}
Caridade, P., Rodrigues, S., Sousa, F., \& Varandas, A. 2005, JPCA, 109, 2356

\bibitem[{Carl {et~al.}(2003)Carl, Sun, Teugels, \& Peeters}]{2003CAR/SUN}
Carl, S., Sun, Q., Teugels, L., \& Peeters, J. 2003, PCCP, 5, 5424

\bibitem[{Carstensen \& Dean(2005)}]{2005CAR/DEA}
Carstensen, H.-H., \& Dean, A.~M. 2005, P Combust Inst, 30, 995

\bibitem[{Carstensen \& Dean(2008)}]{2008CAR/DEA}
---. 2008, JCPA, 113, 367

\bibitem[{{Caselli} {et~al.}(1998){Caselli}, {Hasegawa}, \&
  {Herbst}}]{Caselli1998}
{Caselli}, P., {Hasegawa}, T.~I., \& {Herbst}, E. 1998, \apj, 495, 309

\bibitem[{Catling(2006)}]{Catling2006}
Catling, D.~C. 2006, Science, 311, 38

\bibitem[{{Cavali{\'e}} {et~al.}(2012){Cavali{\'e}}, {Biver}, {Hartogh},
  {Dobrijevic}, {Billebaud}, {Lellouch}, {Sandqvist}, {Brillet}, {Lecacheux},
  {Hjalmarson}, {Frisk}, {Olberg}, \& {Odin Team}}]{Cavalie2012}
{Cavali{\'e}}, T., {Biver}, N., {Hartogh}, P., {et~al.} 2012, \planss, 61, 3

\bibitem[{Cermak {et~al.}(1970)Cermak, Dalgarno, Ferguson, Friedman, \&
  McDaniel}]{2-0029}
Cermak, V., Dalgarno, A., Ferguson, E., Friedman, L., \& McDaniel, E. 1970, Ion
  Molecule Reactions (New York: Wiley)

\bibitem[{Ceursters {et~al.}(2001)}]{Ceursters2001}
Ceursters, B., Nguyen, H.~M.~T., Nguyen, M.~T., Peeters, J., \& Vereecken, L. 2001
PCCP, 3, 3070

\bibitem[{Chakraborty \& Lin(1999)}]{1999CHA/LIN}
Chakraborty, D., \& Lin, M. 1999, JCPA, 103, 601

\bibitem[{Chakraborty {et~al.}(1998)Chakraborty, Park, \& Lin}]{1998CHA/PAR}
Chakraborty, D., Park, J., \& Lin, M. 1998, CP, 231, 39

\bibitem[{Chan {et~al.}(2001)Chan, Heck, \& Pritchard}]{2001CHA/HEC}
Chan, W.-T., Heck, S.~M., \& Pritchard, H.~O. 2001, PCCP, 3, 56

\bibitem[{Chang {et~al.}(2007)Chang, Chen, Xu, \& Lin}]{2007CHA/CHE}
Chang, J.-G., Chen, H.-T., Xu, S., \& Lin, M. 2007, JPCA, 111, 6789

\bibitem[{Chang \& Yu(1995)}]{1995CHA/YU}
Chang, N.-Y., \& Yu, C.-H. 1995, CPL, 242, 232

\bibitem[{{Chapman} \& {Cowling}(1991)}]{Chapman1991}
{Chapman}, S., \& {Cowling}, T.~G. 1991, {The Mathematical Theory of
  Non-uniform Gases} (Cambridge University Press)

\bibitem[{{Charnley}(1997)}]{Charnley1997}
{Charnley}, S.~B. 1997, in IAU Colloq. 161: Astronomical and Biochemical
  Origins and the Search for Life in the Universe, ed. C.~{Batalli Cosmovici},
  S.~{Bowyer}, \& D.~{Werthimer}, 89

\bibitem[{Chau \& Bowers(1976)}]{0-138}
Chau, M., \& Bowers, M.~T. 1976, CPL, 44, 490

\bibitem[{Cheng \& Lampe(1973)}]{0-185}
Cheng, T., \& Lampe, F. 1973, JPhCh, 77, 2841

\bibitem[{Cheng {et~al.}(1973)Cheng, Yu, \& Lampe}]{1-030}
Cheng, T., Yu, T.-Y., \& Lampe, F. 1973, JPhCh, 77,
  2587

\bibitem[{Cheng {et~al.}(1974)Cheng, Yu, \& Lampe}]{0-163}
---. 1974, JPhCh, 78,
  1184

\bibitem[{Choi \& Lin(2005)}]{2005CHO/LIN}
Choi, Y., \& Lin, M. 2005, Int J Chem Kin, 37, 261

\bibitem[{Christie \& Voisey(1967)}]{1967CHR/VOI}
Christie, M.~I., \& Voisey, M. 1967, TrFa, 63,
  2702

\bibitem[{Chuchani {et~al.}(1993)Chuchani, Martin, Rotinov, \&
  Dominguez}]{1993CHU/MAR}
Chuchani, G., Martin, I., Rotinov, A., \& Dominguez, R.~M. 1993, J Phys Org Chem, 6, 54

\bibitem[{Cimas \& Largo(2006)}]{2006CIM/LAR}
Cimas, A., \& Largo, A. 2006, JCPA, 110, 10912

\bibitem[{Claire {et~al.}(2006)}]{Claire2006}
Claire, M.~W., Catling, D.~C., \& Zahnle, K.~J. 2006, Geobiology, 4, 239

\bibitem[{Clark {et~al.}(1978)Clark, Moore, \& Nogar}]{Clark1978}
Clark, J.~H., Moore, C.~B., \& Nogar, N.~S. 1978, JChPh, 68, 1264

\bibitem[{Clary {et~al.}(1985)Clary, Smith, \& Adams}]{4-197}
Clary, D., Smith, D., \& Adams, N. 1985, CPL, 119, 320

\bibitem[{{Cleaves} {et~al.}(2008){Cleaves}, {Chalmers}, {Lazcano}, {Miller},
  \& {Bada}}]{Cleaves2008}
{Cleaves}, H.~J., {Chalmers}, J.~H., {Lazcano}, A., {Miller}, S.~L., \& {Bada},
  J.~L. 2008, OLEB, 38, 105

\bibitem[{Cobos \& Troe(1985)}]{1985COB/TRO}
Cobos, C., \& Troe, J. 1985, JChPh, 83, 1010

\bibitem[{Cohen(1991)}]{1991COH}
Cohen, N. 1991, Int J Chem Kin, 23, 397

\bibitem[{{Cohen} \& {Benson}(1993)}]{Cohen1993}
{Cohen}, N., \& {Benson}, S. 1993, ChRv, 93, 2419

\bibitem[{Cohen \& Westberg(1991)}]{1991COH/WES}
Cohen, N., \& Westberg, K. 1991, JPCRD, 20, 1211

\bibitem[{Colberg \& Friedrichs(2006)}]{2006COL/FRI}
Colberg, M., \& Friedrichs, G. 2006, JCPA, 110,
  160

\bibitem[{Cook \& Metzger(1964)}]{Cook1964}
Cook, G., \& Metzger, P. 1964, JOSA, 54, 968

\bibitem[{Cook {et~al.}(2009)Cook, Davidson, \& Hanson}]{2009COO/DAV}
Cook, R.~D., Davidson, D.~F., \& Hanson, R.~K. 2009, JPCA, 113, 9974

\bibitem[{Corchado \& Espinosa-Garc{\i}a(1997)}]{1997COR/ESP}
Corchado, J.~C., \& Espinosa-Garc{\i}a, J. 1997, JChPh,
  106, 4013

\bibitem[{Corchado {et~al.}(1995)Corchado, Espinosa-Garcia, Hu, Rossi, \&
  Truhlar}]{1995COR/ESP}
Corchado, J.~C., Espinosa-Garc{\i}a, J., Hu, W.-P., Rossi, I., \& Truhlar, D.~G.
  1995, JPhCh, 99, 687

\bibitem[{Corchado {et~al.}(1998)Corchado, Espinosa-Garc{\'\i}a, Roberto-Neto,
  Chuang, \& Truhlar}]{1998COR/ESP}
Corchado, J.~C., Espinosa-Garc{\'\i}a, J., Roberto-Neto, O., Chuang, Y.-Y., \&
  Truhlar, D.~G. 1998, JCPA, 102, 4899

\bibitem[{Cox \& Derwent(1977)}]{Cox1977}
Cox, R., \& Derwent, R. 1977, J Photochem, 6, 23

\bibitem[{Cribb {et~al.}(1992)Cribb, Dove, \& Yamazaki}]{1992CRI/DOV}
Cribb, P.~H., Dove, J.~E., \& Yamazaki, S. 1992, CoFl, 88, 169

\bibitem[{Crosley(1989)}]{1989CRO}
Crosley, D.~R. 1989, JPhCh, 93, 6273

\bibitem[{Curran(2006)}]{2006CUR}
Curran, H. 2006, Int J Chem Kin, 38, 250

\bibitem[{Cvetanovi{\'c}(1987)}]{1987CVE}
Cvetanovi{\'c}, R.~J. 1987, JPCRD,
  16, 261

\bibitem[{Daele {et~al.}(1995)Daele, Laverdet, Le~Bras, \&
  Poulet}]{1995DAE/LAV}
Daele, V., Laverdet, G., Le~Bras, G., \& Poulet, G. 1995, JPhCh, 99, 1470

\bibitem[{Dammeier {et~al.}(2007)Dammeier, Colberg, \&
  Friedrichs}]{2007DAM/COL}
Dammeier, J., Colberg, M., \& Friedrichs, G. 2007, PCCP, 9, 4177

\bibitem[{Davidson {et~al.}(1990)Davidson, Kohse-H{\"o}inghaus, Chang, \&
  Hanson}]{1990DAV/KOH}
Davidson, D.~F., Kohse-H{\"o}inghaus, K., Chang, A.~Y., \& Hanson, R.~K. 1990,
  Int J Chem Kin, 22, 513

\bibitem[{De Cobos \& Troe(1984)}]{1984CRO/TRO}
De Cobos, A.~E.~C., \& Troe, J. 1984, Int J Chem Kinet, 16, 1519

\bibitem[{Dean(1985)}]{1985DEA}
Dean, A.~J. 1985, JPhCh, 89, 4600

\bibitem[{Dean {et~al.}(1991)Dean, Davidson, \& Hanson}]{1991DEA/DAV}
Dean, A.~J., Davidson, D., \& Hanson, R. 1991, JPhCh, 95,
  183

\bibitem[{Dean \& Hanson(1992)}]{1992DEA/HAN}
Dean, A.~J., \& Hanson, R.~K. 1992, Int J Chem Kin,
  24, 517

\bibitem[{Dean \& Kistiakowsky(1971)}]{1970DEA/KIS}
Dean, A.~J., \& Kistiakowsky, G. 1971, JChPh, 54, 1718

\bibitem[{Debrou {et~al.}(1978)Debrou, Fulford, Lewars, \& March}]{4-112}
Debrou, G.~B., Fulford, J.~E., Lewars, E.~G., \& March, R.~E. 1978, IJMIP, 26, 345

\bibitem[{{Demott} {et~al.}(2003){Demott}, {Cziczo}, {Prenni}, {Murphy},
  {Kreidenweis}, {Thomson}, {Borys}, \& {Rogers}}]{DeMott2003}
{Demott}, P.~J., {Cziczo}, D.~J., {Prenni}, A.~J., {et~al.} 2003, PNAS, 100, 14655

\bibitem[{Deppe {et~al.}(1998)Deppe, Friedrichs, Ibrahim, R{\"o}mming, \&
  Wagner}]{1998DEP/FRI}
Deppe, J., Friedrichs, G., Ibrahim, A., R{\"o}mming, H.-J., \& Wagner, H.~G.
  1998, Berich Bunsen Gesell, 102, 1474

\bibitem[{DeSain {et~al.}(2003)DeSain, Klippenstein, Miller, \&
  Taatjes}]{2003DES/KLI}
DeSain, J.~D., Klippenstein, S.~J., Miller, J.~A., \& Taatjes, C.~A. 2003, JCPA, 107, 4415

\bibitem[{{D{\'e}sert} {et~al.}(2008){D{\'e}sert}, {Vidal-Madjar}, {Lecavelier
  Des Etangs}, {Sing}, {Ehrenreich}, {H{\'e}brard}, \& {Ferlet}}]{Desert2008}
{D{\'e}sert}, J.-M., {Vidal-Madjar}, A., {Lecavelier Des Etangs}, A., {et~al.}
  2008, \aap, 492, 585

\bibitem[{Dheandhanoo {et~al.}(1984)Dheandhanoo, Johnsen, \& Biondi}]{4-308}
Dheandhanoo, S., Johnsen, R., \& Biondi, M.~A. 1984, P\&SS, 32, 1301

\bibitem[{Dibeler {et~al.}(1966)Dibeler, Walker, \& Rosenstock}]{Dibeler1966}
Dibeler, V.~H., Walker, J.~A., \& Rosenstock, H.~M. 1966, J Res Nat Bur Stand A, 70, 459

\bibitem[{Ditchburn(1955)}]{Ditchburn1955}
Ditchburn, R. 1955, RSPSA, 229, 44

\bibitem[{Dixon \& Kirby(1968)}]{Dixon1968}
Dixon, R., \& Kirby, G. 1968, TrFa, 64, 2002

\bibitem[{Dombrowsky {et~al.}(1991)Dombrowsky, Hoffmann, Klatt,
  {et~al.}}]{1991DOM/HOF}
Dombrowsky, C., Hoffmann, A., Klatt, M., {et~al.} 1991, Berich Bunsen Gesell, 95, 1685

\bibitem[{Dombrowsky \& Wagner(1992)}]{1992DOM/WAG}
Dombrowsky, C., \& Wagner, H.~G. 1992, Berich Bunsen Gesell, 96, 1048

\bibitem[{Dong {et~al.}(2005)Dong, Ding, \& Sun}]{2005DON/DIN}
Dong, H., Ding, Y.-h., \& Sun, C.-c. 2005, JChPh,
  122, 204321

\bibitem[{Donovan {et~al.}(1971)Donovan, Dorko, \& Harrison}]{1971DON/DOR}
Donovan, T., Dorko, W., \& Harrison, A. 1971, CaJCh,
  49, 828

\bibitem[{Dorko {et~al.}(1979)Dorko, Pchelkin, Wert, \& Mueller}]{1979DOR/PCH}
Dorko, E.~A., Pchelkin, N.~R., Wert, J.~C., \& Mueller, G.~W. 1979, JPhCh, 83, 297

\bibitem[{Dotan \& Lindinger(1982)}]{4-146}
Dotan, I., \& Lindinger, W. 1982, JChPh, 76, 4972

\bibitem[{Dotan {et~al.}(1980)Dotan, Lindinger, Rowe, Fahey, Fehsenfeld, \&
  Albritton}]{4-208}
Dotan, I., Lindinger, W., Rowe, B., {et~al.} 1980, CPL,
  72, 67

\bibitem[{Drossart {et~al.}(1999)}]{Drossart1999}
Drossart, P., Fouchet, T., Crovisier, J., {et~al.} 1999, ``The Universe as Seen
by ISO'', Vol. 427, 169

\bibitem[{Duan \& Page(1995)}]{1995DUA/PAG}
Duan, X., \& Page, M. 1995, JAChS, 117, 5114

\bibitem[{{Dubrovin} {et~al.}(2014){Dubrovin}, {Luque}, {Gordillo-Vazquez},
  {Yair}, {Parra-Rojas}, {Ebert}, \& {Price}}]{Dubrovin2014}
{Dubrovin}, D., {Luque}, A., {Gordillo-Vazquez}, F.~J., {et~al.} 2014, \icarus,
  241, 313

\bibitem[{Duff \& Sharma(1996)}]{1996DUF/SHA}
Duff, J., \& Sharma, R. 1996, GeoRL, 23, 2777

\bibitem[{Dunbar {et~al.}(1972)Dunbar, Shen, \& Olah}]{1-149}
Dunbar, R.~C., Shen, J., \& Olah, G.~A. 1972, JChPh,
  56, 3794

\bibitem[{Dunkin {et~al.}(1970)Dunkin, Fehsenfeld, \& Ferguson}]{2-0089}
Dunkin, D., Fehsenfeld, F., \& Ferguson, E. 1970, JChPh, 53, 987

\bibitem[{Dunkin {et~al.}(1972)Dunkin, Fehsenfeld, \& Ferguson}]{2-0124}
---. 1972, CPL, 15, 257

\bibitem[{Dunkin {et~al.}(1971)Dunkin, McFarland, Fehsenfeld, \&
  Ferguson}]{2-0102}
Dunkin, D., McFarland, M., Fehsenfeld, F., \& Ferguson, E. 1971, JGR, 76, 3820

\bibitem[{Duran {et~al.}(1988)Duran, Amorebieta, \& Colussi}]{1988DUR/AMO}
Duran, R., Amorebieta, V., \& Colussi, A. 1988, JPhCh, 92, 636

\bibitem[{Durup-Ferguson {et~al.}(1984)Durup-Ferguson, Bohringer, Fahey, \&
  Ferguson}]{4-260}
Durup-Ferguson, M., Bohringer, H., Fahey, D.~W., \& Ferguson, E.~E. 1984, JChPh, 79, 265

\bibitem[{{Dyudina} {et~al.}(2007){Dyudina}, {Ingersoll}, {Ewald}, {Porco},
  {Fischer}, {Kurth}, {Desch}, {Del Genio}, {Barbara}, \&
  {Ferrier}}]{Dyudina2007}
{Dyudina}, U.~A., {Ingersoll}, A.~P., {Ewald}, S.~P., {et~al.} 2007, \icarus,
  190, 545

\bibitem[{Edelb{\"u}ttel-Einhaus {et~al.}(1992)Edelb{\"u}ttel-Einhaus,
  Hoyermann, Rohde, \& Seeba}]{1992EDE/HOY}
Edelb{\"u}ttel-Einhaus, J., Hoyermann, K., Rohde, G., \& Seeba, J. 1992 in International Symposium on Combustion (Elsevier), 661

\bibitem[{Edwards {et~al.}(1966)Edwards, Kerr, Lloyd, \&
  Trotman-Dickenson}]{1966EDW/KER}
Edwards, D., Kerr, J., Lloyd, A., \& Trotman-Dickenson, A. 1966, J Chem Soc A, 1500

\bibitem[{England \& Corcoran(1975)}]{1975ENG/COR}
England, C., \& Corcoran, W.~H. 1975, Ind Eng Chem, 14, 55

\bibitem[{{Enskog}(1917)}]{Enskog1917}
{Enskog}, D. 1917, {Kinetische Theorie der Vorgaenge in maessig verduennten
  Gasen. I. Allgemeiner Teil}, Dissertation

\bibitem[{Ercolano \& Storey(2006)}]{Ercolano2006}
Ercolano, B., \& Storey, P.~J. 2006, MNRAS, 372, 1875

\bibitem[{Eremin {et~al.}(1997)Eremin, Ziborov, Shumova, Voiki, \&
  Roth}]{1997ERE/ZIB}
Eremin, A., Ziborov, V., Shumova, V., Voiki, D., \& Roth, P. 1997, Kin Catal, 38, 1

\bibitem[{Espinosa-Garcia {et~al.}(1993)Espinosa-Garcia, Corchado, \&
  Sana}]{1993ESP/COR}
Espinosa-Garcia, J., Corchado, J., \& Sana, M. 1993, J Chim Phys, 90, 1181

\bibitem[{Fahey {et~al.}(1982)Fahey, B{\"o}hringer, Fehsenfeld, \&
  Ferguson}]{4-255}
Fahey, D., B{\"o}hringer, H., Fehsenfeld, F., \& Ferguson, E. 1982, JChPh, 76, 1799

\bibitem[{Fahey {et~al.}(1981)Fahey, Fehsenfeld, Ferguson, \& Viehland}]{4-253}
Fahey, D., Fehsenfeld, F., Ferguson, E., \& Viehland, L. 1981, JChPh, 75, 669

\bibitem[{Fahr \& Nayak(1994)}]{FahrNayak1994}
Fahr, A., \& Nayak, A. 1994, CP, 189, 725

\bibitem[{Fahr \& Nayak(1996)}]{FahrNayak1996}
Fahr, A., \& Nayak, A. 1996, CP, 203, 351

\bibitem[{Fahr \& Stein(1989)}]{1989FAH/STE}
Fahr, A., \& Stein, S. 1989 in International Symposium on Combustion (Elsevier), 1023

\bibitem[{Faigle {et~al.}(1976)Faigle, Isolani, \& Riveros}]{0-031}
Faigle, J. F.~G., Isolani, P.~C., \& Riveros, J.~M. 1976, JAChS, 98, 2049

\bibitem[{Fairbairn(1969)}]{1969FAI}
Fairbairn, A. 1969, RSPSA, 312, 207

\bibitem[{Faravelli {et~al.}(2000)Faravelli, Goldaniga, Zappella, Ranzi,
  Dagaut, \& Cathonnet}]{2000FAR/GOL}
Faravelli, T., Goldaniga, A., Zappella, L., {et~al.} 2000, P Combust Inst, 28, 2601

\bibitem[{Fegley \& Lodders(1994)}]{Fegley1994}
Fegley, B., \& Lodders, K. 1994, Icarus, 110, 117

\bibitem[{Fehsenfeld(1969)}]{2-0054}
Fehsenfeld, F. 1969, CaJCh, 47, 1808

\bibitem[{Fehsenfeld(1975)}]{2-0203}
---. 1975, JChPh, 63, 1686

\bibitem[{Fehsenfeld(1976)}]{2-0218}
---. 1976, ApJ, 209, 638

\bibitem[{Fehsenfeld(1977)}]{2-0229}
---. 1977, P\&SS, 25, 195

\bibitem[{Fehsenfeld {et~al.}(1969)Fehsenfeld, Albritton, Burt,
  \& Schiff}]{2-0050}
Fehsenfeld, F., Albritton, D., Burt, J., \& Schiff, H. 1969,
  CaJCh, 47, 1793

\bibitem[{Fehsenfeld {et~al.}(1978)Fehsenfeld, Dotan, Albritton, Howard, \&
  Ferguson}]{2-0245}
Fehsenfeld, F., Dotan, I., Albritton, D., Howard, C., \& Ferguson, E. 1978,
  JGR:Oceans, 83, 1333

\bibitem[{Fehsenfeld {et~al.}(1970)Fehsenfeld, Dunkin, \& Ferguson}]{2-0090}
Fehsenfeld, F., Dunkin, D., \& Ferguson, E. 1970, P\&SS,
  18, 1267

\bibitem[{Fehsenfeld {et~al.}(1974)Fehsenfeld, Dunkin, \& Ferguson}]{2-0151}
---. 1974, ApJ, 188, 43

\bibitem[{Fehsenfeld \& Ferguson(1970)}]{2-0087}
Fehsenfeld, F., \& Ferguson, E. 1970, JChPh, 53, 2614

\bibitem[{Fehsenfeld \& Ferguson(1971)}]{2-0112}
---. 1971, JGR, 76, 8453

\bibitem[{Fehsenfeld \& Ferguson(1972)}]{2-0115}
---. 1972, JChPh, 56, 3066

\bibitem[{Fehsenfeld \& Ferguson(1974)}]{2-0136}
---. 1974, JChPh, 61, 3181

\bibitem[{Fehsenfeld {et~al.}(1969{\natexlab{b}})Fehsenfeld, Ferguson, \&
  Bohme}]{2-0063}
Fehsenfeld, F., Ferguson, E., \& Bohme, D. 1969{\natexlab{b}}, P\&SS, 17, 1759

\bibitem[{Fehsenfeld {et~al.}(1969{\natexlab{c}})Fehsenfeld, Ferguson, \&
  Mosesman}]{2-0074}
Fehsenfeld, F., Ferguson, E., \& Mosesman, M. 1969{\natexlab{c}}, CPL, 4, 73

\bibitem[{Fehsenfeld {et~al.}(1966{\natexlab{a}})Fehsenfeld, Ferguson, \&
  Schmeltekopf}]{2-0024}
Fehsenfeld, F., Ferguson, E., \& Schmeltekopf, A. 1966{\natexlab{a}}, JChPh, 45, 1844

\bibitem[{Fehsenfeld {et~al.}(1973)Fehsenfeld, Howard, \& Ferguson}]{2-0137}
Fehsenfeld, F., Howard, C.~J., \& Ferguson, E. 1973, JChPh, 58, 5841

\bibitem[{Fehsenfeld {et~al.}(1975{\natexlab{a}})Fehsenfeld, Howard, \&
  Schmeltekopf}]{2-0199}
Fehsenfeld, F., Howard, C.~J., \& Schmeltekopf, A. 1975{\natexlab{a}}, JChPh, 63, 2835

\bibitem[{Fehsenfeld {et~al.}(1975{\natexlab{b}})Fehsenfeld, Lindinger, \&
  Albritton}]{2-0196}
Fehsenfeld, F., Lindinger, W., \& Albritton, D. 1975{\natexlab{b}}, JChPh, 63, 443

\bibitem[{Fehsenfeld {et~al.}(1976)Fehsenfeld, Lindinger, Schiff, Hemsworth, \&
  Bohme}]{2-0224}
Fehsenfeld, F., Lindinger, W., Schiff, H., Hemsworth, R., \& Bohme, D. 1976,
  JChPh, 64, 4887

\bibitem[{Fehsenfeld {et~al.}(1967{\natexlab{a}})Fehsenfeld, Schmeltekopf, \&
  Ferguson}]{2-0016}
Fehsenfeld, F., Schmeltekopf, A., \& Ferguson, E. 1967{\natexlab{a}}, The
  JChPh, 46, 2802

\bibitem[{Fehsenfeld {et~al.}(1966{\natexlab{b}})Fehsenfeld, Schmeltekopf,
  Goldan, Schiff, \& Ferguson}]{2-0008}
Fehsenfeld, F., Schmeltekopf, A., Goldan, P., Schiff, H., \& Ferguson, E.
  1966{\natexlab{b}}, JChPh, 44, 4087

\bibitem[{Fehsenfeld {et~al.}(1967{\natexlab{b}})Fehsenfeld, Schmeltekopf,
  Schiff, \& Ferguson}]{2-0025}
Fehsenfeld, F., Schmeltekopf, A., Schiff, H., \& Ferguson, E.
  1967{\natexlab{b}}, P\&SS, 15, 373

\bibitem[{Feng \& Hershberger(2007)}]{2006FEN/HER}
Feng, W., \& Hershberger, J.~F. 2007, JCPA, 111, 3831

\bibitem[{Ferguson(1968)}]{2-0021}
Ferguson, E. 1968, Adv Electron Electron Phys, 24, 1

\bibitem[{Ferguson \& Fehsenfeld(1968)}]{2-0045}
Ferguson, E., \& Fehsenfeld, F. 1968, JGR, 73, 6215

\bibitem[{Ferguson {et~al.}(1969)Ferguson, Fehsenfeld, \&
  Schmeltekopf}]{2-0034}
Ferguson, E., Fehsenfeld, F., \& Schmeltekopf, A.~L. 1969, Adv Chem Ser, 80, 83

\bibitem[{Fernandes {et~al.}(2005)Fernandes, Giri, Hippler, Kachiani, \&
  Striebel}]{2005FER/GIR}
Fernandes, R.~X., Giri, B.~R., Hippler, H., Kachiani, C., \& Striebel, F. 2005,
  JCPA, 109, 1063

\bibitem[{Fern{\'a}ndez-Ramos {et~al.}(1998)Fern{\'a}ndez-Ramos,
  Mart{\'\i}nez-N{\'u}{\~n}ez, R{\'\i}os, Rodr{\'\i}guez-Otero, V{\'a}zquez, \&
  Est{\'e}vez}]{1998FER/MAR}
Fern{\'a}ndez-Ramos, A., Mart{\'\i}nez-N{\'u}{\~n}ez, E., R{\'\i}os, M.~A.,
  {et~al.} 1998, JAChS, 120, 7594

\bibitem[{Ferradaz {et~al.}(2009)}]{Ferradaz2009}
Ferradaz, T., Benilan, Y., Fraya, A., {et~al.} 2009, P\&SS, 57, 10

\bibitem[{{Ferris}(1992)}]{Ferris1992}
{Ferris}, J.~P. 1992, OLEB, 22, 109

\bibitem[{{Feuchtgruber} {et~al.}(1997){Feuchtgruber}, {Lellouch}, {de Graauw},
  {B{\'e}zard}, {Encrenaz}, \& {Griffin}}]{Feucht1997}
{Feuchtgruber}, H., {Lellouch}, E., {de Graauw}, T., {et~al.} 1997, \nat, 389,
  159

\bibitem[{Field {et~al.}(1957)Field, Franklin, \& Lampe}]{1-056}
Field, F., Franklin, J., \& Lampe, F. 1957, JAChS, 79, 2419

\bibitem[{Fifer(1975)}]{1975FIF}
Fifer, R. 1975, Ber Bunsenges Phys Chem, 10

\bibitem[{Fluegge(1969{\natexlab{a}})}]{0-084c}
Fluegge, R.~A. 1969{\natexlab{a}} in the Bulletin of the American Physical Society, (New York: APS), 14(2), 261

\bibitem[{Fluegge(1969{\natexlab{b}})}]{0-011}
Fluegge, R.~A. 1969{\natexlab{b}}, JChPh, 50, 4373

\bibitem[{Fontijn {et~al.}(2001)Fontijn, Fernandez, Ristanovic, Randall, \&
  Jankowiak}]{2001FON/FER}
Fontijn, A., Fernandez, A., Ristanovic, A., Randall, M.~Y., \& Jankowiak, J.~T.
  2001, JCPA, 105, 3182

\bibitem[{Forst {et~al.}(1957)Forst, Evans, \& Winkler}]{1957FOR/EVA}
Forst, W., Evans, H., \& Winkler, C. 1957, JPhCh,
  61, 320

\bibitem[{{Fouchet} {et~al.}(2000){Fouchet}, {Lellouch}, {B{\'e}zard},
  {Feuchtgruber}, {Drossart}, \& {Encrenaz}}]{Fouchet2000}
{Fouchet}, T., {Lellouch}, E., {B{\'e}zard}, B., {et~al.} 2000, \aap, 355, L13

\bibitem[{Frank(1986)}]{1986FRA}
Frank, P. 1986, in 15th International Symposium on Rarefied Gas Dynamics (Tuebner)

\bibitem[{Frank {et~al.}(1986)Frank, Bhaskaran, \& Just}]{1986FRA/BHA}
Frank, P., Bhaskaran, K., \& Just, T. 1986, JPhCh, 90, 2226

\bibitem[{Frank {et~al.}(1988)Frank, Bhaskaran, \& Just}]{1988FRA/BHA}
Frank, P., Bhaskaran, K., \& Just, T. 1988 in the International Symposium on Combustion (Elsevier), 885

\bibitem[{Freeman {et~al.}(1978{\natexlab{a}})Freeman, Harland, \&
  McEwan}]{2-4004}
Freeman, C., Harland, P., \& McEwan, M. 1978{\natexlab{a}}, IJMIP, 28, 19

\bibitem[{Freeman {et~al.}(1978{\natexlab{b}})Freeman, Harland, \&
  McEwan}]{4-464}
---. 1978{\natexlab{b}}, AJCh, 31, 2157

\bibitem[{Friedrichs {et~al.}(2008)Friedrichs, Colberg, Dammeier, Bentz, \&
  Olzmann}]{2008FRI/COL}
Friedrichs, G., Colberg, M., Dammeier, J., Bentz, T., \& Olzmann, M. 2008,
  PCCP, 10, 6520

\bibitem[{Friedrichs {et~al.}(2002)}]{Friedrichs2002}
Friedrichs, G., DAvidson, D.~F., \& Hanson, R.~K. 2002, Int J Chem Kinet, 37, 374

\bibitem[{Fulle \& Hippler(1997)}]{1997FUL/HIP}
Fulle, D., \& Hippler, H. 1997, JChPh, 106, 8691

\bibitem[{Gannon {et~al.}(2007)Gannon, Glowacki, Blitz, Hughes, Pilling, \&
  Seakins}]{2007GAN/GLO}
Gannon, K.~L., Glowacki, D.~R., Blitz, M.~A., {et~al.} 2007, JPCA, 111, 6679

\bibitem[{Gao \& Macdonald(2006)}]{2006GAO/MAC}
Gao, Y., \& Macdonald, R.~G. 2006, JCPA, 110,
  977

\bibitem[{{Garrod} {et~al.}(2008){Garrod}, {Weaver}, \& {Herbst}}]{Garrod2008}
{Garrod}, R.~T., {Weaver}, S.~L.~W., \& {Herbst}, E. 2008, \apj, 682, 283

\bibitem[{Gear(1971)}]{Gear1971}
Gear, C.~W. 1971, Comm ACM, 14, 185

\bibitem[{Gehring {et~al.}(1969)Gehring, Hoyermann, Wagner, \&
  Wolfrum}]{1969GEH/HOY}
Gehring, M., Hoyermann, K., Wagner, H.~G., \& Wolfrum, J. 1969, Berich Bunsen Gesell, 73, 956

\bibitem[{Geiger {et~al.}(1999)Geiger, Wiesen, \& Becker}]{1999GEI/WIE}
Geiger, H., Wiesen, P., \& Becker, K.~H. 1999, PCCP, 1, 5601

\bibitem[{Geltman(1962)}]{Geltman1962}
Geltman, S. 1962, The Astrophysical Journal, 136, 935

\bibitem[{Gentieu \& Mentall(1970)}]{Gentieu1970}
Gentieu, E., \& Mentall, J. 1970, Science, 169, 681

\bibitem[{Gill {et~al.}(1981)Gill, Johnson, \& Atkinson}]{1981GIL/JOH}
Gill, R., Johnson, W., \& Atkinson, G. 1981, CP, 58, 29

\bibitem[{{Gladstone} {et~al.}(1996){Gladstone}, {Allen}, \&
  {Yung}}]{Gladstone1996}
{Gladstone}, G.~R., {Allen}, M., \& {Yung}, Y.~L. 1996, \icarus, 119, 1

\bibitem[{Gl{\"a}nzer \& Troe(1973)}]{1973GLA/TRO}
Gl{\"a}nzer, K., \& Troe, J. 1973, AcHCh, 56, 577

\bibitem[{Gl{\"a}nzer \& Troe(1975)}]{1975GLA/TRO}
---. 1975, Berich Bunsen Gesell, 79, 465

\bibitem[{Glarborg {et~al.}(1995)Glarborg, Dam-Johansen, \&
  Miller}]{1995GLA/DAM}
Glarborg, P., Dam-Johansen, K., \& Miller, J.~A. 1995, Int J Chem Kinet, 27, 1207

\bibitem[{Glosik {et~al.}(1978)Glosik, Raksit, Twiddy, Adams, \& Smith}]{4-010}
Glosik, J., Raksit, A., Twiddy, N., Adams, N., \& Smith, D. 1978, JPhB, 11, 3365

\bibitem[{Gonzalez {et~al.}(1992)Gonzalez, Theisen, Schlegel, Hase, \&
  Kaiser}]{1992GON/THE}
Gonzalez, C., Theisen, J., Schlegel, H.~B., Hase, W.~L., \& Kaiser, E. 1992,
  JPhCh, 96, 1767

\bibitem[{Gorden \& Ausloos(1961)}]{Gorden1961}
Gorden, R., \& Ausloos, P. 1961, JPhCh, 65, 1033

\bibitem[{{Gorden} \& {Ausloos}(1967)}]{Gorden1967}
---. 1967, JChPh, 46, 4823

\bibitem[{Gordon \& McBride(1999)}]{Gordon1999}
Gordon, S., \& McBride, B.~J. 1999, Thermodynamic Data to 20000 K for Monatomic Gases (Cleveland: NASA)

\bibitem[{Grabowski(1983)}]{4-280}
Grabowski, J.~J. 1983, Doctoral Thesis (University of Colorado)

\bibitem[{Graham \& Johnston(1978{\natexlab{a}})}]{1978GRA/JOH}
Graham, R.~A., \& Johnston, H.~S. 1978{\natexlab{a}}, JPhCh, 82, 254

\bibitem[{Graham \& Johnston(1978{\natexlab{b}})}]{Graham1978}
---. 1978{\natexlab{b}}, JPhCh, 82, 254

\bibitem[{Graham {et~al.}(1973)Graham, James, Keever, Gatland, Albritton,
  \& McDaniel}]{0-113}
Graham, E., James, D., Keever, W., {et~al.} 1973, JChPh, 59, 4648

\bibitem[{Gray \& Herod(1968)}]{1968GRA/HER}
Gray, P., \& Herod, A. 1968, Transactions of the Faraday Society, 64, 2723

\bibitem[{Greenberg {et~al.}(1995)Greenberg, Kouchi, Niessen, Irth, van
  Paradijs, de~Groot, \& Hermsen}]{Greenberg1995}
Greenberg, J.~M., Kouchi, A., Niessen, W., {et~al.} 1995, J Biol Phys, 20, 61

\bibitem[{Griggs(1968)}]{Griggs1968}
Griggs, M. 1968, JChPh, 49, 857

\bibitem[{Grotheer \& Just(1981)}]{1981GRO/JUS}
Grotheer, H., \& Just, T. 1981, CPL, 78, 71

\bibitem[{Grussdorf {et~al.}(1994)Grussdorf, Nolte, Temps, \&
  Wagner}]{1994GRU/NOL}
Grussdorf, J., Nolte, J., Temps, F., \& Wagner, H.~G. 1994, Berich Bunsen Gesell, 98, 546

\bibitem[{Gupta {et~al.}(1967)Gupta, Jones, Harrison, \& Myher}]{0-071}
Gupta, S., Jones, E., Harrison, A.~G., \& Myher, J.~J. 1967, CaJCh, 45, 3107

\bibitem[{{Gurnett} {et~al.}(1990){Gurnett}, {Kurth}, {Cairns}, \&
  {Granroth}}]{Gurnett1990}
{Gurnett}, D.~A., {Kurth}, W.~S., {Cairns}, I.~H., \& {Granroth}, L.~J. 1990,
  NASA STI/Recon Technical Report N, 91, 11642

\bibitem[{Guyon {et~al.}(1976)Guyon, Chupka, \& Berkowitz}]{Guyon1976}
Guyon, P.~M., Chupka, W.~A., \& Berkowitz, J. 1976, JChPh, 64, 1419

\bibitem[{Hack {et~al.}(2005)Hack, Hold, Hoyermann, Wehmeyer, \&
  Zeuch}]{2005HAC/HOL}
Hack, W., Hold, M., Hoyermann, K., Wehmeyer, J., \& Zeuch, T. 2005, PCCP, 7, 1977

\bibitem[{Hahn(1997)}]{Hahn1997}
Hahn, Y. 1997, PhLA, 231, 82

\bibitem[{{Haldane}(1928)}]{Haldane1928}
{Haldane}, J.~B.~S. 1928, Ration Annu, 148, 3

\bibitem[{{Hamill} {et~al.}(1977){Hamill}, {Toon}, \& {Kiang}}]{Hamill1977}
{Hamill}, P., {Toon}, O.~B., \& {Kiang}, C.~S. 1977, JAtS, 34, 1104

\bibitem[{Hammerum \& S{\o}lling(1999)}]{Hammerum1999}
Hammerum, S., \& S{\o}lling, T.~I. 1999, JAChS, 121, 6002

\bibitem[{Hanson \& Salimian(1984)}]{1984HAN/SAL}
Hanson, R.~K., \& Salimian, S. 1984, in Symposium on Combustion Chemistry (Springer), 361

\bibitem[{{Harada} {et~al.}(2010){Harada}, {Herbst}, \& {Wakelam}}]{Harada2010}
{Harada}, N., {Herbst}, E., \& {Wakelam}, V. 2010, \apj, 721, 1570

\bibitem[{{Harding}, {et~al.}(1993)}]{Harding1993}
{Harding}, L.~B., Guadagnini, R., \& Schatz, G.~C. 1993, JPhCh, 97, 5472

\bibitem[{Harding {et~al.}(2005)Harding, Klippenstein, \&
  Georgievskii}]{2005HAR/KLI}
Harding, L.~B., Klippenstein, S.~J., \& Georgievskii, Y. 2005, P Combust Inst, 30, 985

\bibitem[{Harding {et~al.}(2008)Harding, Klippenstein, \& Miller}]{2008HAR/KLI}
Harding, L.~B., Klippenstein, S.~J., \& Miller, J.~A. 2008, JPCA, 112, 522

\bibitem[{Harding \& Wagner(1989)}]{1989HAR/WAG}
Harding, L.~B., \& Wagner, A.~F. 1989 in International Symposium on Combustion (Elsevier), 983

\bibitem[{Hartmann {et~al.}(1990)Hartmann, Karth{\"a}user, Sawerysyn, \&
  Zellner}]{1990HAR/KAR}
Hartmann, D., Karth{\"a}user, J., Sawerysyn, J., \& Zellner, R. 1990, Berich Bunsen Gesell, 94, 639

\bibitem[{{Hasegawa} {et~al.}(1992){Hasegawa}, {Herbst}, \&
  {Leung}}]{Hasegawa1992}
{Hasegawa}, T.~I., {Herbst}, E., \& {Leung}, C.~M. 1992, \apjs, 82, 167

\bibitem[{Hassinen {et~al.}(1990)Hassinen, Kalliorinne, \&
  Koskikallio}]{1990HAS/KAL}
Hassinen, E., Kalliorinne, K., \& Koskikallio, J. 1990, Int J Chem Kin, 22, 741

\bibitem[{Hassinen {et~al.}(1985)Hassinen, Riepponen, Blomqvist, Kalliorinne,
  Evseev, \& Koskikallio}]{1985HAS/RIE}
Hassinen, E., Riepponen, P., Blomqvist, K., {et~al.} 1985, Int J Chem Kin, 17, 1125

\bibitem[{Hastie {et~al.}(1976)Hastie, Freeman, McEwan, \&
  Schiff}]{1976HAS/FRE}
Hastie, D., Freeman, C., McEwan, M., \& Schiff, H. 1976, Int J Chem Kin, 8, 307

\bibitem[{Haworth {et~al.}(2003)Haworth, Mackie, \& Bacskay}]{2003HAW/MAC}
Haworth, N.~L., Mackie, J.~C., \& Bacskay, G.~B. 2003, JPCA, 107, 6792

\bibitem[{He {et~al.}(1993)He, Liu, Lin, \& Melius}]{1993HE/LIU}
He, Y., Liu, X., Lin, M., \& Melius, C. 1993, Int J Chem Kin, 25, 845

\bibitem[{He {et~al.}(1988)He, Sanders, \& Lin}]{1988HE/SAN}
He, Y., Sanders, W., \& Lin, M. 1988, JPhCh, 92,
  5474

\bibitem[{H\'{e}brard {et~al.}(2013)}]{Hebrard2013}
H\'{e}brard, E., Dobrijevic, M., Loison, J.~C., {et~al.} 2013, A\&A, 552, A132

\bibitem[{Hedin(1987)}]{Hedin1987}
Hedin, A.~E. 1987, JGR: Space Physics, 92, 4649

\bibitem[{Hedin(1991)}]{Hedin1991}
---. 1991, JGR: Space Physics, 96, 1159

\bibitem[{Heicklen \& Johnston(1962)}]{1962HEI/JOH}
Heicklen, J., \& Johnston, H.~S. 1962, JAChS, 84, 4394

\bibitem[{Heimerl {et~al.}(1969)Heimerl, Johnsen, \& Biondi}]{0-016}
Heimerl, J., Johnsen, R., \& Biondi, M.~A. 1969, JChPh, 51, 5041

\bibitem[{Helling \& Fomins(2013)}]{Helling2013}
Helling, Ch., \& Fomins, A. 2013, RSPTA, 371, 20110581

\bibitem[{{Helling} {et~al.}(2013){Helling}, {Jardine}, {Diver}, \&
  {Witte}}]{Helling2013b}
{Helling}, Ch., {Jardine}, M., {Diver}, D., \& {Witte}, S. 2013, \planss, 77,
  152

\bibitem[{{Helling} {et~al.}(2014){Helling}, {Woitke}, {Rimmer}, {Kamp}, {Thi},
  \& {Meijerink}}]{Helling2014}
{Helling}, Ch., {Woitke}, P., {Rimmer}, P.~B., {et~al.} 2014, Life, 4, 142

\bibitem[{Hemsworth {et~al.}(1974)Hemsworth, Payzant, Schiff, \&
  Bohme}]{2-6015}
Hemsworth, R., Payzant, J., Schiff, H., \& Bohme, D. 1974, CPL, 26, 417

\bibitem[{Hemsworth {et~al.}(1973)Hemsworth, Rundle, Bohme, Schiff, Dunkin, \&
  Fehsenfeld}]{2-6009}
Hemsworth, R., Rundle, H., Bohme, D., {et~al.} 1973, JChPh, 59, 61

\bibitem[{Henis {et~al.}(1973)Henis, Stewart, \& Gaspar}]{0-155}
Henis, J., Stewart, G., \& Gaspar, P. 1973, JChPh,
  58, 3639

\bibitem[{Hennig \& Wagner(1994)}]{1994HEN/WAG}
Hennig, G., \& Wagner, H. 1994, Berich Bunsen Gesell, 98, 749

\bibitem[{{Henry} {et~al.}(2000){Henry}, {Marcy}, {Butler}, \&
  {Vogt}}]{Henry2000}
{Henry}, G.~W., {Marcy}, G.~W., {Butler}, R.~P., \& {Vogt}, S.~S. 2000, \apjl,
  529, L41

\bibitem[{Henry(1970)}]{Henry1970}
Henry, R.~J. 1970, ApJ, 161, 1153

\bibitem[{Henry \& McElroy(1968)}]{Henry1968}
Henry, R.~J., \& McElroy, M.~B. 1968, Adv Space Res, 1, 251

\bibitem[{Herbst {et~al.}(1975)Herbst, Payzant, Schiff, \& Bohme}]{2-6022}
Herbst, E., Payzant, J., Schiff, H., \& Bohme, D. 1975, ApJ, 201, 603

\bibitem[{Herman \& Mentall(1982)}]{Herman1982}
Herman, J., \& Mentall, J. 1982, JGR: Oceans, 87, 8967

\bibitem[{Herron(1988)}]{1988HER}
Herron, J.~T. 1988, JPCRD, 17, 967

\bibitem[{Herron(1999)}]{1999HER}
---. 1999, JPCRD, 28, 1453

\bibitem[{Hidaka {et~al.}(1990)Hidaka, Nakamura, Tanaka, Inami, \&
  Kawano}]{1990HID/NAK}
Hidaka, Y., Nakamura, T., Tanaka, H., Inami, K., \& Kawano, H. 1990,
  Int J Chem Kin, 22, 701

\bibitem[{Hidaka {et~al.}(1989)Hidaka, Oki, Kawano, \&
  Higashihara}]{Hidaka1989}
Hidaka, Y., Oki, T., Kawano, H., \& Higashihara, T. 1989, JPhCh, 93, 7134

\bibitem[{Hidaka {et~al.}(2000)Hidaka, Sato, \& Yamane}]{2000HID/SAT}
Hidaka, Y., Sato, K., \& Yamane, M. 2000, CoFl, 123, 1

\bibitem[{Hinshelwood \& Askey(1927)}]{1927HIN/ASK}
Hinshelwood, C., \& Askey, P. 1927, RSPSA, 215

\bibitem[{Hiraoka \& Kebarle(1980)}]{4-043}
Hiraoka, K., \& Kebarle, P. 1980, CaJCh, 58, 2262

\bibitem[{Hoehlein \& Freeman(1970)}]{1970HOH/FRE}
Hoehlein, G., \& Freeman, G. 1970, JAChS,
  92, 6118

\bibitem[{Homann \& Wellmann(1983)}]{1983HOM/WEL}
Homann, K., \& Wellmann, C. 1983, Berich Bunsen Gesell, 87, 609

\bibitem[{Hopkinson {et~al.}(1979)Hopkinson, Mackay, \& Bohme}]{4-220}
Hopkinson, A., Mackay, G., \& Bohme, D. 1979, CaJCh,
  57, 2996

\bibitem[{Horne \& Norrish(1970)}]{1970HOR/NOR}
Horne, D., \& Norrish, R. 1970, RSPSA, 315, 301

\bibitem[{Howard(1979)}]{1979HOW}
Howard, C.~J. 1979, JChPh, 71, 2352

\bibitem[{Howard {et~al.}(1974)Howard, Fehsenfeld, \& McFarland}]{2-0163}
Howard, C.~J., Fehsenfeld, F., \& McFarland, M. 1974, JChPh, 60, 5086

\bibitem[{Howorka {et~al.}(1974)Howorka, Lindinger, \& Varney}]{0-160}
Howorka, F., Lindinger, W., \& Varney, R.~N. 1974, JChPh, 61, 1180

\bibitem[{Hoyermann {et~al.}(1999)Hoyermann, Olzmann, Seeba, \&
  Viskolcz}]{1999HOY/OLZ}
Hoyermann, K., Olzmann, M., Seeba, J., \& Viskolcz, B. 1999, JPCA, 103, 5692

\bibitem[{Hsu {et~al.}(1997)Hsu, Lin, Mebel, \& Melius}]{1997HSU/LIN}
Hsu, C.-C., Lin, M., Mebel, A., \& Melius, C. 1997, JPCA, 101, 60

\bibitem[{Hu \& Seager(2014)}]{Hu2014}
Hu, R., \& Seager, S. 2014, ApJ, 784, 63

\bibitem[{{Hu} {et~al.}(2012){Hu}, {Seager}, \& {Bains}}]{Hu2012}
{Hu}, R., {Seager}, S., \& {Bains}, W. 2012, \apj, 761, 166

\bibitem[{Hu {et~al.}(2013)Hu, Seager, \& Bains}]{Hu2013}
Hu, R., Seager, S., \& Bains, W. 2013, ApJ, 769, 6

\bibitem[{{Hu} {et~al.}(2015){Hu}, {Seager}, \& {Yung}}]{Hu2015}
{Hu}, R., {Seager}, S., \& {Yung}, Y.~L. 2015, arXiv:1505.02221

\bibitem[{{Huebner} \& {Carpenter}(1979)}]{Huebner1979}
{Huebner}, W.~F., \& {Carpenter}, C.~W. 1979, NASA STI/Recon Technical Report
  N, 80, 24243

\bibitem[{{Huebner} {et~al.}(1992){Huebner}, {Keady}, \& {Lyon}}]{Huebner1992}
{Huebner}, W.~F., {Keady}, J.~J., \& {Lyon}, S.~P. 1992, \apss, 195, 1

\bibitem[{{Huebner} \& {Mukherjee}(2015)}]{Huebner2015}
{Huebner}, W.~F., \& {Mukherjee}, J. 2015, \planss, 106, 11

\bibitem[{Huffman(1969)}]{Huffman1969}
Huffman, R.~E. 1969, CaJCh, 47, 1823

\bibitem[{Huffman {et~al.}(1963)Huffman, Tanaka, \& Larrabee}]{Huffman1963}
Huffman, R.~E., Tanaka, Y., \& Larrabee, J. 1963, JChPh, 39, 910

\bibitem[{Humpfer {et~al.}(1995)Humpfer, Oser, \& Grotheer}]{1995HUM/OSE}
Humpfer, R., Oser, H., \& Grotheer, H.-H. 1995, Int J Chem Kin, 27, 577

\bibitem[{Hunt {et~al.}(1965)Hunt, Kerr, \& Trotman-Dickenson}]{1965HUN/KER}
Hunt, M., Kerr, J., \& Trotman-Dickenson, A. 1965, J Chem Soc, 5074

\bibitem[{Huntress {et~al.}(1980)Huntress, Anicich, McEwan, \&
  Karpas}]{4-472}
Huntress, W.~T., Anicich, V., McEwan, M., \& Karpas, Z. 1980, ApJS, 44, 481

\bibitem[{{Huntress} \& {Baldeschwieler}(1969)}]{0-014}
{Huntress}, W.~T., \& {Baldeschwieler}, J.~D. 1969, \nat, 223, 468

\bibitem[{Huntress \& Elleman(1970)}]{0-020}
Huntress, W.~T., \& Elleman, D. 1970, JAChS, 92, 3565

\bibitem[{Huntress {et~al.}(1971)Huntress, Mosesman, \& Elleman}]{0-022}
Huntress, W.~T., Mosesman, M.~M., \& Elleman, D.~D. 1971, JChPh, 54, 843

\bibitem[{Huntress {et~al.}(1973)Huntress, Pinizzotto, \& Laudenslager}]{1-037}
Huntress, W.~T., Pinizzotto, R.~F., \& Laudenslager, J.~B. 1973, JAChS, 95, 4107

\bibitem[{Husain \& Lee(1988)}]{1988HUS/LEE}
Husain, D., \& Lee, Y. 1988, Int J Chem Kin, 20,
  223

\bibitem[{Husain \& Marshall(1986)}]{1986HUS/MAR}
Husain, D., \& Marshall, P. 1986, Int J Chem Kin,
  18, 83

\bibitem[{Husain \& Young(1975)}]{1975HUS/YOU}
Husain, D., \& Young, A.~N. 1975, FaTr II, 71, 525

\bibitem[{Huynh \& Truong(2008)}]{2008HUY/TRU}
Huynh, L.~K., \& Truong, T.~N. 2008, Theoretical Chemistry Accounts, 120, 107

\bibitem[{Huynh \& Violi(2008)}]{2008HUY/VIO}
Huynh, L.~K., \& Violi, A. 2008, J Org Chem, 73, 94

\bibitem[{Ibragimova(1986)}]{1986IBR}
Ibragimova, L. 1986, Kin Catal, 27

\bibitem[{Ikeda \& Mackie(1996)}]{1996IKE/MAC}
Ikeda, E., \& Mackie, J.~C. 1996 in International Symposium on Combustion (Elsevier), 597

\bibitem[{Ikezoe {et~al.}(1987)Ikezoe, Matsuoka, \& Takebe}]{Ikezoe1987}
Ikezoe, Y., Matsuoka, S., \& Takebe, M. 1987, Gas phase ion-molecule reaction
  rate constants through 1986 (Ion reaction research group of the Mass
  spectroscopy society of Japan)

\bibitem[{Imai \& Toyama(1962)}]{1962IMA/TOY}
Imai, N., \& Toyama, O. 1962, Bulletin of the Chemical Society of Japan, 35, 860

\bibitem[{{Indriolo} {et~al.}(2009){Indriolo}, {Fields}, \&
  {McCall}}]{Indriolo2009}
{Indriolo}, N., {Fields}, B.~D., \& {McCall}, B.~J. 2009, \apj, 694, 257

\bibitem[{Ing {et~al.}(2003)Ing, Sheng, \& Bozzelli}]{2003ING/SHE}
Ing, W.-C., Sheng, C.~Y., \& Bozzelli, J.~W. 2003, Fuel Process Technol, 83, 111

\bibitem[{Inn(1975)}]{Inn1975}
Inn, E.~C. 1975, JAtS, 32, 2375

\bibitem[{Jachimowski(1977)}]{1977JAC}
Jachimowski, C.~J. 1977, CoFl, 29, 55

\bibitem[{Jaffe {et~al.}(1973)Jaffe, Karpas, \& Klein}]{0-142}
Jaffe, S., Karpas, Z., \& Klein, F.~S. 1973, JChPh,
  58, 2190

\bibitem[{Jaffe \& Klein(1974)}]{0-143}
Jaffe, S., \& Klein, F. 1974, IJMIP, 14, 459

\bibitem[{Jamieson {et~al.}(1970)Jamieson, Brown, \& Tanner}]{1970JAM/BRO}
Jamieson, J., Brown, G., \& Tanner, J. 1970, CaJCh, 48,
  3619

\bibitem[{Jarrold {et~al.}(1983)Jarrold, Bass, Kemper, van Koppen, \&
  Bowers}]{4-080}
Jarrold, M.~F., Bass, L.~M., Kemper, P.~R., van Koppen, P.~A., \& Bowers, M.~T.
  1983, JChPh, 78, 3756

\bibitem[{Jasper {et~al.}(2007)Jasper, Klippenstein, Harding, \&
  Ruscic}]{2007JAS/KLI}
Jasper, A.~W., Klippenstein, S.~J., Harding, L.~B., \& Ruscic, B. 2007, JCPA, 111, 3932

\bibitem[{Javoy {et~al.}(2003)Javoy, Naudet, Abid, \& Paillard}]{2003JAV/NAU}
Javoy, S., Naudet, V., Abid, S., \& Paillard, C. 2003, ExTFS, 27, 371

\bibitem[{Jebens {et~al.}(1992)}]{Jebens1992}
Jebens, D.~S., Lakkaraju, H.~S., McKay, C.~P., \& Borucki, W.~J. 1992, GRL, 19, 273

\bibitem[{Jensen(1982)}]{1982JEN}
Jensen, D.~E. 1982, FaTr I, 78, 2835

\bibitem[{Johnsen {et~al.}(1970)Johnsen, Brown, \& Biondi}]{0-021}
Johnsen, R., Brown, H., \& Biondi, M.~A. 1970, JChPh, 52, 5080

\bibitem[{Johnsen {et~al.}(1974)Johnsen, Castell, \& Biondi}]{0-137}
Johnsen, R., Castell, F., \& Biondi, M.~A. 1974, JChPh, 61, 5404

\bibitem[{{Johnson} {et~al.}(2008){Johnson}, {Cleaves}, {Dworkin}, {Glavin},
  {Lazcano}, \& {Bada}}]{Johnson2008}
{Johnson}, A.~P., {Cleaves}, H.~J., {Dworkin}, J.~P., {et~al.} 2008, Science,
  322, 404

\bibitem[{Johnston(1951)}]{1951JOH}
Johnston, H.~S. 1951, JAChS, 73, 4542

\bibitem[{Jones {et~al.}(1981{\natexlab{a}})Jones, Birkinshaw, \&
  Twiddy}]{4-138}
Jones, J., Birkinshaw, K., \& Twiddy, N. 1981{\natexlab{a}}, CPL, 77, 484

\bibitem[{Jones {et~al.}(1981{\natexlab{b}})Jones, Birkinshaw, \&
  Twiddy}]{4-135}
---. 1981{\natexlab{b}}, JPhB, 14, 2705

\bibitem[{Jones {et~al.}(1979)Jones, Lister, \& Twiddy}]{4-425}
Jones, J., Lister, D., \& Twiddy, N. 1979, JPhB, 12, 2723

\bibitem[{Joshi {et~al.}(2005)Joshi, You, Barckholtz, \& Wang}]{2005JOS/YOU}
Joshi, A., You, X., Barckholtz, T.~A., \& Wang, H. 2005, JPCA, 109, 8016

\bibitem[{Ju {et~al.}(2007)Ju, Han, \& Varandas}]{2007JU/HAN}
Ju, L.-P., Han, K.-L., \& Varandas, A.~J. 2007, Int J Chem Kin, 39, 148

\bibitem[{Kappes \& Staley(1981)}]{4-059}
Kappes, M.~M., \& Staley, R.~H. 1981, JAChS,
  103, 1286

\bibitem[{Karkach \& Osherov(1999)}]{1999KAR/OSH}
Karkach, S.~P., \& Osherov, V.~I. 1999, JChPh, 110, 11918

\bibitem[{Karpas {et~al.}(1978)Karpas, Anicich, \& Huntress}]{4-107}
Karpas, Z., Anicich, V., \& Huntress, W. 1978, CPL, 59, 84

\bibitem[{Karpas {et~al.}(1979)Karpas, Anicich, \& Huntress}]{4-465}
---. 1979, JChPh, 70, 2877

\bibitem[{Karpas \& Klein(1975)}]{0-145}
Karpas, Z., \& Klein, F.~S. 1975, IJMIP, 16, 289

\bibitem[{Kasper \& Franklin(1972)}]{0-181}
Kasper, S., \& Franklin, J. 1972, JChPh, 56, 1156

\bibitem[{{Kasting}(1993)}]{Kasting1993}
{Kasting}, J.~F. 1993, Science, 259, 920

\bibitem[{Katayama {et~al.}(1973)Katayama, Huffman, \& O'Bryan}]{Katayama1973}
Katayama, D., Huffman, R., \& O'Bryan, C. 1973, JChPh, 59, 4309

\bibitem[{Kato \& Cvetanovic(1967)}]{1967KAT/CVE}
Kato, A., \& Cvetanovic, R. 1967, CaJCh, 45, 1845

\bibitem[{Keller {et~al.}(1998)Keller, Anicich, \& Cravens}]{Keller1998}
Keller, C., Anicich, V., \& Cravens, T. 1998, P\&SS, 46,
  1157

\bibitem[{Keller-Rudek {et~al.}(2013)}]{Keller2013}
Keller-Rudek, H., Moortgat, G.~K., Sander, R., \& S\"{o}rensen, R. 2013, ESSD, 5, 365

\bibitem[{Kelly \& Heicklen(1978)}]{1978KEL/HEI}
Kelly, N., \& Heicklen, J. 1978, J Photochem, 8, 83

\bibitem[{Kemper \& Bowers(1984)}]{4-311}
Kemper, P.~R., \& Bowers, M. 1984, Int J Chem Kin,
  16, 707

\bibitem[{Kemper {et~al.}(1983)Kemper, Bowers, Parent, Mauclaire, Derai, \&
  Marx}]{4-393}
Kemper, P.~R., Bowers, M.~T., Parent, D.~C., {et~al.} 1983, JChPh, 79, 160

\bibitem[{Kern {et~al.}(1988)Kern, Singh, \& Wu}]{1988KER/SIN}
Kern, R., Singh, H., \& Wu, C. 1988, Int J Chem Kin, 20, 731

\bibitem[{Kiefer {et~al.}(1988)Kiefer, Mitchell, Kern, \& Yong}]{1988KIE/MIT}
Kiefer, J., Mitchell, K., Kern, R., \& Yong, J. 1988, JPhCh, 92, 677

\bibitem[{Kiefer {et~al.}(2005)Kiefer, Santhanam, Srinivasan, Tranter, Klippenstein, \& Oehlschlaeger}]{2005KIE/SAN}
Kiefer, J., Santhanam, S., Srinivasan, N.~K., {et~al.} 2005, Proceedings of the Combustion 
Institute, 30, 1129

\bibitem[{Kim {et~al.}(1975)Kim, Theard, \& Huntress}]{1-036}
Kim, J., Theard, L., \& Huntress, W.-T. 1975, JChPh, 62, 45

\bibitem[{Kim {et~al.}(2010)}]{Kim2010}
Kim, S.~J., Geballe, T.~R., Kim, J., {et~al.} 2010, Icarus, 208, 837

\bibitem[{Kim \& Fox(1994)}]{Kim1994}
Kim, Y.~H., \& Fox, J.~L. 1994, Icarus, 112, 310

\bibitem[{Kim {et~al.}(2014)}]{Kim2014}
Kim, Y.~H., Fox, J.~L., Black, J.~H., \& Moses, J.~I. 2014, JGRA, 119, 384

\bibitem[{Klatt {et~al.}(1995)Klatt, Spindler, \& Wagner}]{1995KLA/SPI}
Klatt, M., Spindler, B., \& Wagner, H.~G. 1995, ZPC, 191, 241

\bibitem[{Klemm(1965)}]{1965KLE}
Klemm, R. 1965, CaJCh, 43, 2633

\bibitem[{{Knutson} {et~al.}(2008){Knutson}, {Charbonneau}, {Allen}, {Burrows},
  \& {Megeath}}]{Knutson2008}
{Knutson}, H.~A., {Charbonneau}, D., {Allen}, L.~E., {Burrows}, A., \&
  {Megeath}, S.~T. 2008, \apj, 673, 526

\bibitem[{Knyazev {et~al.}(1996a)}]{Knyazev1996}
Knyazev, V.~D., Benscura, \'{A}, Stoliarov, S.~I., {et~al.} 1996, JPhCh, 100, 11346

\bibitem[{Knyazev {et~al.}(1996b)Knyaev, Stoliarov, \& Slagle}]{1996KNY/STO}
Knyazev, V.~D., Stoliarov, S.~I., \& Slagle, I.~R. 1996b in International Symposium on Combustion (Elsevier), 513

\bibitem[{Koch \& Skibowski(1971)}]{Koch1971}
Koch, E.-E., \& Skibowski, d.~M. 1971, CPL, 9, 429

\bibitem[{Koike {et~al.}(2000)Koike, Kudo, Maeda, \& Yamada}]{2000KOI/KUD}
Koike, T., Kudo, M., Maeda, I., \& Yamada, H. 2000, Int J Chem Kin, 32, 1

\bibitem[{{Kooij}(1893)}]{Kooij1893}
{Kooij}, D.~M. 1893, ZPC, 12, 155

\bibitem[{Korovkina(1976)}]{1976KOR}
Korovkina, T. 1976, High Energ Chem, 10, 75

\bibitem[{Kovalenko {et~al.}(2007)}]{Kovalenko2007}
{Kovalenko}, L.~J., {Jucks}, K.~W., {Salawitch}, R.~J., et al. 2007, JGL, 34, 19801 

\bibitem[{Kretschmer \& Petersen(1963)}]{0-067}
Kretschmer, C., \& Petersen, H. 1963, JChPh, 39, 1772

\bibitem[{Kronebusch \& Berkowitz(1976)}]{Kronebusch1976}
Kronebusch, P., \& Berkowitz, J. 1976, IJMIP, 22, 283

\bibitem[{Kruse \& Roth(1997)}]{1997KRU/ROT}
Kruse, T., \& Roth, P. 1997, JCPA, 101, 2138

\bibitem[{Kruse \& Roth(1999)}]{1999KRU/ROT}
---. 1999, Int J Chem Kin, 31, 11

\bibitem[{Kukui {et~al.}(1995)Kukui, Jungkamp, \& Schindler}]{1995KUK/JUN}
Kukui, A., Jungkamp, T., \& Schindler, R. 1995, Berich Bunsen Gesell, 99, 1565

\bibitem[{Kumakura {et~al.}(1978{\natexlab{a}})Kumakura, Arakawa, \&
  Sugiura}]{4-268}
Kumakura, M., Arakawa, K., \& Sugiura, T. 1978{\natexlab{a}}, B Chem Soc Jpn, 51, 49

\bibitem[{Kumakura {et~al.}(1978{\natexlab{b}})Kumakura, Arakawa, \&
  Sugiura}]{4-114}
---. 1978{\natexlab{b}}, IJMIP, 26, 303

\bibitem[{Kumakura {et~al.}(1979)Kumakura, Arakawa, \& Sugiura}]{4-121}
---. 1979, IJMIP, 29, 21

\bibitem[{Laidler \& McKenney(1964)}]{1964LAI/MCK}
Laidler, K., \& McKenney, D. 1964, RSPSA, 278, 517

\bibitem[{Laidler \& Wojciechowski(1961)}]{1961LAI/WOJ}
Laidler, K., \& Wojciechowski, B. 1961, RSPSA, 260, 103

\bibitem[{Lamb {et~al.}(1984)Lamb, Mozurkewich, \& Benson}]{1984LAM/MOZ}
Lamb, J.~J., Mozurkewich, M., \& Benson, S.~W. 1984, JPhCh, 88, 6441

\bibitem[{Lambert {et~al.}(1968)Lambert, Christie, Golesworthy, \&
  Linnett}]{1968LAM/CHR}
Lambert, R., Christie, M., Golesworthy, R., \& Linnett, J. 1968, RSPSA 302, 167

\bibitem[{Lambert {et~al.}(1967)Lambert, Christie, \& Linnett}]{1967LAM/CHR}
Lambert, R., Christie, M., \& Linnett, J. 1967, Chem Commun (London), 388

\bibitem[{Langer \& Ljungstr{\"o}m(1994)}]{1994LAN/LJU}
Langer, S., \& Ljungstr{\"o}m, E. 1994, Int J Chem Kin, 26, 367

\bibitem[{Langer \& Ljungstr{\"o}m(1995)}]{1995LAN/LJU}
---. 1995, FaTr, 91, 405

\bibitem[{Laudenslager {et~al.}(1974)Laudenslager, Huntress, \&
  Bowers}]{0-144}
Laudenslager, J.~B., Huntress, W.~T., \& Bowers, M.~T. 1974, JChPh, 61, 4600

\bibitem[{Laufer \& Fahr(2004)}]{2004LAU/FAH}
Laufer, A.~H., \& Fahr, A. 2004, ChRv, 104, 2813

\bibitem[{Lavendy {et~al.}(1984)Lavendy, Gandara, \& Robbe}]{Lavendy1984}
Lavendy, H., Gandara, G., \& Robbe, J. 1984, JMoSp,
  106, 395

\bibitem[{Lavendy {et~al.}(1987)Lavendy, Robbe, \& Gandara}]{Lavendy1987}
Lavendy, H., Robbe, J., \& Gandara, G. 1987, JPhB, 20, 3067

\bibitem[{Lavvas {et~al.}(2008{\natexlab{a}})Lavvas, Coustenis, \&
  Vardavas}]{Lavvas2008a}
Lavvas, P., Coustenis, A., \& Vardavas, I. 2008{\natexlab{a}}, P\&SS, 56, 27

\bibitem[{Lavvas {et~al.}(2008{\natexlab{b}})Lavvas, Coustenis, \&
  Vardavas}]{Lavvas2008b}
---. 2008{\natexlab{b}}, P\&SS, 56, 67

\bibitem[{{Lavvas} {et~al.}(2014){Lavvas}, {Koskinen}, \& {Yelle}}]{Lavvas2014}
{Lavvas}, P., {Koskinen}, T., \& {Yelle}, R.~V. 2014, \apj, 796, 15

\bibitem[{Lawson {et~al.}(1976)Lawson, Bonner, Mather, Todd, \& March}]{0-042}
Lawson, G., Bonner, R.~F., Mather, R.~E., Todd, J.~F., \& March, R.~E. 1976, FaTr I, 72, 545

\bibitem[{Lee {et~al.}(2015)Lee, Helling, Dobbs-Dixon, Juncher}]{Lee2015}
Lee, G., Helling, Ch., Dobbs-Dixon, I., \& Juncher, D. 2015, arXiv:1505.06576

\bibitem[{Lee \& Bozzelli(2003)}]{2003LEE/BOZ}
Lee, J., \& Bozzelli, J.~W. 2003, Int J Chem Kin, 35, 20

\bibitem[{Lee {et~al.}(1973)Lee, Carlson, Judge, \& Ogawa}]{Lee1973}
Lee, L., Carlson, R., Judge, D., \& Ogawa, M. 1973, JQSRT, 13, 1023

\bibitem[{Lellouch {et~al.}(1994)}]{Lellouch1994}
{Lellouch}, E., {Romani}, P.~N., \& Rosenqvist, J. 1994, Icarus, 108, 112

\bibitem[{Levush {et~al.}(1969)Levush, Abadzhev, \& Shevchuk}]{1969LEV/ABA}
Levush, S., Abadzhev, S., \& Shevchuk, V. 1969, Neftekhimiya, 9, 215

\bibitem[{Levy(1956)}]{1956LEV}
Levy, J.~B. 1956, JAChS, 78, 1780

\bibitem[{Li \& Wang(2004)}]{2004LI/WAN}
Li, Q.-S., \& Wang, C.~Y. 2004, JCoCh, 25, 251

\bibitem[{Li {et~al.}(2004)Li, Zhang, \& Zhang}]{2004LI/ZHA}
Li, Q.~S., Zhang, Y., \& Zhang, S. 2004, JCPA, 108, 2014

\bibitem[{Li \& Williams(1996)}]{1996LI/WIL}
Li, S., \& Williams, F. 1996 in International Symposium on Combustion (Elsevier), 1017

\bibitem[{Li {et~al.}(2006)Li, Zhang, \& Wang}]{2006LI/ZHA}
Li, S., Zhang, Q., \& Wang, W. 2006, CPL, 428, 262

\bibitem[{{Liang} {et~al.}(2003){Liang}, {Parkinson}, {Lee}, {Yung}, \&
  {Seager}}]{Liang2003}
{Liang}, M.-C., {Parkinson}, C.~D., {Lee}, A.~Y.-T., {Yung}, Y.~L., \&
  {Seager}, S. 2003, \apjl, 596, L247

\bibitem[{Lias(1988)}]{Lias1988}
Lias, S. 1988, JPCRD, 17, 1

\bibitem[{Lias {et~al.}(1970)Lias, Collin, Rebbert, \& Ausloos}]{Lias1970}
Lias, S., Collin, G., Rebbert, R., \& Ausloos, P. 1970, JChPh, 52, 1841

\bibitem[{Lichtin {et~al.}(1984)Lichtin, Berman, \& Lin}]{1984LIC/BER}
Lichtin, D., Berman, M., \& Lin, M. 1984, CPL, 108, 18

\bibitem[{Liddy {et~al.}(1977{\natexlab{a}})Liddy, Freeman, \& McEwan}]{2-4002}
Liddy, J., Freeman, C., \& McEwan, M. 1977{\natexlab{a}}, MNRAS, 180, 683

\bibitem[{Liddy {et~al.}(1977{\natexlab{b}})Liddy, Freeman, \& McEwan}]{2-4001}
---. 1977{\natexlab{b}}, IJMIP, 23, 153

\bibitem[{Lifshitz \& Ben-Hamou(1983)}]{1983LIF/BEN}
Lifshitz, A., \& Ben-Hamou, H. 1983, JPhCh, 87, 1782

\bibitem[{Lifshitz \& Frenklach(1980)}]{1980LIF/FRE}
Lifshitz, A., \& Frenklach, M. 1980, Int J Chem Kin, 12, 159

\bibitem[{Lifshitz \& Tamburu(1994)}]{1994LIF/TAM}
Lifshitz, A., \& Tamburu, C. 1994, JPhCh, 98, 1161

\bibitem[{Lifshitz \& Tamburu(1998)}]{1998LIF/TAM}
---. 1998, Int J Chem Kin, 30, 341

\bibitem[{Lifshitz {et~al.}(1997)Lifshitz, Tamburu, \& Carroll}]{1997LIF/TAM}
Lifshitz, A., Tamburu, C., \& Carroll, H.~F. 1997, Int J Chem Kin, 29, 839

\bibitem[{Lifshitz {et~al.}(1993)Lifshitz, Tamburu, Frank, \&
  Just}]{1993LIF/TAM}
Lifshitz, A., Tamburu, C., Frank, P., \& Just, T. 1993, JPhCh, 97, 4085

\bibitem[{Lifshitz \& Tassa(1973)}]{0-133}
Lifshitz, C., \& Tassa, R. 1973, IJMIP, 12, 433

\bibitem[{Lifshitz {et~al.}(1977)Lifshitz, Wu, Haartz, \& Tiernan}]{4-281}
Lifshitz, C., Wu, R., Haartz, J., \& Tiernan, T. 1977, JChPh, 67, 2381

\bibitem[{Lifshitz {et~al.}(1978)Lifshitz, Wu, Tiernan, \& Terwilliger}]{4-284}
Lifshitz, C., Wu, R., Tiernan, T., \& Terwilliger, D. 1978, JChPh, 68, 247

\bibitem[{Lin {et~al.}(1992)Lin, He, \& Melius}]{1992LIN/HE}
Lin, M., He, Y., \& Melius, C. 1992, Int J Chem Kin, 24, 1103

\bibitem[{Lin {et~al.}(1993)Lin, He, \& Melius}]{1993LIN/HE}
---. 1993, JPhCh, 97, 9124

\bibitem[{{Lindemann} {et~al.}(1922){Lindemann}, {Arrhenius}, {Langmuir},
  {Dhar}, {Perrin}, \& Lewis}]{Lindemann1922}
{Lindemann}, F.~A., {Arrhenius}, S., {Langmuir}, I., {et~al.} 1922,
  TrFa, 17, 598

\bibitem[{Linder {et~al.}(1996)Linder, Duan, \& Page}]{1996LIN/DUA}
Linder, D.~P., Duan, X., \& Page, M. 1996, JChPh, 104, 6298

\bibitem[{Lindinger(1973)}]{0-153}
Lindinger, W. 1973, PhRvA, 7, 328

\bibitem[{Lindinger(1976)}]{2-0222}
---. 1976, JChPh, 64, 3720

\bibitem[{Lindinger {et~al.}(1979)Lindinger, Albritton, \&
  Fehsenfeld}]{4-241}
Lindinger, W., Albritton, D., \& Fehsenfeld, F. 1979, JChPh, 70, 2038

\bibitem[{Lindinger {et~al.}(1975{\natexlab{a}})Lindinger, Albritton,
  Fehsenfeld, \& Ferguson}]{2-0198}
Lindinger, W., Albritton, D., Fehsenfeld, F., \& Ferguson, E.
  1975{\natexlab{a}}, JGR, 80, 3725

\bibitem[{Lindinger {et~al.}(1975{\natexlab{b}})Lindinger, Albritton,
  Fehsenfeld, \& Ferguson}]{2-0201}
---. 1975{\natexlab{b}}, JChPh, 63, 3238

\bibitem[{Lindinger {et~al.}(1975{\natexlab{c}})Lindinger, Albritton,
  Fehsenfeld, Schmeltekopf, \& Ferguson}]{2-0182}
Lindinger, W., Albritton, D., Fehsenfeld, F., Schmeltekopf, A., \& Ferguson, E.
  1975{\natexlab{c}}, JChPh, 62, 3549

\bibitem[{Lindinger {et~al.}(1975{\natexlab{d}})Lindinger, Albritton, Howard,
  Fehsenfeld, \& Ferguson}]{2-0195}
Lindinger, W., Albritton, D., Howard, C.~J., Fehsenfeld, F., \& Ferguson, E.
  1975{\natexlab{d}}, JChPh, 63, 5220

\bibitem[{Lindinger {et~al.}(1975{\natexlab{e}})Lindinger, Albritton,
  McFarland, Fehsenfeld, Schmeltekopf, \& Ferguson}]{2-0188}
Lindinger, W., Albritton, D., McFarland, M., {et~al.} 1975{\natexlab{e}}, JChPh, 62, 4101

\bibitem[{Lindinger {et~al.}(1974)Lindinger, Fehsenfeld, Schmeltekopf, \&
  Ferguson}]{2-0171}
Lindinger, W., Fehsenfeld, F., Schmeltekopf, A., \& Ferguson, E. 1974, JGR, 79, 4753

\bibitem[{Lindinger {et~al.}(1981)Lindinger, Howorka, Lukac, Kuhn, Villinger,
  Alge, \& Ramler}]{4-410}
Lindinger, W., Howorka, F., Lukac, P., {et~al.} 1981, PhRvA, 23,
  2319

\bibitem[{Lindinger {et~al.}(1975{\natexlab{f}})Lindinger, McFarland,
  Fehsenfeld, Albritton, Schmeltekopf, \& Ferguson}]{2-0186}
Lindinger, W., McFarland, M., Fehsenfeld, F., {et~al.} 1975{\natexlab{f}}, JChPh, 63, 2175

\bibitem[{Lindley {et~al.}(1979)Lindley, Calvert, \& Shaw}]{1979LIN/CAL}
Lindley, C.~R., Calvert, J.~G., \& Shaw, J.~H. 1979, CPL, 67, 57

\bibitem[{{Little} {et~al.}(1999){Little}, {Anger}, {Ingersoll}, {Vasavada},
  {Senske}, {Breneman}, {Borucki}, \& {The Galileo SSI Team}}]{Little1999}
{Little}, B., {Anger}, C.~D., {Ingersoll}, A.~P., {et~al.} 1999, \icarus, 142,
  306

\bibitem[{Liu {et~al.}(2002)Liu, Ding, Li, Fu, Huang, Sun, \&
  Tang}]{2002LIU/DIN}
Liu, G.-x., Ding, Y.-h., Li, Z.-s., {et~al.} 2002, PCCP, 4, 1021

\bibitem[{Lloyd(1974)}]{1974LLO}
Lloyd, A.~C. 1974, Int J Chem Kin, 6, 169

\bibitem[{Loison {et~al.}(2015)}]{Loison2015}
{Loison}, J.~C., {H\'{e}brard}, E., {Dobrijevic}, M., {Hickson}, K.~M.,
{Caralp}, F., {et~al.} 2015, Icarus, 247, 218

\bibitem[{Lombos {et~al.}(1967)Lombos, Sauvageau, \& Sandorfy}]{Lombos1967}
Lombos, B., Sauvageau, P., \& Sandorfy, C. 1967, JMoSp, 24, 253

\bibitem[{Louge \& Hanson(1984)}]{1984LOU/HAN}
Louge, M.~Y., \& Hanson, R.~K. 1984, CoFl, 58, 291

\bibitem[{Lukirskii {et~al.}(1964)Lukirskii, Brytov, \&
  Zimkina}]{Lukirskii1964}
Lukirskii, A., Brytov, I., \& Zimkina, T. 1964, OptSp, 17,
  234

\bibitem[{{Luque} \& {Ebert}(2009)}]{Luque2009}
{Luque}, A., \& {Ebert}, U. 2009, NatGe, 2, 757

\bibitem[{Macdonald(2007)}]{2007MAC}
Macdonald, R.~G. 2007, PCCP, 9, 4301

\bibitem[{Mackay {et~al.}(1976{\natexlab{a}})Mackay, Betowski, Payzant, Schiff,
  \& Bohme}]{2-6027}
Mackay, G.~I., Betowski, L., Payzant, J., Schiff, H., \& Bohme, D.
  1976{\natexlab{a}}, JPhCh, 80, 2919

\bibitem[{Mackay \& Bohme(1978)}]{2-6030}
Mackay, G.~I., \& Bohme, D.~K. 1978, IJMIP, 26, 327

\bibitem[{Mackay {et~al.}(1976{\natexlab{b}})Mackay, Hemsworth, \&
  Bohme}]{2-6024}
Mackay, G.~I., Hemsworth, R.~S., \& Bohme, D.~K. 1976{\natexlab{b}}, CaJCh, 54, 1624

\bibitem[{Mackay {et~al.}(1978)Mackay, Hopkinson, \& Bohme}]{4-217}
Mackay, G.~I., Hopkinson, A., \& Bohme, D. 1978, JAChS, 100, 7460

\bibitem[{Mackay {et~al.}(1982)Mackay, Rakshit, \& Bohme}]{4-225}
Mackay, G.~I., Rakshit, A.~B., \& Bohme, D.~K. 1982, CaJCh, 60, 2594

\bibitem[{Mackay {et~al.}(1981)Mackay, Schiff, \& Bohme}]{4-069}
Mackay, G.~I., Schiff, H., \& Bohme, D. 1981, CaJCh, 59,
  1771

\bibitem[{Mackay {et~al.}(1977)Mackay, Tanaka, \& Bohme}]{2-6033}
Mackay, G.~I., Tanaka, K., \& Bohme, D.~K. 1977, IJMIP, 24, 125

\bibitem[{Mackay {et~al.}(1979)Mackay, Tanner, Hopkinson, \& Bohme}]{4-218}
Mackay, G.~I., Tanner, S.~D., Hopkinson, A.~C., \& Bohme, D.~K. 1979, CaJCh, 57, 1518

\bibitem[{Mackay {et~al.}(1980)Mackay, Vlachos, Bohme, \& Schiff}]{4-052}
Mackay, G.~I., Vlachos, G., Bohme, D.-K., \& Schiff, H. 1980, IJMIP, 36, 259

\bibitem[{{Madhusudhan} {et~al.}(2011){Madhusudhan}, {Harrington}, {Stevenson},
  {Nymeyer}, {Campo}, {Wheatley}, {Deming}, {Blecic}, {Hardy}, {Lust},
  {Anderson}, {Collier-Cameron}, {Britt}, {Bowman}, {Hebb}, {Hellier},
  {Maxted}, {Pollacco}, \& {West}}]{Madhu2011}
{Madhusudhan}, N., {Harrington}, J., {Stevenson}, K.~B., {et~al.} 2011, \nat,
  469, 64

\bibitem[{{Madhusudhan} {et~al.}(2012){Madhusudhan}, {Lee}, \&
  {Mousis}}]{Madhu2012}
{Madhusudhan}, N., {Lee}, K.~K.~M., \& {Mousis}, O. 2012, \apjl, 759, L40

\bibitem[{{Madhusudhan} \& {Seager}(2009)}]{Madhu2009}
{Madhusudhan}, N., \& {Seager}, S. 2009, \apj, 707, 24

\bibitem[{Madronich(1987)}]{Madronich1987}
Madronich, S. 1987, JGR: Atmospheres, 92, 9740

\bibitem[{Magnotta \& Johnston(1980)}]{Magnotta1980}
Magnotta, F., \& Johnston, H.~S. 1980, GeoRL, 7, 769

\bibitem[{Mahmud {et~al.}(1987)Mahmud, Marshall, \& Fontijn}]{1987MAH/MAR}
Mahmud, K., Marshall, P., \& Fontijn, A. 1987, JPhCh,
  91, 1568

\bibitem[{{Manion} {et~al.}(2013){Manion}, {Huie}, {Levin}, L., {Tsang},
  {McGivern}, {Hudgens}, {Knyazev}, {Atkinson}, {Chai}, Tereza, {Lin},
  {Allison}, {Mallard}, {Westley}, {Herron}, {Hampson}, \&
  {Frizzell}}]{NIST2013}
{Manion}, J.~A., {Huie}, R.~E., {Levin}, R.~D., {et~al.} 2013, {NIST Standard
  Reference Database 17, Version 7.0 (Web Version), Release 1.6.8, Data version
  2013.03}

\bibitem[{Mansergas \& Anglada(2006)}]{Mansergas2006}
Mansergas, A., \& Anglada, J.~M. 2006, JCPA, 110, 4001

\bibitem[{Marinov {et~al.}(1998)Marinov, Pitz, Westbrook, Vincitore, Castaldi,
  Senkan, \& Melius}]{1998MAR/PIT}
Marinov, N.~M., Pitz, W.~J., Westbrook, C.~K., {et~al.} 1998, CoFl, 114, 192

\bibitem[{M{\"a}rk \& Oskam(1971)}]{0-103}
M{\"a}rk, T.~D., \& Oskam, H. 1971, PhRvA, 4, 1445

\bibitem[{Marmo(1953)}]{Marmo1953}
Marmo, F. 1953, JOSA, 43, 1186

\bibitem[{Marx {et~al.}(1973)Marx, Mauclaire, Fehsenfeld, Dunkin, \&
  Ferguson}]{0-112}
Marx, R., Mauclaire, G., Fehsenfeld, F., Dunkin, D., \& Ferguson, E. 1973, JChPh, 58, 3267

\bibitem[{Massie \& Hunten(1981)}]{Massie1981}
{Massie}, S.~T., \& {Hunten}, D.~M. 1981, JGR, 86, 9859

\bibitem[{Masuoka \& Samson(1981)}]{Masuoka1981}
Masuoka, T., \& Samson, J.~A. 1981, JChPh, 74, 1093

\bibitem[{Matsui \& Nomaguchi(1978)}]{1978MAT/NOM}
Matsui, Y., \& Nomaguchi, T. 1978, CoFl, 32, 205

\bibitem[{Matsumoto {et~al.}(1975)Matsumoto, Okada, Misaki, Taniguchi, \&
  Hayakawa}]{0-053}
Matsumoto, A., Okada, S., Misaki, T., Taniguchi, S., \& Hayakawa, T. 1975,
  B Chem Soc Jpn, 48, 794

\bibitem[{Matsunaga \& Watanabe(1967)}]{Matsunaga1967}
Matsunaga, F., \& Watanabe, K. 1967, Sci Light, 16, 31

\bibitem[{Matsuoka \& Ikezoe(1988)}]{4-516}
Matsuoka, S., \& Ikezoe, Y. 1988, JPhCh, 92, 1126

\bibitem[{Mauclaire {et~al.}(1978)Mauclaire, Derai, \& Marx}]{4-111}
Mauclaire, G., Derai, R., \& Marx, R. 1978, IJMIP, 26, 289

\bibitem[{Mayer \& Schieler(1968)}]{1968MAY/SCH}
Mayer, S., \& Schieler, L. 1968, JPhCh, 72, 2628

\bibitem[{Mayer {et~al.}(1966)Mayer, Schieler, \& Johnston}]{1966MAY/SCH}
Mayer, S., Schieler, L., \& Johnston, H.~S. 1966, JChPh, 45, 385

\bibitem[{Mayer {et~al.}(1967)Mayer, Schieler, \& Johnston}]{1967MAY/SCH}
---. 1967 in International Symposium on Combustion (Elsevier), 837

\bibitem[{Mayer \& Lampe(1974{\natexlab{a}})}]{1-130}
Mayer, T., \& Lampe, F. 1974{\natexlab{a}}, JPhCh,
  78, 2645

\bibitem[{Mayer \& Lampe(1974{\natexlab{b}})}]{1-129}
---. 1974{\natexlab{b}}, JPhCh, 78, 2433

\bibitem[{McAllister(1973)}]{1-103}
McAllister, T. 1973, IJMIP, 10, 419

\bibitem[{McAllister \& Pitman(1976)}]{0-032}
McAllister, T., \& Pitman, P. 1976, IJMIP, 19, 423

\bibitem[{{McBride} {et~al.}(1993){McBride}, {Gordon}, \& {Reno}}]{McBride1993}
{McBride}, B.~J., {Gordon}, S., \& {Reno}, M.~A. 1993, {Coefficients for
  calculating thermodynamic and transport properties of individual species},
  Tech. rep.

\bibitem[{McElroy \& McConnell(1971)}]{McElroy1971}
McElroy, M.~B., \& McConnell, J.~C. 1971, JGR, 76,
  6674

\bibitem[{McEwan {et~al.}(1981)McEwan, Anicich, \& Huntress}]{4-415}
McEwan, M., Anicich, V., \& Huntress, W. 1981, IJMIP, 37, 273

\bibitem[{McEwan {et~al.}(1983)McEwan, Anicich, Huntress, Kemper, \&
  Bowers}]{4-305}
McEwan, M., Anicich, V., Huntress, W., Kemper, P., \& Bowers, M. 1983,
  IJMIP, 50, 179

\bibitem[{McFarland {et~al.}(1972)McFarland, Dunkin, Fehsenfeld, Schmeltekopf,
  \& Ferguson}]{2-0105}
McFarland, M., Dunkin, D., Fehsenfeld, F., Schmeltekopf, A., \& Ferguson, E.
  1972, JChPh, 56, 2358

\bibitem[{McKenney {et~al.}(1963)McKenney, Wojciechowski, \&
  Laidler}]{1963MCK/WOJ}
McKenney, D., Wojciechowski, B., \& Laidler, K. 1963, CaJCh, 41, 1993

\bibitem[{McKnight(1970)}]{0-122}
McKnight, L. 1970, PhRvA, 2, 762

\bibitem[{McNesby \& Okabe(1964)}]{McNesby1964}
McNesby, J.~R., \& Okabe, H. 1964, Adv Photochem, 3, 157

\bibitem[{{McNesby} {et~al.}(1962){McNesby}, {Tanaka}, \&
  {Okabe}}]{McNesby1962}
{McNesby}, J.~R., {Tanaka}, I., \& {Okabe}, H. 1962, \jcp, 36, 605

\bibitem[{Meaburn \& Gordon(1968)}]{1968MEA/GOR}
Meaburn, G., \& Gordon, S. 1968, JPhCh, 72, 1592

\bibitem[{Meagher \& Anderson(2000)}]{2000MEA/AND}
Meagher, N.~E., \& Anderson, W.~R. 2000, JCPA,
  104, 6013

\bibitem[{Mebel \& Lin(1997)}]{1997MEB/LIN}
Mebel, A., \& Lin, M. 1997, IRPC, 16, 249

\bibitem[{Mebel \& Lin(1999)}]{1999MEB/LIN}
---. 1999, JCPA, 103, 2088

\bibitem[{Mebel {et~al.}(1996)Mebel, Lin, Morokuma, \& Melius}]{1996MEB/LIN}
Mebel, A., Lin, M., Morokuma, K., \& Melius, C. 1996, Int J Chem Kin, 28, 693

\bibitem[{Melton \& Rudolph(1960)}]{1-126}
Melton, C., \& Rudolph, P. 1960, JChPh, 32, 1128

\bibitem[{Mentall {et~al.}(1971)Mentall, Gentieu, Krauss, \&
  Neumann}]{Mentall1971}
Mentall, J., Gentieu, E., Krauss, M., \& Neumann, D. 1971, JChPh, 55, 5471

\bibitem[{Meot-Ner {et~al.}(1986)Meot-Ner, Karpas, \& Deakyne}]{4-452}
Meot-Ner, M., Karpas, Z., \& Deakyne, C.~A. 1986, JAChS, 108, 3913

\bibitem[{Mertens \& Hanson(1996)}]{1996MER/HAN}
Mertens, J.~D., \& Hanson, R.~K. 1996 in International Symposium on Combustion (Elsevier), 
551

\bibitem[{Mertens {et~al.}(1991)Mertens, Kohse-H{\"o}inghaus, Hanson, \&
  Bowman}]{1991MER/KOH}
Mertens, J.~D., Kohse-H{\"o}inghaus, K., Hanson, R.~K., \& Bowman, C.~T. 1991,
  Int J Chem Kin, 23, 655

\bibitem[{Metcalfe {et~al.}(1983)Metcalfe, Booth, McAndrew, \&
  Wooley}]{1983MET/BOO}
Metcalfe, E., Booth, D., McAndrew, H., \& Wooley, W. 1983, Fire Mater,
  7, 185

\bibitem[{Metzger \& Cook(1964)}]{Metzger1964}
Metzger, P., \& Cook, G. 1964, JChPh, 41, 642

\bibitem[{Meyer {et~al.}(1969)Meyer, Olschewski, Troe, \& Wagner}]{1969MEY/OLS}
Meyer, E., Olschewski, H., Troe, J., \& Wagner, H.~G. 1969 in International Symposium on Combustion (Elsevier), 345

\bibitem[{Meyer \& Hershberger(2005)}]{2005MEY/HER}
Meyer, J.~P., \& Hershberger, J.~F. 2005, JCPA, 109, 4772

\bibitem[{Michael {et~al.}(1999)Michael, Kumaran, \& Su}]{1999MIC/KUM}
Michael, J., Kumaran, S., \& Su, M.-C. 1999, JPhCh A, 103, 5942

\bibitem[{Mick {et~al.}(1993)Mick, Burmeister, \& Roth}]{1993MIC/BUR}
Mick, H.-J., Burmeister, M., \& Roth, P. 1993, AIAA, 31, 671

\bibitem[{Miller \& Glarborg(1999)}]{1999MIL/GLA}
Miller, J.~A., \& Glarborg, P. 1999, Int J Chem Kin, 31, 757

\bibitem[{Miller \& Melius(1988)}]{1988MIL/MEL}
Miller, J.~A., \& Melius, C.~F. 1988 in International Symposium on Combustion (Elsevier), 919

\bibitem[{Miller \& Melius(1989)}]{1989MIL/MEL}
Miller, J.~A., \& Melius, C.~F. 1989 in International Symposium on Combustion (Elsevier), 1031

\bibitem[{Miller \& Melius(1992)}]{1992MIL/MEL}
Miller, J.~A., \& Melius, C.~F. 1992, Int J Chem Kin, 24, 421

\bibitem[{Miller {et~al.}(2005)Miller, Pilling, \& Troe}]{2005MIL/PIL}
Miller, J.~A., Pilling, M.~J., \& Troe, J. 2005, P Combust Inst, 30, 43

\bibitem[{Miller {et~al.}(2004)Miller, McCunn, Krisch, Butler, \&
  Shu}]{2004MIL/MCC}
Miller, J.~L., McCunn, L.~R., Krisch, M.~J., Butler, L.~J., \& Shu, J. 2004,
  JChPh, 121, 1830

\bibitem[{{Miller}(1953)}]{Miller1953}
{Miller}, S.~L. 1953, Science, 117, 528

\bibitem[{{Miller} \& {Urey}(1959)}]{Miller1959}
{Miller}, S.~L., \& {Urey}, H.~C. 1959, Science, 130, 245

\bibitem[{Miller {et~al.}(1984)Miller, Wetterskog, \& Paulson}]{4-204}
Miller, T.~M., Wetterskog, R.~E., \& Paulson, J.~F. 1984, JChPh, 80, 4922

\bibitem[{{Miyakawa} {et~al.}(2002){Miyakawa}, {Yamanashi}, {Kobayashi},
  {Cleaves}, \& {Miller}}]{Miyakawa2002}
{Miyakawa}, S., {Yamanashi}, H., {Kobayashi}, K., {Cleaves}, H.~J., \&
  {Miller}, S.~L. 2002, PNAS, 99, 14628

\bibitem[{Miyoshi {et~al.}(1993)Miyoshi, Ohmori, Tsuchiya, \&
  Matsui}]{1993MIY/OHM}
Miyoshi, A., Ohmori, K., Tsuchiya, K., \& Matsui, H. 1993, CPL, 204, 241

\bibitem[{Molina \& Molina(1981)}]{Molina1981}
Molina, L.~T., \& Molina, M.~J. 1981, J Photochem, 15, 97

\bibitem[{Molina {et~al.}(1999)}]{Molina1999}
{Molina-Cuberos}, G.~J., {Lopez-Moreno}, J.~J., {Rodrigo}, R., {Lara}, L.~M.,
\& {O'Brien}, K. 1999, P\&SS, 47, 1347

\bibitem[{Monks {et~al.}(1993)Monks, Romani, Nesbitt, Scanlon, \&
  Stief}]{1993MON/ROM}
Monks, P., Romani, P., Nesbitt, F., Scanlon, M., \& Stief, L. 1993, JGR: Planets, 98, 17115

\bibitem[{Moortgat {et~al.}(1977)Moortgat, {\v{S}}Slemr, \&
  Warneck}]{1977MOO/SLE}
Moortgat, G., {\v{S}}Slemr, F., \& Warneck, P. 1977, Int J Chem Kin, 9, 249

\bibitem[{Moortgat \& Warneck(1975)}]{Moortgat1975}
Moortgat, G., \& Warneck, P. 1975, ZNatA, 30, 835

\bibitem[{Morris \& Niki(1973)}]{1973MOR/NIK}
Morris, E., \& Niki, H. 1973, Int J Chem Kin, 5, 47

\bibitem[{Morrissey \& Schubert(1963)}]{1963MOR/SCH}
Morrissey, R.~J., \& Schubert, C. 1963, CoFl, 7, 263

\bibitem[{Moses(2014)}]{Moses2014}
Moses, J.~I. 2014, RSPTA, 372, 20130073

\bibitem[{Moses(2015)}]{Moses2015}
{Moses}, J.~I., {Armstrong}, E.~S., {Fletcher}, L.~N., {et al.} 2015, Icarus, in press

\bibitem[{{Moses} {et~al.}(2000{\natexlab{a}}){Moses}, {B{\'e}zard},
  {Lellouch}, {Gladstone}, {Feuchtgruber}, \& {Allen}}]{Moses2000a}
{Moses}, J.~I., {B{\'e}zard}, B., {Lellouch}, E., {et~al.} 2000{\natexlab{a}},
  \icarus, 143, 244

\bibitem[{{Moses} {et~al.}(2005){Moses}, {Fouchet}, {B{\'e}zard}, {Gladstone},
  {Lellouch}, \& {Feuchtgruber}}]{Moses2005}
{Moses}, J.~I., {Fouchet}, T., {B{\'e}zard}, B., {et~al.} 2005, JGR: Planets, 110, 8001

\bibitem[{{Moses} {et~al.}(2000{\natexlab{b}}){Moses}, {Lellouch},
  {B{\'e}zard}, {Gladstone}, {Feuchtgruber}, \& {Allen}}]{Moses2000b}
{Moses}, J.~I., {Lellouch}, E., {B{\'e}zard}, B., {et~al.} 2000{\natexlab{b}},
  \icarus, 145, 166

\bibitem[{{Moses} {et~al.}(2013){Moses}, {Madhusudhan}, {Visscher}, \&
  {Freedman}}]{Moses2013}
{Moses}, J.~I., {Madhusudhan}, N., {Visscher}, C., \& {Freedman}, R.~S. 2013,
  \apj, 763, 25

\bibitem[{{Moses} {et~al.}(2011){Moses}, {Visscher}, {Fortney}, {Showman},
  {Lewis}, {Griffith}, {Klippenstein}, {Shabram}, {Friedson}, {Marley}, \&
  {Freedman}}]{Moses2011}
{Moses}, J.~I., {Visscher}, C., {Fortney}, J.~J., {et~al.} 2011, \apj, 737, 15

\bibitem[{Moshkina {et~al.}(1980)Moshkina, Polyak, Sokolova, Masterovoi, \&
  Nalbandyan}]{1980MOS/POL}
Moshkina, R., Polyak, S., Sokolova, N., Masterovoi, I., \& Nalbandyan, A. 1980,
  Int J Chem Kin, 12, 315

\bibitem[{Mount \& Moos(1978)}]{Mount1978}
Mount, G., \& Moos, H. 1978, ApJ, 224, L35

\bibitem[{Mousavipour \& Saheb(2007)}]{2007MOU/SAH}
Mousavipour, S.~H., \& Saheb, V. 2007, B Chem Soc Jpn, 80, 1901

\bibitem[{Moylan {et~al.}(1985)Moylan, Jasinski, \& Brauman}]{4-446}
Moylan, C.~R., Jasinski, J.~M., \& Brauman, J.~I. 1985, JAChS, 107, 1934

\bibitem[{Mulvihill \& Phillips(1975)}]{1975MUL/PHI}
Mulvihill, J.~N., \& Phillips, L.~F. 1975 in International Symposium on Combustion (Elsevier), 1113

\bibitem[{Munson \& Field(1969)}]{1-109}
Munson, M.~S., \& Field, F.~H. 1969, JAChS, 91, 3413

\bibitem[{Munson {et~al.}(1964)Munson, Franklin, \& Field}]{1-035}
Munson, M.~S., Franklin, J., \& Field, F. 1964, JPhCh, 68, 3098

\bibitem[{Murrell \& Rodriguez(1986)}]{1986MUR/ROD}
Murrell, J., \& Rodriguez, J. 1986, JMoSt, 139, 267

\bibitem[{Musin \& Lin(1998)}]{1998MUS/LIN}
Musin, R., \& Lin, M. 1998, JCPA, 102, 1808

\bibitem[{Myer \& Samson(1970)}]{Myer1970}
Myer, J.~A., \& Samson, J.~A. 1970, JChPh, 52, 266

\bibitem[{Myerson(1973)}]{1973MYE}
Myerson, A.~L. 1973in , Elsevier, 219--228

\bibitem[{Nadtochenko {et~al.}(1979)Nadtochenko, Sarkisov, \&
  Vedeneev}]{1979NAD/SAR}
Nadtochenko, V., Sarkisov, O., \& Vedeneev, V. 1979, Doklady Akademii Nauk
  SSSR, 244, 152

\bibitem[{Nakata {et~al.}(1965)Nakata, Watanabe, \& Matsunaga}]{Nakata1965}
Nakata, R., Watanabe, K., \& Matsunaga, F. 1965, Sci Light, 14, 54V71

\bibitem[{Nakayama {et~al.}(1959)Nakayama, Kitamura, \&
  Watanabe}]{Nakayama1959}
Nakayama, T., Kitamura, M.~Y., \& Watanabe, K. 1959, JChPh, 30, 1180

\bibitem[{Nakayama \& Watanabe(1964)}]{Nakayama1964}
Nakayama, T., \& Watanabe, K. 1964, JChPh, 40, 558

\bibitem[{Natarajan \& Bhaskaran(1981)}]{1981NAT/BHA}
Natarajan, K., \& Bhaskaran, K. 1981, Experimental and analytical
  investigation of high temperature ignition of ethanol, Tech rep, Indian
  Inst of Tech, Madras Dept of Mechanical Engineering

\bibitem[{Natarajan {et~al.}(1986)Natarajan, Thielen, Hermanns, \&
  Roth}]{1986NAT/THI}
Natarajan, K., Thielen, K., Hermanns, H., \& Roth, P. 1986, Berich Bunsen Gesell, 90, 533

\bibitem[{Navarro-Gonz\'{a}lez {et~al.}(2001)}]{Navarro2001}
Navarro-Gonz\'{a}lez, R., Villagr\'{a}n-Muniz, M., Sobral, H., Molina, L.~T.,
\& Molina, M.~J. 2001, GRL, 28, 3867

\bibitem[{Nee \& Lee(1984)}]{Nee1984}
Nee, J.~B., \& Lee, L. 1984, JChPh, 81, 31

\bibitem[{Neilson {et~al.}(1978)Neilson, Bowers, Chau, Davidson, \&
  Aue}]{4-287}
Neilson, P.~V., Bowers, M.~T., Chau, M., Davidson, W.~R., \& Aue, D.~H. 1978,
  JAChS, 100, 3649

\bibitem[{Neiman \& Feklisov(1961)}]{1961NEI/FEK}
Neiman, M., \& Feklisov, G. 1961, Zh Fiz Khim, 35, 1064

\bibitem[{Nissen(2013)}]{Nissen2013}
Nissen, P.~E. 2013, A\&A, 552, 10

\bibitem[{Nguyen {et~al.}(2004)Nguyen, Zhang, Peeters, Truong, \&
  Nguyen}]{2004NGU/ZHA}
Nguyen, H. M.~T., Zhang, S., Peeters, J., Truong, T.~N., \& Nguyen, M.~T. 2004,
  CPL, 388, 94

\bibitem[{Nielsen {et~al.}(1991)Nielsen, Sidebottom, Donlon, \&
  Treacy}]{1991NIE/SID}
Nielsen, O.~J., Sidebottom, H.~W., Donlon, M., \& Treacy, J. 1991, CPL, 178, 163

\bibitem[{Nizamov \& Dagdigian(2003)}]{2003NIZ/DAG}
Nizamov, B., \& Dagdigian, P.~J. 2003, JCPA,
  107, 2256

\bibitem[{Oehlschlaeger {et~al.}(2004)Oehlschlaeger, Davidson, \&
  Hanson}]{2004OEH/DAV}
Oehlschlaeger, M.~A., Davidson, D.~F., \& Hanson, R.~K. 2004, JPCA, 108, 4247

\bibitem[{Ohmori {et~al.}(1990)Ohmori, Miyoshi, Matsui, \&
  Washida}]{1990OHM/MIY}
Ohmori, K., Miyoshi, A., Matsui, H., \& Washida, N. 1990, JPhCh, 94, 3253

\bibitem[{Okabe(1970)}]{Okabe1970}
Okabe, H. 1970, JChPh, 53, 3507

\bibitem[{Okabe(1980)}]{Okabe1980}
---. 1980, JChPh, 72, 6642

\bibitem[{Okabe(1981)}]{Okabe1981}
---. 1981, JChPh, 75, 2772

\bibitem[{Okabe(1983)}]{Okabe1983}
---. 1983, JChPh, 78, 1312

\bibitem[{Okabe \& Becker(1963)}]{Okabe1963}
Okabe, H., \& Becker, D. 1963, JChPh, 39, 2549

\bibitem[{Okabe {et~al.}(1978)}]{Okabe1978}
Okabe, H., {et~al.} 1978, Photochemistry of Small Molecules, Vol. 431 (New York: Wiley)

\bibitem[{Okada {et~al.}(1972)Okada, Matsumoto, Dohmaru, Taniguchi, \&
  Hayakawa}]{0-149}
Okada, S., Matsumoto, A., Dohmaru, T., Taniguchi, S., \& Hayakawa, T. 1972,
  Mass Spectroscopy (Japan), 20, 311

\bibitem[{O'Neal \& Benson(1962)}]{1962ONE/BEN}
O'Neal, E., \& Benson, S.~W. 1962, JChPh, 36, 2196

\bibitem[{Opansky \& Leone(1996a)}]{Opansky1996a}
Opansky, B.~J., \& Leone, S.~R. 1996a, JPhCh, 100, 4888

\bibitem[{Opansky \& Leone(1996b)}]{Opansky1996b}
Opansky, B.~J., \& Leone, S.~R. 1996b, JPhCh, 100, 19904

\bibitem[{Oparin (1957)}]{Oparin1957}
Oparin, A.~I. 1957, The Origin of Life on the Earth.

\bibitem[{{Orville}(1968)}]{Orville1968}
{Orville}, R.~E. 1968, JAtS, 25, 852

\bibitem[{Osborn(2003)}]{2003OSB}
Osborn, D.~L. 2003, JCPA, 107, 3728

\bibitem[{Owens {et~al.}(1985)Owens, Hales, Filkin, Miller, Steed, \&
  Jesson}]{Owens1985}
Owens, A., Hales, C., Filkin, D., {et~al.} 1985, JGR: Atmospheres, 90, 2283

\bibitem[{Padial {et~al.}(1985)Padial, Collins, \& Schneider}]{Padial1985}
Padial, N., Collins, L., \& Schneider, B. 1985, ApJ, 298, 369

\bibitem[{Pang {et~al.}(2008)Pang, Xie, Zhang, Ding, \& Tang}]{2008PAN/XIE}
Pang, J.-L., Xie, H.-B., Zhang, S.-W., Ding, Y.-H., \& Tang, A.-Q. 2008, JCPA, 112, 5251

\bibitem[{Paraskevopoulos \& Winkler(1967)}]{1967PAR/WIN}
Paraskevopoulos, G., \& Winkler, C.~A. 1967, JPhCh,
  71, 947

\bibitem[{Park \& Hershberger(1993)}]{1993PAR/HER}
Park, J., \& Hershberger, J.~F. 1993, JChPh, 99, 3488

\bibitem[{Park \& Lin(1997)}]{1997PAR/LIN}
Park, J., \& Lin, M. 1997, JCPA, 101, 5

\bibitem[{Parkes(1972{\natexlab{a}})}]{0-177}
Parkes, D.~A. 1972{\natexlab{a}}, FaTr I, 68, 613

\bibitem[{Parkes(1972{\natexlab{b}})}]{0-119}
---. 1972{\natexlab{b}}, FaTr I, 68, 627

\bibitem[{{Parra-Rojas} {et~al.}(2013){Parra-Rojas}, {Luque}, \&
  {Gordillo-V{\'a}Zquez}}]{Parra2013}
{Parra-Rojas}, F.~C., {Luque}, A., \& {Gordillo-V{\'a}Zquez}, F.~J. 2013,
JGR:Space Physics, 118, 5190

\bibitem[{Patel {et~al.}(2015)Patel, Percivalle, Ritson, Duffy, \&
  Sutherland}]{Patel2015}
Patel, B.~H., Percivalle, C., Ritson, D.~J., Duffy, C.~D., \& Sutherland, J.~D.
  2015, NatCh, 7, 301

\bibitem[{Patrick \& Golden(1984)}]{1984PAT/GOL}
Patrick, R., \& Golden, D.~M. 1984, Int J Chem Kin,
  16, 1567

\bibitem[{Patterson \& Greene(1962)}]{1962PAT/GRE}
Patterson, W., \& Greene, E. 1962, JChPh, 36, 1146

\bibitem[{Payzant {et~al.}(1976)Payzant, Tanaka, Betowski, \& Bohme}]{2-6023}
Payzant, J., Tanaka, K., Betowski, L., \& Bohme, D. 1976, JAChS, 98, 894

\bibitem[{Peeters {et~al.}(1995)Peeters, Boullart, \& Devriendt}]{1995PEE/BOU}
Peeters, J., Boullart, W., \& Devriendt, K. 1995, JPhCh, 99, 3583

\bibitem[{Petrov {et~al.}(2009)Petrov, Turetskii, \& Bulgakov}]{2009PET/TUR}
Petrov, Y.~P., Turetskii, S., \& Bulgakov, A. 2009, Kin Catal, 50,
  344

\bibitem[{Petty {et~al.}(1993)Petty, Harrison, \& Moore}]{1993PET/HAR}
Petty, J.~T., Harrison, J.~A., \& Moore, C.~B. 1993, JPhCh, 97, 11194

\bibitem[{Phillips {et~al.}(1977)Phillips, Lee, \& Judge}]{Phillips1977}
Phillips, E., Lee, L., \& Judge, D. 1977, JQSRT, 18, 309

\bibitem[{Pitts {et~al.}(1982)}]{Pitts1982}
Pitts, W.~M., Pasternack, L. \& McDonald, J.~R. 1982, CP, 68, 417

\bibitem[{Porter \& Noyes(1959)}]{Porter1959}
Porter, R.~P., \& Noyes, W.~A. 1959, JAChS, 81, 2307

\bibitem[{Pouilly {et~al.}(1983)Pouilly, Robbe, Schamps, \&
  Roueff}]{Pouilly1983}
Pouilly, B., Robbe, J., Schamps, J., \& Roueff, E. 1983, JPhB, 16, 437

\bibitem[{Powner {et~al.}(2009)}]{Powner2009}
Powner, M.~W., Gerland, B, \& Sutherland, J.~D. 2009, Nature, 459.7244, 239

\bibitem[{Prasad \& Huntress(1980)}]{Prasad1980}
Prasad, S.~S., \& Huntress, W.~T. 1980, ApJ, 43, 1

\bibitem[{{Price} {et~al.}(1997){Price}, {Penner}, \& {Prather}}]{Price1997}
{Price}, C., {Penner}, J., \& {Prather}, M. 1997, \jgr, 102, 5929

\bibitem[{Pshezhetskii {et~al.}(1959)Pshezhetskii, Morozov, Kamenetskaya,
  Siryatskaya, \& Gribova}]{1959PSH/MOR}
Pshezhetskii, S.~Y., Morozov, N., Kamenetskaya, S., Siryatskaya, V., \&
  Gribova, E. 1959, Russ J Phys Chem, 33, 402

\bibitem[{Quandt \& Hershberger(1995)}]{1995QUA/HER}
Quandt, R.~W., \& Hershberger, J.~F. 1995, JPhCh,
  99, 16939

\bibitem[{{Queloz} {et~al.}(2000){Queloz}, {Eggenberger}, {Mayor}, {Perrier},
  {Beuzit}, {Naef}, {Sivan}, \& {Udry}}]{Queloz2000}
{Queloz}, D., {Eggenberger}, A., {Mayor}, M., {et~al.} 2000, \aap, 359, L13

\bibitem[{Raksit(1982)}]{4-147}
Raksit, A.~B. 1982, IJMIP, 41, 185

\bibitem[{Raksit(1986)}]{4-372}
---. 1986, IJMSI, 69, 45

\bibitem[{Raksit \& Bohme(1984)}]{4-233}
Raksit, A.~B., \& Bohme, D.~K. 1984, IJMSI, 57, 211

\bibitem[{Raksit \& Bohme(1985)}]{4-234}
---. 1985, IJMSI, 63, 217

\bibitem[{Raksit {et~al.}(1984)Raksit, Schiff, \& Bohme}]{4-232}
Raksit, A.~B., Schiff, H., \& Bohme, D. 1984, IJMSI, 56, 321

\bibitem[{Raksit \& Warneck(1979)}]{4-176}
Raksit, A.~B., \& Warneck, P. 1979, ZNatA, 34,
  1410

\bibitem[{Raksit \& Warneck(1980{\natexlab{a}})}]{4-034}
---. 1980{\natexlab{a}}, JChPh, 73, 2673

\bibitem[{Raksit \& Warneck(1980{\natexlab{b}})}]{4-125}
---. 1980{\natexlab{b}}, IJMIP, 35, 23

\bibitem[{Raksit \& Warneck(1980{\natexlab{c}})}]{4-177}
---. 1980{\natexlab{c}}, FaTr II, 76, 1084

\bibitem[{Raksit \& Warneck(1981)}]{4-137}
---. 1981, JChPh, 74, 2853

\bibitem[{Ray {et~al.}(1996)Ray, Da{\"e}le, Vassalli, Poulet, \&
  Le~Bras}]{1996RAY/DAE}
Ray, A., Da{\"e}le, V., Vassalli, I., Poulet, G., \& Le~Bras, G. 1996, JPhCh, 100, 5737

\bibitem[{Reitel'boim {et~al.}(1978)Reitel'boim, Romanovich, \&
  Vedeneev}]{1978REI/ROM}
Reitel'boim, M., Romanovich, L., \& Vedeneev, B. 1978, Kin Catal, 19, 1131

\bibitem[{Ribas {et~al.}(2005)}]{Ribas2005}
Ribas, I., Guinan, E.~F., G\"{u}del, M., \& Audard, M. 2005, \apj, 622, 680

\bibitem[{Ribas {et~al.}(2010)}]{Ribas2010}
Ribas, I., Porto de Mello, G.~F., Ferreira, L.~D., H\'{e}brard, E., Selsis, F., et~al. 2010, \apj 714, 384

\bibitem[{Rim \& Hershberger(1999)}]{1999RIM/HER}
Rim, K.~T., \& Hershberger, J.~F. 1999, JCPA,
  103, 3721

\bibitem[{{Rimmer} {et~al.}(2012){Rimmer}, {Herbst}, {Morata}, \&
  {Roueff}}]{Rimmer2012}
{Rimmer}, P.~B., {Herbst}, E., {Morata}, O., \& {Roueff}, E. 2012, \aap, 537,

\bibitem[{{Rimmer} \& {Helling}(2013)}]{Rimmer2013}
{Rimmer}, P.~B., \& {Helling}, Ch. 2013, \apj, 774, 108

\bibitem[{{Rimmer} {et~al.}(2014){Rimmer}, {Helling}, \& {Bilger}}]{Rimmer2014}
{Rimmer}, P.~B., {Helling}, Ch., \& {Bilger}, C. 2014, IJAsB, 13, 173

\bibitem[{Robertson {et~al.}(1983)Robertson, Hils, Chatham, \&
  Gallagher}]{4-079}
Robertson, R., Hils, D., Chatham, H., \& Gallagher, A. 1983, ApPhL, 43, 544

\bibitem[{Roble \& Ridley(1994)}]{Roble1994}
Roble, R., \& Ridley, E. 1994, GeoRL, 21, 417

\bibitem[{Roche {et~al.}(1971)Roche, Sutton, Bohme, \& Schiff}]{2-6005}
Roche, A., Sutton, M., Bohme, D., \& Schiff, H. 1971, JChPh, 55, 5480

\bibitem[{Rogers(1990)}]{Rogers1990}
Rogers, J.~D. 1990, JPhCh, 94, 4011

\bibitem[{R{\"o}hrig {et~al.}(1994)R{\"o}hrig, R{\"o}mming, \&
  Wagner}]{1994ROH/ROM}
R{\"o}hrig, M., R{\"o}mming, H.-J., \& Wagner, H.~G. 1994, Berich Bunsen Gesell, 98, 1332

\bibitem[{R{\"o}hrig \& Wagner(1994)}]{1994ROH/WAG}
R{\"o}hrig, M., \& Wagner, H.~G. 1994 in International Symposium on Combustion (Elsevier), 
975

\bibitem[{Romani {et~al.} (2008)}]{Romani2008}
Romani, P.~N., Jennings, D.~E., Bjoraker, P.~V., {et~al.} 2008, Icarus, 198, 420

\bibitem[{Roose {et~al.}(1978)Roose, Hanson, \& Kruger}]{1978ROO/HAN}
Roose, T., Hanson, R., \& Kruger, C. 1978, in 11th International Symposium on
  Shock Tubes and Waves, 245

\bibitem[{Roscoe \& Roscoe(1973)}]{1973ROS/ROS}
Roscoe, J.~M., \& Roscoe, S.~G. 1973, CaJCh, 51, 3671

\bibitem[{Ross {et~al.}(1997)Ross, Sutherland, Kuo, \& Klemm}]{1997ROS/SUT}
Ross, S.~K., Sutherland, J.~W., Kuo, S.-C., \& Klemm, R.~B. 1997, JPCA, 101, 1104

\bibitem[{{Rottman} {et~al.}(2006){Rottman}, {Woods}, \&
  {McClintock}}]{Rottman2006}
{Rottman}, G.~J., {Woods}, T.~N., \& {McClintock}, W. 2006, AdSpR, 37, 201

\bibitem[{Rowe {et~al.}(1980)Rowe, Fahey, Fehsenfeld, \& Albritton}]{4-133}
Rowe, B., Fahey, D., Fehsenfeld, F., \& Albritton, D. 1980, JChPh, 73, 194

\bibitem[{Rowe {et~al.}(1981)Rowe, Fahey, Ferguson, \& Fehsenfeld}]{4-068}
Rowe, B., Fahey, D., Ferguson, E., \& Fehsenfeld, F. 1981, JChPh, 75, 3325

\bibitem[{{Rustgi}(1964)}]{Rustgi1964}
{Rustgi}, O.~P. 1964, JOSA, 54, 464

\bibitem[{Saeys {et~al.}(2006)Saeys, Reyniers, Van~Speybroeck, Waroquier, \&
  Marin}]{2006SAE/REY}
Saeys, M., Reyniers, M.-F., Van~Speybroeck, V., Waroquier, M., \& Marin, G.~B.
  2006, CPPC, 7, 188

\bibitem[{Safrany \& Jaster(1968)}]{1968SAF/JAS}
Safrany, D.~R., \& Jaster, W. 1968, JPhCh, 72, 3305

\bibitem[{Sahetchian {et~al.}(1987)Sahetchian, Heiss, \& Rigny}]{1987SAH/HEI}
Sahetchian, K., Heiss, A., \& Rigny, R. 1987, JPhCh,
  91, 2382

\bibitem[{Saito {et~al.}(1984)Saito, Kakumoto, \& Murakami}]{1984SAI/KAK}
Saito, K., Kakumoto, T., \& Murakami, I. 1984, CPL, 110,
  478

\bibitem[{Saito {et~al.}(1990)Saito, Mochizuki, Yoshinobu, \&
  Imamura}]{1990SAI/MOC}
Saito, K., Mochizuki, Y., Yoshinobu, I., \& Imamura, A. 1990, CPL, 167, 347

\bibitem[{Salahub \& Sandorfy(1971)}]{Salahub1971}
Salahub, D., \& Sandorfy, C. 1971, CPL, 8, 71

\bibitem[{Samson \& Cairns(1964)}]{Samson1964}
Samson, J.~A., \& Cairns, R. 1964, JGR, 69, 4583

\bibitem[{Samson \& Cairns(1965)}]{Samson1965}
---. 1965, JOSA, 55, 1035

\bibitem[{Sander {et~al.}(2005)Sander, Kerkweg, J{\"o}ckel, \&
  Lelieveld}]{Sander2005}
Sander, R., Kerkweg, A., J{\"o}ckel, P., \& Lelieveld, J. 2005, ACP, 5, 445

\bibitem[{{Sander} {et~al.}(2011){Sander}, {Friedl}, {Barker}, {Golden},
  {Kurylo}, {Wine}, {Abbatt}, {Burkholder}, {Kolb}, \& {Moortgat}}]{Sander2011}
{Sander}, S.~P., {Friedl}, R.~R., {Barker}, J.~R., {et~al.} 2011, Chemical Kinetics
and Photochemical Data for use in Atmospheric Studies 17, (JPL Publications)

\bibitem[{Sanders {et~al.}(1980)Sanders, Butler, Pasternack, \&
  McDonald}]{1980SAN/BUT}
Sanders, N., Butler, J., Pasternack, L., \& McDonald, J. 1980, CP, 48, 203

\bibitem[{Sanders {et~al.}(1987)Sanders, Lin, \& Lin}]{1987SAN/LIN}
Sanders, W., Lin, C., \& Lin, M. 1987, CST, 51, 103

\bibitem[{Sato \& Hidaka(2000)}]{2000SAT/HID}
Sato, K., \& Hidaka, Y. 2000, CoFl, 122, 291

\bibitem[{Saxon {et~al.}(1983)Saxon, Lengsfield~III, \& Liu}]{Saxon1983}
Saxon, R.~P., Lengsfield~III, B.~H., \& Liu, B. 1983, JChPh, 78, 312

\bibitem[{Sayah {et~al.}(1988)Sayah, Li, Caballero, \& Jackson}]{1988SAY/LI}
Sayah, N., Li, X., Caballero, J., \& Jackson, W.~M. 1988, J Photoch Photobio A, 45, 177

\bibitem[{Scattergood {et~al.}(1989)Scattergood, McKay, Borucki, Giver, van
  Ghyseghem, Parris, \& Miller}]{Scattergood1989}
Scattergood, T.~W., McKay, C.~P., Borucki, W.~J., {et~al.} 1989, Icarus, 81,
  413

\bibitem[{Schacke {et~al.}(1974)Schacke, Schmatjko, \& Wolfrum}]{1974SCH/SCH}
Schacke, H., Schmatjko, K., \& Wolfrum, J. 1974, Arch Procesow Spalania, 5

\bibitem[{Scherzer {et~al.}(1987)Scherzer, L{\"o}ser, \& Stiller}]{1987SCH/LOS}
Scherzer, K., L{\"o}ser, U., \& Stiller, W. 1987, ZCh, 27, 300

\bibitem[{Schiff \& Bohme(1979)}]{4-466}
Schiff, H., \& Bohme, D.~K. 1979, ApJ, 232, 740

\bibitem[{Schildcrout \& Franklin(1970)}]{1-024}
Schildcrout, S.~M., \& Franklin, J. 1970, JAChS, 92, 251

\bibitem[{Schlesinger \& Miller(1983)}]{Schlesinger1983}
Schlesinger, G., \& Miller, S.~L. 1983, JMolE, 19, 376

\bibitem[{Schoen(1962)}]{Schoen1962}
Schoen, R.~I. 1962, JChPh, 37, 2032

\bibitem[{Schulz {et~al.}(1985)Schulz, Klotz, \& Spangenberg}]{1985SCH/KLO}
Schulz, G., Klotz, H.-D., \& Spangenberg, H.-J. 1985, ZCh, 25, 88

\bibitem[{Schurath {et~al.}(1969)Schurath, Tiedemann, \&
  Schindler}]{Schurath1969}
Schurath, U., Tiedemann, P., \& Schindler, R.~N. 1969, JPhCh, 73, 456

\bibitem[{{Schwarz} {et~al.}(2015){Schwarz}, {Brogi}, {de Kok}, {Birkby}, \&
  {Snellen}}]{Schwarz2015}
{Schwarz}, H., {Brogi}, M., {de Kok}, R., {Birkby}, J., \& {Snellen}, I. 2015,
arXiv:1502.04713

\bibitem[{Seager {et~al.}(2013{\natexlab{a}})Seager, Bains, \&
  Hu}]{Seager2013a}
Seager, S., Bains, W., \& Hu, R. 2013{\natexlab{a}}, ApJ, 775, 104

\bibitem[{Seager {et~al.}(2013{\natexlab{b}})Seager, Bains, \&
  Hu}]{Seager2013b}
---. 2013{\natexlab{b}}, ApJ, 777, 95

\bibitem[{Seery(1969)}]{1969SEE}
Seery, D. 1969 in the International Symposium on Combustion, 12

\bibitem[{Seetula {et~al.}(1986)Seetula, Blomqvist, Kalliorinne, \&
  Koskikallio}]{1986SEE/BLO}
Seetula, J., Blomqvist, K., Kalliorinne, K., \& Koskikallio, J. 1986, Acta
  Chem Scand, 40, 653

\bibitem[{Seinfeld \& Pandis(2006)}]{Seinfeld2006}
Seinfeld, J.~H., \& Pandis, S.~N. 2006, Atmospheric Chemistry and Physics: From Air Pollution to Climate Change (2nd ed.; New Jersey: Wiley)

\bibitem[{Selwyn {et~al.}(1977)Selwyn, Podolske, \& Johnston}]{Selwyn1977}
Selwyn, G., Podolske, J., \& Johnston, H.~S. 1977, GeoRL, 4, 427

\bibitem[{Sen {et~al.}(1998)}]{Sen1998}
{Sen}, B., {Toon}, G.~C., {Osterman}, G.~B. 1998, JGR, 103, 3571

\bibitem[{Senosiain {et~al.}(2006)Senosiain, Klippenstein, \&
  Miller}]{2006SEN/KLI}
Senosiain, J.~P., Klippenstein, S.~J., \& Miller, J.~A. 2006, JPCA, 110, 5772

\bibitem[{Setser \& Rabinovitch(1962)}]{1962SET/RAB}
Setser, D., \& Rabinovitch, B. 1962, CaJCh, 40, 1425

\bibitem[{{Shardanand} \& {Rao}(1977)}]{Shardanand1977}
{Shardanand}, \& {Rao}, A.~D.~P. 1977, \jqsrt, 17, 433

\bibitem[{Shaw(1977)}]{1977SHA}
Shaw, R. 1977, Int J Chem Kin, 9, 929

\bibitem[{Shaw(1978)}]{1978SHA}
---. 1978, JPCRD, 7, 1179

\bibitem[{Sheng {et~al.}(2002)Sheng, Bozzelli, Dean, \& Chang}]{2002SHE/BOZ}
Sheng, C.~Y., Bozzelli, J.~W., Dean, A.~M., \& Chang, A.~Y. 2002, JPCA, 106, 7276

\bibitem[{{Showman} {et~al.}(2008){Showman}, {Cooper}, {Fortney}, \&
  {Marley}}]{Showman2008}
{Showman}, A.~P., {Cooper}, C.~S., {Fortney}, J.~J., \& {Marley}, M.~S. 2008,
  \apj, 682, 559

\bibitem[{Shul {et~al.}(1987{\natexlab{a}})Shul, Upschulte, Passarella, Keesee,
  \& Castleman}]{4-454}
Shul, R., Upschulte, B., Passarella, R., Keesee, R., \& Castleman, A.
  1987{\natexlab{a}}, JPhCh, 91, 2556

\bibitem[{Shul {et~al.}(1987{\natexlab{b}})Shul, Upschulte, Passarella, Keesee,
  \& Castleman}]{4-476}
---. 1987{\natexlab{b}}, JPhCh, 91, 2556

\bibitem[{Sieck(1978)}]{4-004}
Sieck, L.~W. 1978, Int J Chem Kin, 10, 335

\bibitem[{Sieck \& Futrell(1968)}]{1-075}
Sieck, L.~W., \& Futrell, J. 1968, JChPh, 48, 1409

\bibitem[{Sieck \& Searles(1970)}]{1-158}
Sieck, L.~W., \& Searles, S.~K. 1970, JChPh, 53, 2601

\bibitem[{Sims {et~al.}(1993)Sims, Queffelec, Travers, Rowe, Herbert,
  Karth{\"a}user, \& Smith}]{1993SIM/QUE}
Sims, I.~R., Queffelec, J.-L., Travers, D., {et~al.} 1993, CPL, 211, 461

\bibitem[{Sivaramakrishnan {et~al.}(2009)Sivaramakrishnan, Michael, \&
  Klippenstein}]{2010SIV/MIC}
Sivaramakrishnan, R., Michael, J., \& Klippenstein, S. 2009, JPCA, 114, 755

\bibitem[{Skinner \& Ruehrwein(1959)}]{1959SKI/RUE}
Skinner, G.~B., \& Ruehrwein, R.~A. 1959, JPhCh,
  63, 1736

\bibitem[{Slack \& Fishburne(1970)}]{1970SLA/FIS}
Slack, M., \& Fishburne, E. 1970, JChPh, 52, 5830

\bibitem[{Slanger \& Black(1982)}]{Slanger1982}
Slanger, T.~G., \& Black, G. 1982, JChPh, 77, 2432

\bibitem[{Smith \& Adams(1977{\natexlab{a}})}]{2-2006}
Smith, D.~L., \& Adams, N. 1977{\natexlab{a}}, ApJ, 217, 741

\bibitem[{Smith \& Adams(1977{\natexlab{b}})}]{2-2003}
---. 1977{\natexlab{b}}, CPL, 47, 145

\bibitem[{Smith \& Adams(1978)}]{2-2009}
---. 1978, CPL, 54, 535

\bibitem[{Smith \& Adams(1980)}]{4-050}
---. 1980, CPL, 76, 418

\bibitem[{Smith \& Adams(1981)}]{4-183}
---. 1981, MNRAS, 197, 377

\bibitem[{Smith {et~al.}(1982)Smith, Adams, \& Alge}]{4-074}
Smith, D.~L., Adams, N.~G., \& Alge, E. 1982, JChPh, 77,
  1261

\bibitem[{Smith {et~al.}(1978)Smith, Adams, \& Miller}]{2-2011}
Smith, D.~L., Adams, N.~G., \& Miller, T. 1978, JChPh, 69,
  308

\bibitem[{Smith {et~al.}(1976)Smith, Smith, \& Futrell}]{0-034}
Smith, R.~D., Smith, D.~L., \& Futrell, J.~H. 1976, IJMIP, 19, 369

\bibitem[{Song {et~al.}(2003)Song, Golden, Hanson, Bowman, Senosiain, Musgrave,
  \& Friedrichs}]{2003SON/GOL}
Song, S., Golden, D.~M., Hanson, R.~K., {et~al.} 2003, Int J Chem Kin, 35, 304

\bibitem[{Song {et~al.}(2005)Song, Hou, \& Wang}]{2005SON/HOU}
Song, X., Hou, H., \& Wang, B. 2005, PCCP, 7,
  3980

\bibitem[{Spokes \& Benson(1967)}]{1967SPO/BEN}
Spokes, G.~N., \& Benson, S.~W. 1967, JAChS,
  89, 6030

\bibitem[{Sridharan \& Kaufman(1983)}]{1983SRI/KAU}
Sridharan, U., \& Kaufman, F. 1983, CPL, 102, 45

\bibitem[{Srinivasan {et~al.}(2007)Srinivasan, Su, \& Michael}]{2007SRI/SU}
Srinivasan, N., Su, M.-C., \& Michael, J. 2007, JPCA, 111, 3951

\bibitem[{Srinivasan {et~al.}(2005)Srinivasan, Su, Sutherland, \&
  Michael}]{2005SRI/SU}
Srinivasan, N., Su, M.-C., Sutherland, J., \& Michael, J. 2005, JPCA, 109, 1857

\bibitem[{Stevenson \& Schissler(1955)}]{0-065}
Stevenson, D.~P., \& Schissler, D.~O. 1955, JChPh, 23, 1353

\bibitem[{Stief {et~al.}(1972)Stief, Donn, Glicker, Gentieu, \&
  Mentall}]{Stief1972}
Stief, L., Donn, B., Glicker, S., Gentieu, E., \& Mentall, J. 1972, ApJ, 171, 21

\bibitem[{Stockwell \& Calvert(1978)}]{Stockwell1978}
Stockwell, W.~R., \& Calvert, J.~G. 1978, J Photochem, 8, 193

\bibitem[{Stothard {et~al.}(1995)Stothard, Humpfer, \& Grotheer}]{1995STO/HUM}
Stothard, N., Humpfer, R., \& Grotheer, H.-H. 1995, CPL,
  240, 474

\bibitem[{Strausz {et~al.}(1970)Strausz, Duholke, \& Gunning}]{0-090}
Strausz, O.~P., Duholke, W., \& Gunning, H.~E. 1970, JAChS, 92, 4128

\bibitem[{Streit(1982)}]{4-327}
Streit, G.~E. 1982, JPhCh, 86, 2321

\bibitem[{Striebel {et~al.}(2004)Striebel, Jusinski, Fahr, Halpern,
  Klippenstein, \& Taatjes}]{2004STR/JUS}
Striebel, F., Jusinski, L.~E., Fahr, A., {et~al.} 2004, PCCP, 6, 2216

\bibitem[{Strobel(1983)}]{Strobel1983}
Strobel, D.~F. 1983, IRPC, 3, 145

\bibitem[{Su {et~al.}(2002)Su, Kumaran, Lim, Michael, Wagner, Harding, \&
  Fang}]{2002SU/KUM}
Su, M.-C., Kumaran, S., Lim, K., {et~al.} 2002, JPCA, 106, 8261

\bibitem[{Sumathi \& Nguyen(1998)}]{1998SUM/NGU}
Sumathi, R., \& Nguyen, M.~T. 1998, JCPA, 102,
  8013

\bibitem[{Sumathi \& Peyerimhoff(1996)}]{1996SUM/PEY}
Sumathi, R., \& Peyerimhoff, S. 1996, CPL, 263, 742

\bibitem[{Sun {et~al.}(2001)Sun, DeSain, Scott, Hung, Thompson, Glass, \&
  Curl}]{2001SUN/DES}
Sun, F., DeSain, J., Scott, G., {et~al.} 2001, JPCA, 105, 6121

\bibitem[{Sun {et~al.}(2006)Sun, He, Hong, Chang, An, \& Wang}]{2006SUN/HE}
Sun, H., He, H.-Q., Hong, B., {et~al.} 2006, IJQC, 106, 894

\bibitem[{Sun \& Weissler(1955)}]{Sun1955}
Sun, H., \& Weissler, G. 1955, JChPh, 23, 1160

\bibitem[{Sun {et~al.}(2004)Sun, Zhang, Ai, Zhang, \& Sun}]{2004SUN/ZHA}
Sun, Y., Zhang, Q.-y., Ai, X.-c., Zhang, J.-p., \& Sun, C.-c. 2004, JMoSt, 686, 123

\bibitem[{Suzaki {et~al.}(2006)Suzaki, Kanno, Tonokura, Koshi, Tsuchiya, \&
  Tezaki}]{2006SUZ/KAN}
Suzaki, K., Kanno, N., Tonokura, K., {et~al.} 2006, CPL,
  425, 179

\bibitem[{Suzaki {et~al.}(2007)Suzaki, Tsuchiya, Koshi, \&
  Tezaki}]{2007SUZ/TSU}
Suzaki, K., Tsuchiya, K., Koshi, M., \& Tezaki, A. 2007, JPCA, 111, 3776

\bibitem[{{Swain} {et~al.}(2009){Swain}, {Tinetti}, {Vasisht}, {Deroo},
  {Griffith}, {Bouwman}, {Chen}, {Yung}, {Burrows}, {Brown}, {Matthews},
  {Rowe}, {Kuschnig}, \& {Angerhausen}}]{Swain2009}
{Swain}, M.~R., {Tinetti}, G., {Vasisht}, G., {et~al.} 2009, \apj, 704, 1616

\bibitem[{Szabo \& Derrick(1971)}]{1-034}
Szabo, I., \& Derrick, P. 1971, IJMIP, 7, 55

\bibitem[{Szekely {et~al.}(1985)Szekely, Hanson, \& Bowman}]{1984SZE/HAN}
Szekely, A., Hanson, R.~K., \& Bowman, C.~T. 1985 in International Symposium 
on Combustion (Elsevier), 647

\bibitem[{Tabayashi \& Bauer(1979)}]{1979TAB/BAU}
Tabayashi, K., \& Bauer, S. 1979, CoFl, 34, 63

\bibitem[{Takahashi {et~al.}(2007)Takahashi, Yamamoto, Inomata, \&
  Kogoma}]{2007TAK/YAM}
Takahashi, K., Yamamoto, O., Inomata, T., \& Kogoma, M. 2007, Int J Chem Kin, 39, 97

\bibitem[{Takahashi(1972)}]{1972TAK}
Takahashi, S. 1972, Mem Def Acad: Phys \& Chem Eng, 12

\bibitem[{Tanaka {et~al.}(1976)Tanaka, Betowksi, Mackay, \& Bohme}]{2-6026}
Tanaka, K., Betowksi, L., Mackay, G., \& Bohme, D. 1976, JCP, 65, 3203

\bibitem[{Tanaka {et~al.}(1953)Tanaka, Inn, \& Watanabe}]{Tanaka1953}
Tanaka, Y., Inn, E.~C., \& Watanabe, K. 1953, JChPh,
  21, 1651

\bibitem[{Tang {et~al.}(2008)Tang, Wang, \& Wang}]{2008TAN/WAN}
Tang, Y., Wang, R., \& Wang, B. 2008, JCPA, 112,
  5295

\bibitem[{Tanner {et~al.}(1979{\natexlab{b}})Tanner, Mackay, \& Bohme}]{4-219}
Tanner, S.~D., Mackay, G.~I., \& Bohme, D.~K. 1979{\natexlab{b}}, CaJCh, 57, 2350

\bibitem[{Tanner {et~al.}(1979{\natexlab{c}})Tanner, Mackay, \& Bohme}]{4-467}
---. 1979{\natexlab{c}}, CaJCh, 57, 2350

\bibitem[{Tanner {et~al.}(1981)Tanner, Mackay, \& Bohme}]{4-070}
---. 1981, CaJCh, 59, 1615

\bibitem[{Tanner {et~al.}(1979{\natexlab{a}})Tanner, Mackay, Hopkinson, \&
  Bohme}]{4-007}
Tanner, S.~D., Mackay, G.~I., Hopkinson, A., \& Bohme, D.~K. 1979{\natexlab{a}},
  IJMIP, 29, 153

\bibitem[{Tanzawa \& Gardiner(1980)}]{1980TAN/GAR}
Tanzawa, T., \& Gardiner, W. 1980, JPhCh, 84,
  236

\bibitem[{Tao {et~al.}(2001)Tao, Ding, Li, Huang, \& Sun}]{2001TAO/DIN}
Tao, Y.-g., Ding, Y.-h., Li, Z.-s., Huang, X.-r., \& Sun, C.-C. 2001, JCPA, 105, 9598

\bibitem[{{Taylor} {et~al.}(1979){Taylor}, {Scarf}, {Russell}, \&
  {Brace}}]{Taylor1979}
{Taylor}, W.~W.~L., {Scarf}, F.~L., {Russell}, C.~T., \& {Brace}, L.~H. 1979,
  \nat, 279, 614

\bibitem[{Thaxton {et~al.}(1997)Thaxton, Hsu, \& Lin}]{1997THA/HSU}
Thaxton, A.~G., Hsu, C.-C., \& Lin, M. 1997, Int J Chem Kin, 29, 245

\bibitem[{Theard \& Huntress(1974)}]{0-166}
Theard, L.~P., \& Huntress, W.~T. 1974, JChPh, 60, 2840

\bibitem[{Thielen \& Roth(1986)}]{1986THI/ROT}
Thielen, K., \& Roth, P. 1986, AIAA, 24, 1102

\bibitem[{Thomas {et~al.}(1978)Thomas, Barassin, \& Burke}]{4-108}
Thomas, R., Barassin, A., \& Burke, R. 1978, IJMSI, 28, 275

\bibitem[{Thompson {et~al.}(1963)Thompson, Harteck, \& Reeves}]{Thompson1963}
Thompson, B., Harteck, P., \& Reeves, R. 1963, JGR, 68, 6431

\bibitem[{Thweatt {et~al.}(2004)Thweatt, Erickson, \&
  Hershberger}]{2004THW/ERI}
Thweatt, W.~D., Erickson, M.~A., \& Hershberger, J.~F. 2004, JPCA, 108, 74

\bibitem[{Thynne \& Gray(1963{\natexlab{a}})}]{1963THY/GRA}
Thynne, J., \& Gray, P. 1963{\natexlab{a}}, TrFa, 59, 1149

\bibitem[{Thynne \& Gray(1963{\natexlab{b}})}]{1962THY/GRA}
---. 1963{\natexlab{b}}, TrFa, 59, 1149

\bibitem[{{Tian} {et~al.}(2005){Tian}, {Toon}, {Pavlov}, \& {De
  Sterck}}]{Tian2005}
{Tian}, F., {Toon}, O.~B., {Pavlov}, A.~A., \& {De Sterck}, H. 2005, Science,
  308, 1014

\bibitem[{Tich{\`y} {et~al.}(1979)Tich{\`y}, Raksit, Lister, Twiddy, Adams, \&
  Smith}]{4-008}
Tich{\`y}, M., Raksit, A., Lister, D., {et~al.} 1979, IJMIP, 29, 231

\bibitem[{Tomeczek \& Grado{\'n}(2003)}]{2003TOM/GRA}
Tomeczek, J., \& Grado{\'n}, B. 2003, CoFl, 133, 311

\bibitem[{Tonkyn \& Weisshaar(1986)}]{4-484}
Tonkyn, R., \& Weisshaar, J.~C. 1986, JPhCh, 90,
  2305

\bibitem[{{Toon} \& {Farlow}(1981)}]{Toon1981}
{Toon}, O.~B., \& {Farlow}, N.~H. 1981, AREPS, 9, 19

\bibitem[{{Toon} {et~al.}(1989){Toon}, {McKay}, {Ackerman}, \&
  {Santhanam}}]{Toon1989}
{Toon}, O.~B., {McKay}, C.~P., {Ackerman}, T.~P., \& {Santhanam}, K. 1989,
  \jgr, 94, 16287

\bibitem[{Trenwith(1960)}]{1960TRE}
Trenwith, A. 1960, J Chem Soc, 3722

\bibitem[{Troe(1983)}]{Troe1983}
Troe, J. 1983, Berich Bunsen Gesell, 87, 161

\bibitem[{Troe(2005)}]{2005TRO}
---. 2005, JCPA, 109, 8320

\bibitem[{Tsang(1987)}]{1987TSA}
Tsang, W. 1987, JPCRD, 16, 471

\bibitem[{Tsang(1992)}]{1992TSA}
---. 1992, JPCRD, 21, 753

\bibitem[{Tsang(2004)}]{2004TSA}
---. 2004, Int J Chem Kin, 36, 456

\bibitem[{Tsang \& Hampson(1986)}]{1986TSA/HAM}
Tsang, W., \& Hampson, R. 1986, JPCRD, 15, 1087

\bibitem[{Tsang \& Herron(1991)}]{1991TSA/HER}
Tsang, W., \& Herron, J.~T. 1991, JPCRD, 20, 609

\bibitem[{Tsuboi \& Hashimoto(1981)}]{1981TSU/HAS}
Tsuboi, T., \& Hashimoto, K. 1981, CoFl, 42, 61

\bibitem[{Tsuboi {et~al.}(1981)Tsuboi, Katoh, Kikuchi, \&
  Hashimoto}]{1981TSU/KAT}
Tsuboi, T., Katoh, M., Kikuchi, S., \& Hashimoto, K. 1981, JaJAP, 20, 985

\bibitem[{{Tu} {et~al.}(2015){Tu}, {Johnstone}, {G{\"u}del}, \&
  {Lammer}}]{Tu2015}
{Tu}, L., {Johnstone}, C.~P., {G{\"u}del}, M., \& {Lammer}, H. 2015, arXiv:1504.04546

\bibitem[{Tuazon {et~al.}(1984)Tuazon, Carter, Atkinson, Winer, \&
  Pitts}]{1984TUA/CAR}
Tuazon, E.~C., Carter, W.~P., Atkinson, R., Winer, A.~M., \& Pitts, J.~N. 1984,
  EnST, 18, 49

\bibitem[{Tzeng {et~al.}(2009)Tzeng, Chen, Wang, Lee, Xu, \& Lin}]{2009TZE/CHE}
Tzeng, S.-Y., Chen, P.-H., Wang, N.~S., {et~al.} 2009, JPCA, 113, 6314

\bibitem[{Vaghjiani(1995)}]{1995VAG}
Vaghjiani, G.~L. 1995, Int J Chem Kin, 27, 777

\bibitem[{Vakhtin {et~al.}(2001)}]{Vakhtin2001}
Vakhtin, A.~B., Hard, D.~E., Smith, I.~W.~M., \& Leone, S.~R. 2001, CPL, 344, 317

\bibitem[{Van~Dishoeck(1984)}]{Van1984}
van Dishoeck, E.~F. 1984, Dissertation

\bibitem[{van Dishoeck(1987)}]{Van1987}
van Dishoeck, E.~F. 1987, JChPh, 86, 196

\bibitem[{Vandooren {et~al.}(1994)Vandooren, Bian, \&
  Van~Tiggelen}]{1994VAN/BIA}
Vandooren, J., Bian, J., \& Van~Tiggelen, P. 1994, CoFl, 98,
  402

\bibitem[{Vandooren \& Van~Tiggelen(1977)}]{1977VAN/VAN}
Vandooren, J., \& Van~Tiggelen, P. 1977in , Elsevier, 1133--1144

\bibitem[{Vardanyan {et~al.}(1974)Vardanyan, Sachyan, Philiposyan, \&
  Nalbandyan}]{1974VAR/SAC}
Vardanyan, I., Sachyan, G., Philiposyan, A., \& Nalbandyan, A. 1974, CoFl, 22, 153

\bibitem[{{Velinov} \& {Mateev}(2008)}]{Velinov2008}
{Velinov}, P.~I.~Y., \& {Mateev}, L.~N. 2008, Journal of Atmospheric and
  Solar-Terrestrial Physics, 70, 574

\bibitem[{{Venot} {et~al.}(2012){Venot}, {H{\'e}brard}, {Ag{\'u}ndez},
  {Dobrijevic}, {Selsis}, {Hersant}, {Iro}, \& {Bounaceur}}]{Venot2012}
{Venot}, O., {H{\'e}brard}, E., {Ag{\'u}ndez}, M., {et~al.} 2012, \aap, 546,
  A43

\bibitem[{Verner {et~al.}(1996)Verner, Ferland, Korista, \&
  Yakovlev}]{Verner1996}
Verner, D., Ferland, G., Korista, K., \& Yakovlev, D. 1996, ApJ, 465, 487

\bibitem[{Verner \& Yakovlev(1995)}]{Verner1995}
Verner, D., \& Yakovlev, D. 1995, A\&AS, 109, 125

\bibitem[{Verner {et~al.}(1993)Verner, Yakovlev, Band, \&
  Trzhaskovskaya}]{Verner1993}
Verner, D., Yakovlev, D., Band, I., \& Trzhaskovskaya, M. 1993, ADNDT, 55, 233

\bibitem[{Veyret {et~al.}(1982)Veyret, Rayez, \& Lesclaux}]{1982VEY/RAY}
Veyret, B., Rayez, J.~C., \& Lesclaux, R. 1982, JPhCh, 86, 3424

\bibitem[{Viggiano {et~al.}(1980)Viggiano, Albritton, Fehsenfeld, Adams, Smith,
  \& Howorka}]{4-045}
Viggiano, A., Albritton, D., Fehsenfeld, F., {et~al.} 1980, ApJ, 236, 492

\bibitem[{Viggiano \& Paulson(1983)}]{4-203}
Viggiano, A., \& Paulson, J.~F. 1983, JChPh, 79, 2241

\bibitem[{Vijayan(1980)}]{Vijayan1980}
Vijayan, M. 1980, FEBS Lett, 112, 135

\bibitem[{Villinger {et~al.}(1982)Villinger, Futrell, Howorka, Duric, \&
  Lindinger}]{4-402}
Villinger, H., Futrell, J., Howorka, F., Duric, N., \& Lindinger, W. 1982,
  JChPh, 76, 3529

\bibitem[{{Visscher} \& {Moses}(2011)}]{Visscher2011}
{Visscher}, C., \& {Moses}, J.~I. 2011, \apj, 738, 72

\bibitem[{Visscher {et~al.}(2010)Visscher, Moses, \& Saslow}]{Visscher2010}
Visscher, C., Moses, J.~I., \& Saslow, S.~A. 2010, Icarus, 209, 602

\bibitem[{Vogt {et~al.}(1978)Vogt, Williamson, \& Beauchamp}]{4-166}
Vogt, J., Williamson, A.~D., \& Beauchamp, J. 1978, JAChS, 100, 3478

\bibitem[{Vuitton {et~al.}(2012)}]{Vuitton2012}
Vuitton, V., Yelle, R.~V., Lavvas, P., \& Klippenstein, S.~J. 2012, ApJ,
744, 11

\bibitem[{Wagner \& Bowman(1987)}]{1987WAG/BOW}
Wagner, A.~F., \& Bowman, J.~M. 1987, JPhCh, 91, 5314

\bibitem[{Wagner {et~al.}(1971)Wagner, Warnatz, \& Zetzsch}]{1971WAG/WAR}
Wagner, H.~G., Warnatz, J., \& Zetzsch, C. 1971, An Asoc Quim Argent, 59

\bibitem[{Wagner-Redeker {et~al.}(1985)Wagner-Redeker, Kemper, Jarrold, \&
  Bowers}]{4-313}
Wagner-Redeker, W., Kemper, P.~R., Jarrold, M.~F., \& Bowers, M.~T. 1985, 
JChPh, 83, 1121

\bibitem[{Waite {et~al.}(2007)}]{Waite2007}
{Waite}, J.~H., {Young}, D.~T., {Cravens}, T.~E., {Coates}, A.~J., 
{Crary}, F.~J., {et~al.} 2007, Science, 316.5826, 870

\bibitem[{Wakamatsu \& Hidaka(2008)}]{2008WAK/HID}
Wakamatsu, H., \& Hidaka, Y. 2008, Int J Chem Kin, 40, 320

\bibitem[{{Wakelam} {et~al.}(2012){Wakelam}, {Herbst}, {Loison}, {Smith},
  {Chandrasekaran}, {Pavone}, {Adams}, {Bacchus-Montabonel}, {Bergeat},
  {B{\'e}roff}, {Bierbaum}, {Chabot}, {Dalgarno}, {van Dishoeck}, {Faure},
  {Geppert}, {Gerlich}, {Galli}, {H{\'e}brard}, {Hersant}, {Hickson},
  {Honvault}, {Klippenstein}, {Le Picard}, {Nyman}, {Pernot}, {Schlemmer},
  {Selsis}, {Sims}, {Talbi}, {Tennyson}, {Troe}, {Wester}, \&
  {Wiesenfeld}}]{Wakelam2012}
{Wakelam}, V., {Herbst}, E., {Loison}, J.-C., {et~al.} 2012, \apjs, 199, 21

\bibitem[{Walker \& Kelly(1972)}]{Walker1972}
Walker, T., \& Kelly, H. 1972, CPL, 16, 511

\bibitem[{Wallington \& Japar(1989)}]{1989WAL/JAP}
Wallington, T.~J., \& Japar, S.~M. 1989, JAtC, 9,
  399

\bibitem[{{Walsh} \& {Millar}(2011)}]{Walsh2011}
{Walsh}, C., \& {Millar}, T.~J. 2011, in IAU Symposium, Vol. 280, IAU
  Symposium, ed. J.~{Cernicharo} \& R.~{Bachiller}, 56

\bibitem[{Wang {et~al.}(2002)Wang, Zhang, \& Li}]{2002WAN/ZHA}
Wang, C.~Y., Zhang, S., \& Li, Q.~S. 2002, Theor Chem Acc, 108,
  341

\bibitem[{{Wang} {et~al.}(2001){Wang}, {Cui}, {He}, {Fan}, \&
  {Wang}}]{2001SU/PIN}
{Wang}, S., {Cui}, J.-P., {He}, Y.-Z., {Fan}, B.-C., \& {Wang}, J. 2001,
  ChPhL, 18, 289

\bibitem[{Warnatz(1984)}]{1984WAR}
Warnatz, J. 1984, in Combustion Chemistry (Springer), 197

\bibitem[{Warneck(1972)}]{1-079}
Warneck, P. 1972, Berich Bunsen Gesell, 76, 421

\bibitem[{Watanabe(1954)}]{Watanabe1954}
Watanabe, K. 1954, JChPh, 22, 1564

\bibitem[{Watanabe \& Jursa(1964)}]{Watanabe1964}
Watanabe, K., \& Jursa, A. 1964, JChPh, 41, 1650

\bibitem[{Watanabe {et~al.}(1967)Watanabe, Matsunaga, \& Sakai}]{Watanabe1967}
Watanabe, K., Matsunaga, F.~M., \& Sakai, H. 1967, ApOpt, 6, 391

\bibitem[{Watanabe \& Sood(1965)}]{Watanabe1965}
Watanabe, K., \& Sood, S. 1965, Sci Light, 14, 36

\bibitem[{Watanabe \& Zelikoff(1953)}]{Watanabe1953}
Watanabe, K., \& Zelikoff, M. 1953, JOSA, 43, 753

\bibitem[{Weaver {et~al.}(1977)Weaver, Meagher, \& Heicklen}]{Weaver1977}
Weaver, J., Meagher, J., \& Heicklen, J. 1977, J Photochem, 6,
  111

\bibitem[{Westenberg \& De~Haas(1969)}]{1969WES/DEH}
Westenberg, A.~A., \& De~Haas, N. 1969, JPhCh, 73,
  1181

\bibitem[{Whyte \& Phillips(1983)}]{1983WHY/PHI}
Whyte, A., \& Phillips, L. 1983, CPL, 98, 590

\bibitem[{Wight \& Beauchamp(1980)}]{4-171}
Wight, C.~A., \& Beauchamp, J. 1980, JPhCh, 84,
  2503

\bibitem[{Wijnen(1960)}]{1960WIJ}
Wijnen, M. 1960, JAChS, 82, 3034

\bibitem[{Williamson \& Beauchamp(1975)}]{0-208}
Williamson, A.~D., \& Beauchamp, J. 1975, JAChS, 97, 5714

\bibitem[{Williamson \& Bayes(1967)}]{1967WIL/BAY}
Williamson, D.~G., \& Bayes, K.~D. 1967, JAChS, 89, 3390

\bibitem[{Wilson(1972)}]{1972WIL}
Wilson, W.~E. 1972, JPCRD, 1, 535

\bibitem[{Woods \& Haynes(1994)}]{1994WOO/HAY}
Woods, I., \& Haynes, B. 1994 in International Symposium on Combustion (Elsevier), 909

\bibitem[{Wooldridge {et~al.}(1996)Wooldridge, Hanson, \& Bowman}]{1996WOO/HAN}
Wooldridge, M.~S., Hanson, R.~K., \& Bowman, C.~T. 1996, Int J Chem Kin, 28, 361

\bibitem[{Wu {et~al.}(1990)Wu, Wang, Lin, \& Fifer}]{1990WU/WAN}
Wu, C., Wang, H., Lin, M., \& Fifer, R. 1990, JPhCh, 94, 3344

\bibitem[{Wu {et~al.}(2007)Wu, Lee, Xu, \& Lin}]{2007WU/LEE}
Wu, C.-W., Lee, Y.-P., Xu, S., \& Lin, M. 2007, JPCA, 111, 6693

\bibitem[{Wu {et~al.}(2003)Wu, Liu, Li, \& Sun}]{2003WU/LIU}
Wu, J.-Y., Liu, J.-Y., Li, Z.-S., \& Sun, C.-c. 2003, JChPh, 118, 10986

\bibitem[{Xu \& Lin(2007)}]{2007XU/LIN}
Xu, S., \& Lin, M. 2007, JCPA, 111, 6730

\bibitem[{Xu \& Lin(2004)}]{2004XU/LIN}
Xu, Z.-F., \& Lin, M. 2004, Int J Chem Kin, 36, 205

\bibitem[{Xu \& Sun(1999)}]{1999XU/SUN}
Xu, Z.-F., \& Sun, C.-C. 1999, JMoSt, 459, 37

\bibitem[{Xu \& Sun(1998)}]{1998XU/SUN}
Xu, Z.-F., \& Sun, J.-Z. 1998, JCPA, 102, 1194

\bibitem[{Yang {et~al.}(2008)Yang, Li, \& Zhang}]{2008YAN/LI}
Yang, J., Li, Q.~S., \& Zhang, S. 2008, JCoCh, 29, 247

\bibitem[{Yang {et~al.}(2009)Yang, Goldsmith, \& Tranter}]{2009YAN/GOL}
Yang, X., Goldsmith, C.~F., \& Tranter, R.~S. 2009, JPCA, 113, 8307

\bibitem[{Yang {et~al.}(2005)Yang, Zhang, Pei, Shao, W, \& Gao}]{2005YAN/ZHA}
Yang, Y., Zhang, W., Pei, S., {et~al.} 2005, JMoSt, 725, 133

\bibitem[{Yasunaga {et~al.}(2008)Yasunaga, Kubo, Hoshikawa, Kamesawa, \&
  Hidaka}]{2008YAS/KUB}
Yasunaga, K., Kubo, S., Hoshikawa, H., Kamesawa, T., \& Hidaka, Y. 2008,
  Int J Chem Kin, 40, 73

\bibitem[{Yee~Quee \& Thynne(1968)}]{1968YEE/THY}
Yee~Quee, M., \& Thynne, J. 1968, Berich Bunsen Gesell, 72, 211

\bibitem[{Yelle {et~al.}(2001)}]{Yelle2001}
Yelle, R.~V., Griffith, C.~A. \& Young, L.~A. 2001, Icarus, 152, 331

\bibitem[{Yelle {et~al.}(1996)}]{Yelle1996}
Yelle, R.~V., Young, L.~A., Vervack, R.~J., {et~al.} 1996, JGR, 101, 2149

\bibitem[{You {et~al.}(2007)You, Wang, Goos, Sung, \&
  Klippenstein}]{2007YOU/WAN}
You, X., Wang, H., Goos, E., Sung, C.-J., \& Klippenstein, S.~J. 2007, 
JCPA, 111, 4031

\bibitem[{Young(1958)}]{1958YOU}
Young, J. 1958, J Chem Soc, 2909

\bibitem[{Young {et~al.}(1971)Young, Lee-Ruff, \& Bohme}]{2-6002}
Young, L.~B., Lee-Ruff, E., \& Bohme, D. 1971, CaJCh, 49, 979

\bibitem[{Yumura \& Asaba(1981)}]{1981YUM/ASA}
Yumura, M., \& Asaba, T. 1981 in International Symposium on Combustion (Elsevier), 863

\bibitem[{Yung {et~al.}(1984)Yung, Allen, \& Pinto}]{Yung1984}
Yung, Y.~L., Allen, M., \& Pinto, J.~P. 1984, ApJS, 55, 465

\bibitem[{{Yung} \& {Demore}(1999)}]{Yung1999}
{Yung}, Y.~L., \& {Demore}, W.~B., eds. 1999, {Photochemistry of Planetary Atmospheres}
(New York: OUP)

\bibitem[{Zabarnick \& Heicklen(1985)}]{1985ZAB/HEI}
Zabarnick, S., \& Heicklen, J. 1985, Int J Chem Kin, 17, 455

\bibitem[{Zahnle(1986)}]{Zahnle1986}
Zahnle, K. 1986, JGR: Atmospheres, 91, 2819

\bibitem[{Zahnle {et~al.}(2006)Zahnle, Claire, \& Catling}]{Zahnle2006}
Zahnle, K., Claire, M., \& Catling, D. 2006, Geobiology, 4, 271

\bibitem[{Zahnle {et~al.}(1995)Zahnle, Mac~Low, Lodders, \&
  Fegley}]{Zahnle1995}
Zahnle, K., Mac~Low, M.-M., Lodders, K., \& Fegley, B. 1995, GeoRL, 22, 1593

\bibitem[{{Zahnle} {et~al.}(2009){Zahnle}, {Marley}, {Freedman}, {Lodders}, \&
  {Fortney}}]{Zahnle2009}
{Zahnle}, K., {Marley}, M.~S., {Freedman}, R.~S., {Lodders}, K., \& {Fortney},
  J.~J. 2009, \apjl, 701, L20

\bibitem[{Zalotai {et~al.}(1983)Zalotai, Hunyadi-zolt{\'a}n, B{\'e}rces, \&
  Marta}]{1983ZAL/HUN}
Zalotai, L., Hunyadi-zolt{\'a}n, Z., B{\'e}rces, T., \& Marta, F. 1983,
  Int J Chem Kin, 15, 505

\bibitem[{{Zarka} \& {Pedersen}(1986)}]{Zarka1986}
{Zarka}, P., \& {Pedersen}, B.~M. 1986, \nat, 323, 605

\bibitem[{Zaslonko {et~al.}(1997)Zaslonko, Petrov, \& Smirnov}]{1997ZAS/PET}
Zaslonko, I., Petrov, Y.~P., \& Smirnov, V. 1997, Kinetics and catalysis, 38,
  321

\bibitem[{Zaslonko {et~al.}(1993)Zaslonko, Smirnov, \& Tereza}]{1993ZAS/SMI}
Zaslonko, I., Smirnov, V., \& Tereza, A. 1993, Kin Catal, 34, 531

\bibitem[{Zel'dovich \& Raizer(1966)}]{Zeldovich1966}
Zel'dovich, Y.~B., \& Raizer, Y.~P. 1996, Physics of Shock Waves and 
High Temperature Hydrodynamic Phenomena, (New York, NY: Academic Press)

\bibitem[{Zelikoff {et~al.}(1953)Zelikoff, Watanabe, \& Inn}]{Zelikoff1953}
Zelikoff, M., Watanabe, K., \& Inn, E.~C. 1953, JChPh, 21, 1643

\bibitem[{Zhang \& Bauer(1997)}]{1997ZHA/BAU}
Zhang, Y.-X., \& Bauer, S. 1997, JPCB, 101, 8717

\bibitem[{Zhang {et~al.}(2004)Zhang, Zhang, \& Li}]{2004ZHA/ZHA}
Zhang, Y.-X., Zhang, S., \& Li, Q.~S. 2004, CP, 296, 79

\bibitem[{Zhang {et~al.}(2005)Zhang, Zhang, \& Li}]{2005ZHA/ZHA}
---. 2005, CP, 308, 109

\bibitem[{Zhu \& Lin(2003)}]{2003ZHU/LIN}
Zhu, R., \& Lin, M. 2003, JChPh, 119, 10667

\bibitem[{Zhu \& Lin(2007)}]{2007ZHU/LIN}
---. 2007, JCPA, 111, 6766

\bibitem[{Zhu \& Lin(2009)}]{2009ZHU/LIN}
---. 2009, CPL, 478, 11

\bibitem[{Zhu \& Lin(2005)}]{2005ZHU/LIN}
Zhu, R., \& Lin, M.-C. 2005, Int J Chem Kin, 37,
  593

\bibitem[{Zielinska \& Wincel(1970)}]{0-026}
Zielinska, T.~J., \& Wincel, H. 1970, Nukleonika, 15, 343

\bibitem[{Zsom {et~al.}(2012)}]{Zsom2012}
Zsom, A., Kaltenegger, L., \& Goldblatt, C. 2012, Icarus, 221, 603


\end{thebibliography}

\clearpage

\begin{table*}
\centering
\caption{Initial Conditions for the Chemistry at the Lower Boundary in terms of $n(X)/n_{\rm gas}$}\label{tab:initial-conditions}
\begin{tabular}{lcccc}
Species & HD209458b\tablenotemark{a} & Earth\tablenotemark{b} & Early Earth\tablenotemark{c}  & Jupiter\tablenotemark{d} \\
\hline\hline
H       & 9.2092(-1)  &             &              &             \\
He      & 7.8383(-2)  &             &              & 1.3600(-1)  \\
C       & 2.4787(-4)  &             &              &             \\
N       & 6.2262(-5)  &             &              &             \\
O       & 4.5105(-4)  &             &              &             \\
Ar      & 2.3133(-6)  &             & 9.1150(-3)   &             \\
K       & 9.8766(-8)  &             &              &             \\
Cl      & 2.9122(-7)  &             &              &             \\
Fe      & 2.9122(-5)  &             &              &             \\
Mg      & 3.6663(-5)  &             &              &             \\
Na      & 1.6004(-6)  &             &              &             \\
Si      & 2.9800(-5)  &             &              &             \\
Ti      & 8.2077(-8)  &             &              &             \\
CO      &             & 1.1300(-7)  & 4.9005(-5)   & 8.0000(-10) \\
H$_2$   &             & 1.0000(-6)  & 9.8010(-4)   & 8.6219(-1)  \\
N$_2$   &             & 7.9172(-1)  & 7.8408(-1)   &             \\
NO      &             & 2.4000(-11) &              &             \\
O$_2$   &             & 1.9793(-1)  &              &             \\
CO$_2$  &             & 3.5000(-4)  & 1.9602(-1)   &             \\
H$_2$O  &             & 1.0000(-2)  & 9.8010(-3)   &             \\
N$_2$O  &             & 3.0200(-7)  &              &             \\
NH$_3$  &             & 2.4000(-10) &              &             \\
CH$_4$  &             & 1.9390(-6)  &              & 1.8100(-3)  \\
\end{tabular}
\tablenotetext{a}{$\,$Solar metallicity, from \citet{Asplund2009}.}
\tablenotetext{b}{$\,$Surface mixing ratios based on the US Standard Atmosphere 1976.}
\tablenotetext{c}{$\,$Based on Early Earth models \citep{Kasting1993}.}
\tablenotetext{d}{$\,$\citet{Moses2005}.}
\end{table*}

\begin{table}
\centering
\caption{Mixing Ratios for Laboratory Simulations\tablenotemark{*}}\label{tab:glycine}
\begin{tabular}{lcccccc}
Model & H$_2$O & CH$_4$ & NH$_3$ & H$_2$ & CO$_2$ & N$_2$ \\
\hline
A& 0.80  & 0.08   & 0.08   & 0.04  & 0.0    & 0.0   \\
B& 0.80  & 0.06   & 0.06   & 0.04  & 0.02   & 0.02  \\
C& 0.80  & 0.04   & 0.04   & 0.04  & 0.04   & 0.04  \\
D& 0.80  & 0.02   & 0.02   & 0.04  & 0.06   & 0.06  \\
E& 0.84  & 0.00   & 0.00   & 0.00  & 0.08   & 0.08  \\
\end{tabular}
\tablenotetext{*}{$\,$Not including the injected FeO and HCOOH}
\end{table}

\begin{table*}
\caption{Bond Constants for Benson Additivity} \label{tab:Benson}
\begin{tabular}{lccccccc} 
{\bf Species} & $a_1$    & $a_2$    & $a_3$    & $a_4$    & $a_5$    & $a_6$    & $a_7$     \\
\hline\hline
C--H          &          &          &          &          &          &          &           \\
\hline
 300-1000K    & 1.1E0    & -3.0E-3  & 1.2E-5   & 5.8E-8   & 4.8E-12  & -2.5E3   & -2.5E-2   \\
1000-6000K    & 4.0E-1   & 2.3E-3   & -1.9E-6  & 1.4E-10  & -8.3E-15 & -2.0E3   & 2.8E0     \\
C--C          &          &          &          &          &          &          &           \\
\hline
 300-1000K    & -2.4E0   & 1.3E-3   & -1.4E-5  & -2.8E-7  & -5.0E-13 & 3.0E3    & 3.0E0     \\
1000-6000K    & 1.8E0    & 1.5E-3   & 5.7E-7   & 4.0E-11  & -2.5E-15 & 2.5E2    & -1.8E1    \\
C--O          &          &          &          &          &          &          &           \\
\hline
 300-1000K    & -6.8E-1  & 8.3E-3   & -1.0E-5  & -2.0E-7  & -2.3E-12 & 9.0E2    & 4.8E-1    \\
1000-6000K    & 1.7E0    & 1.3E-3   & 8.3E-6   & 5.5E-11  & -3.8E-15 & -6.5E2   & -1.1E1    \\
O--H          &          &          &          &          &          &          &           \\
\hline
 300-1000K    & 5.0E-2   & -7.3E-3  & 1.8E-5   & 1.2E-7   & 8.5E-12  & -9.0E3   & -4.2E0    \\
1000-6000K    & -4.5E-1  & -3.1E-3  & -6.7E-6  & -1.6E10  & -5.4E-14 & -7.5E3   & 5.2E0     \\
\end{tabular}
\end{table*}

\begin{deluxetable}{cclr}
\tablewidth{0pt}
\tablecaption{Neutral Species included in the \textsc{Stand2015} Network\label{tab:list-of-species}}
\tablehead{
\colhead{Network Formula} & \colhead{Standard Formula} & \colhead{Name} & \colhead{Thermochem Data}
}
\startdata
H                  & H                   & Atomic hydrogen              &   Burcat\\
C                  & C                   & Atomic carbon                &   Burcat\\
C($^1$D)           & C($^1$D)            & Singlet D carbon             &   Burcat\\
C($^1$S)           & C($^1$S)            & Singlet S carbon             &   Burcat\\
N                  & N                   & Atomic nitrogen              &   Burcat\\
O                  & O                   & Atomic oxygen                &   Burcat\\
O($^1$D)           & O($^1$D)            & Singlet D oxygen             &   Burcat\\
O($^1$S)           & O($^1$S)            & Singlet S oxygen             &   Burcat\\
\enddata
\end{deluxetable}

\begin{table}
\caption{List of ions included the \textsc{Stand2015} Network \label{tab:list-of-ions}}
\begin{tabular}{ccccccc}
\hline\hline
e$^-$ & C$^+$ & C$^-$ & H$^+$ & H$^-$ & K$^+$ & N$^+$\\ 
O$^+$ & O$^+$($^2$D) & O$^+$($^3$P) & O$^-$ & Ar$^+$ & Cl$^+$ & C$_{2}^+$\\ 
C$_{2}^-$ & C$_{3}^+$ & C$_{4}^+$ & CH$^+$ & CN$^+$ & CN$^-$ & CO$^+$\\ 
Fe$^+$ & H$_{2}^+$ & H$_{3}^+$ & HN$^+$ & HO$^+$ & HO$^-$ & He$^+$\\ 
Mg$^+$ & Na$^+$ & N$_{2}^+$ & N$_{3}^+$ & N$_{3}^-$ & NO$^+$ & NO$^-$\\ 
O$_{2}^+$ & O$_2^+$(X$^2\Pi_g$) & O$_{2}^-$ & O$_{3}^-$ & Si$^+$ & Ti$^+$ & Ar$_{2}^+$\\ 
ArH$^+$ & C$_{2}$H$^+$ & C$_{2}$H$^-$ & C$_{2}$N$^+$ & C$_{2}$O$^+$ & C$_{3}$H$^+$ & C$_{3}$N$^+$\\ 
C$_{3}$O$^+$ & C$_{4}$H$^+$ & C$_{5}$N$^+$ & CH$_{2}^+$ & CH$_{3}^+$ & CH$_{4}^+$ & CH$_{5}^+$\\
\hline
\end{tabular}
\end{table}

\begin{deluxetable}{cclcccc}
\tablewidth{0pt}
\tablecaption{\textsc{Stand2015} Chemical Kinetics Network \label{tab:STAND-network}}
\tablehead{
\colhead{\#} & \colhead{Type} & \colhead{Reaction} & \colhead{$\alpha$} & \colhead{$\beta$} & \colhead{$\gamma$} & \colhead{Ref}
}
\startdata
1    & 2d       & C$_{2}$ $\leftrightharpoons$ C + C                                                                & 2.49e-08 & 0.00    & 71600              & 1 \\
2    & 2d       & C$_{2}$ $\leftrightharpoons$ C + C                                                                & 6.01e+11 & -1.00   & 71600              & \nodata \\
3    & 2d       & CH $\leftrightharpoons$ C + H                                                                     & 3.16e-10 & 0.00    & 33700              & 2 \\
4    & 2d       & CH $\leftrightharpoons$ C + H                                                                     & 7.63e+09 & -1.00   & 33700              & \nodata \\
5    & 2d       & CN $\leftrightharpoons$ C + N                                                                     & 1.00e-09 & 0.00    & 71000              & 3 \\
6    & 2d       & CN $\leftrightharpoons$ C + N                                                                     & 2.42e+10 & 0.00    & 71000              & \nodata \\
7    & 2d       & CO $\leftrightharpoons$ C + O                                                                     & 1.52e-04 & -3.10   & 129000             & 4 \\
8    & 2d       & CO $\leftrightharpoons$ C + O                                                                     & 3.67e+15 & -4.10   & 129000             & \nodata \\
\enddata
\end{deluxetable}

\end{document}